\documentclass[preprint2]{aastex63}
\usepackage[utf8]{inputenc}
\usepackage{graphicx}
\usepackage{float}
\usepackage{subfigure}
\usepackage{CJK}
\usepackage{amsmath}

\accepted{to the Monthly Notices of the Royal Astronomical Society}

\shortauthors{L.-Y. Lu et al.}
\shorttitle{eDIG of CHANG-ES galaxies}

\begin{document}
\begin{CJK*}{UTF8}{gbsn}

\title{eDIG-CHANGES I: Extended H$\alpha$ Emission from the Extraplanar Diffuse Ionized Gas (eDIG) around CHANG-ES Galaxies}

\author[0000-0002-3286-5346]{Li-Yuan Lu (芦李源)}
\affiliation{Department of Astronomy, Xiamen University, 422 Siming South Road, Xiamen, Fujian, People’s Republic of China}

\author[0000-0001-6239-3821]{Jiang-Tao Li (李江涛)}
\affiliation{Purple Mountain Observatory, Chinese Academy of Sciences, 10 Yuanhua Road, Nanjing 210023, People’s Republic of China}
\affiliation{Department of Astronomy, University of Michigan, 311 West Hall, 1085 S. University Ave, Ann Arbor, MI, 48109-1107, U.S.A.}

\author[0000-0001-7936-0831]{Carlos J. Vargas}
\affiliation{Department of Astronomy and Steward Observatory, University of Arizona, Tucson, AZ, U.S.A.}

\author{Rainer Beck} 
\affiliation{Max-Planck-Institut f$\ddot{u}$r Radioastronomie, Auf dem H$\ddot{u}$gel 69, 53121 Bonn, Germany}

\author[0000-0001-6276-9526]{Joel N. Bregman}
\affiliation{Department of Astronomy, University of Michigan, 311 West Hall, 1085 S. University Ave, Ann Arbor, MI, 48109-1107, U.S.A.}

\author[0000-0001-8206-5956]{Ralf-J\"{u}rgen Dettmar}
\affiliation{Ruhr\, University Bochum, Faculty of Physics and Astronomy, Astronomical Institute, D-44780 Bochum, Germany}

\author{Jayanne English}
\affiliation{Department of Physics and Astronomy, University of Manitoba, Winnipeg, Manitoba R3T 2N2, Canada}

\author[0000-0002-2853-3808]{Taotao Fang (方陶陶)}
\affiliation{Department of Astronomy, Xiamen University, 422 Siming South Road, Xiamen, Fujian, People’s Republic of China}

\author{George H. Heald}
\affiliation{CSIRO, Space and Astronomy, PO Box 1130, Bentley, WA 6102, Australia}

\author[0000-0002-1253-2763]{Hui Li}
\altaffiliation{NASA Hubble Fellow}
\affiliation{Department of Astronomy, Columbia University, New York, NY 10027, U.S.A.}

\author[0000-0002-2941-646X]{Zhijie Qu}
\affiliation{Department of Astronomy \& Astrophysics, The University of Chicago, Chicago, IL 60637, U.S.A.}

\author[0000-0003-2048-4228]{Richard J. Rand}
\affiliation{Department of Physics and Astronomy, University of New Mexico, 210 Yale Blvd NE, Albuquerque, NM 87106, U.S.A.}

\author{Michael Stein}
\affiliation{Ruhr-University Bochum, Faculty of Physics and Astronomy, Astronomical Institute, D-44780 Bochum, Germany}

\author[0000-0002-9279-4041]{Q. Daniel Wang}
\affiliation{Department of Astronomy, University of Massachusetts, Amherst, MA 01003, U.S.A.}

\author{Jing Wang (王菁)}
\affiliation{Kavli Institute for Astronomy and Astrophysics, Peking University, Beijing 100871, People’s Republic of China}

\author[0000-0002-3502-4833]{Theresa Wiegert}
\affiliation{Instituto de Astrof$\acute{i}$sica de Andaluc$\acute{i}$a (IAA-CSIC), Glorieta de la Astronom$\acute{i}$a, 18008, Granada, Spain}

\author{Yun Zheng (郑\hbox{\kern-.2em\scalebox{0.8}[1]{氵}\kern-.3em\scalebox{0.8}[1]{云}})}
\affiliation{Research Center for Intelligent Computing Platforms, Zhejiang Laboratory, Hangzhou 311100, People’s Republic of China}
\affiliation{Kavli Institute for Astronomy and Astrophysics, Peking University, Beijing 100871, People’s Republic of China}

\correspondingauthor{Jiang-Tao Li}
\email{pandataotao@gmail.com}

\begin{abstract}
The extraplanar diffuse ionized gas (eDIG) represents the cool/warm ionized gas reservoir around galaxies. We present a spatial analysis of H$\alpha$ images of 22 nearby edge-on spiral galaxies from the CHANG-ES sample (the eDIG-CHANGES project), taken with the APO~3.5m telescope, in order to study their eDIG. We conduct an exponential fit to the vertical H$\alpha$ intensity profiles of the sample galaxies, of which 16 can be decomposed into a thin disk plus an extended thick disk component. The median value of the H$\alpha$ scale height of the extended component is $\langle h_{\rm H\alpha}\rangle =1.13\pm 0.14$ kpc. We further examine the dependence of $h_{\rm H\alpha}$ on the stellar mass, star formation rate (SFR), and SFR surface density (${\rm SFR_{SD}}$) of the galaxies. We find a tight sublinear correlation between $h_{\rm H\alpha}$ and the SFR, expressed in $h_{\rm H\alpha} \propto {\rm SFR}^\alpha$, where $\alpha \approx 0.29$. Moreover, the offset of individual galaxies from the best-fit SFR-$h_{\rm H\alpha}$ relation, expressed in $h_{\rm H\alpha}/{\rm SFR}^\alpha$, shows significant anti-correlation with ${\rm SFR_{SD}}$. This indicates that galaxies with more intense star formation tend to have disproportionately extended eDIG compared to those with less intense star formation. Combined with data from the literature, we also find that the correlations between the eDIG properties and the galaxies' properties extend to broader ranges. We further compare the vertical extension of the eDIG to multi-wavelength measurements of other circum-galactic medium (CGM) phases. We find the eDIG to be slightly more extended than the neutral gas traced by the \ion{H}{1} $21$~cm line. The H$\alpha$ emission is affected by the distribution of both the neutral gas and the ionizing photons, so the slightly more extended eDIG indicates the existence of some extended ionizing sources, in addition to the leaking photons from the disk star formation regions, possibly the UV background. Most galaxies have an X-ray scale height smaller than the H$\alpha$ scale height, suggesting that the majority of the X-ray emission detected in shallow observations are actually from the thick disk instead of the extended CGM. The H$\alpha$ scale height is comparable to the L-band radio continuum scale height, both slightly larger than that at higher frequencies (C-band), where the cooling is stronger and the thermal contribution may be larger. The comparable H$\alpha$ and L-band scale height indicates that the thermal and non-thermal electrons have similar spatial distributions, a natural result if both of them are transported outwards by a galactic wind. This further indicates that the thermal gas, the cosmics rays (CRs), and the magnetic field may be close to energy equipartition.
\end{abstract}


\section{Introduction} \label{sec:Intro}


The extraplanar diffuse ionized gas (eDIG) plays a key role in the recycling of gas between the galactic disk and halo \citep{Rand96, Rossa00, Haffner09, Putman12, Levy19}.
The eDIG has been observed in the Milky Way (known as Reynolds layer, e.g., \citealt{Reynolds84}) and many other late-type galaxies (e.g., \citealt{Dettmar90, Rand90, Hoopes96, Hoopes99, Collins01}).
This gas component typically has a characteristic electron density of~$\sim 0.1\ {\rm cm^{-3}}$, a temperature of~$8000-10000$ K, and radiates strong emission in optical emission lines such as H$\alpha$ \citep{Reynolds85, Haffner99}.


The eDIG could be ionized via various mechanisms. First, it could be photo-ionized by OB stars from star formation regions embedded in the galactic disk (e.g., \citealt{Reynolds84, Zurita02, Wood04, Haffner09, Barnes15}) or an extended Hot Low-mass Evolved Stars (HOLMES) component (e.g., \citealt{FloresFajardo11}). Second, it could be shock ionized by disk outflows, with additional turbulent mixing with the preexisting cold gas envelope or the entrained cold gas clouds (e.g., \citealt{Chevalier85,Slavin93,Binette09}). Third, the diffuse H$\alpha$ emission we observe today may also originate from the gas that was photo-ionized in the past but is currently cooling and recombining (e.g., \citealt{Dong11}). Fourth, the observed extended H$\alpha$ emission around the disk may also be partialy contributed to by the dust-scattered light of the escaping photons from disk regions with star formation (e.g., \citealt{Ferrara96, Wood99, Dong11}).


The eDIG is ubiquitous in actively star forming (SF) galaxies \citep{Lehnert95, Rossa00}.
In most cases, the vertical extension of the global diffuse H$\alpha$ emission of the eDIG, characterized by the exponential scale height, is typically only $\sim 1-2$ kpc \citep{Rossa03, Jo18, Levy19}.
In some galaxies, the eDIG has been detected even up to $\sim 5 {\rm~kpc}$ from the galactic plane (e.g., NGC~891; \citealt{Rand97}), and some large scale (mostly filamentary) emission line structures extend up to a few tens of kpc above the disk (e.g., \citealt{HodgesKluck20}).
Morphologically, eDIG around disk galaxies could be classified into at least three components: (1) a diffuse ionized gas envelope surrounding the galaxy, with a typical vertical extension of a few kpc; (2) some $\sim {\rm~kpc}$-scale fine structures at the disk-halo interface such as filaments or bubbles; and (3) some large-scale (typically $>10 {\rm~kpc}$) filamentary structures connecting the galaxy to the environment.


There are various methods for studying the eDIG around nearby galaxies, including, but not limited to, narrow-band imaging covering some prominent emission lines (e.g., \citealt{Rossa00,Rossa03}), long-slit spectroscopy (e.g., \citealt{Collins01,Rand97,Boettcher19}), integral field unit (IFU; e.g., \citealt{Boettcher16}), or imaging Fabry-Perot spectroscopy observations (e.g., \citealt{Heald06}), etc.
Our group is conducting a systematic study of the eDIG around the CHANG-ES galaxies (the eDIG-CHANGES project; this will be introduced in \S\ref{subsec:Sample}).
While our multi-slit spectroscopy observations will be presented in future papers, here we focus on the initial APO 3.5m H$\alpha$ imaging observations of the sample galaxies (original data obtained from \citealt{Vargas19}).
We estimate the scale height of the vertical H$\alpha$ intensity profiles of 22 nearby edge-on galaxies and examine the dependence of this vertical extension on various galactic parameters.
We introduce the data analysis in \S\ref{sec:Method}, then compare our H$\alpha$ measurements to multi-wavelength galaxy properties and other samples, and discuss the results in \S\ref{sec:Discussion}.
The major conclusions are summarized in \S\ref{sec:Summary}.
Throughout the paper, we quote the errors at 1$\sigma$ confidence level.

\section{Spatial Analysis of the H$\alpha$ Images}\label{sec:Method}

\subsection{Sample}\label{subsec:Sample}

Our sample is comprised of 22 galaxies selected from a comprehensive radio continuum and multi-wavelength survey of 35 nearby edge-on spiral galaxies (CHANG-ES: \emph{Continuum Halos in Nearby Galaxies - an EVLA Survey}; \citealt{Irwin12a,Irwin12b,Wiegert15}). The H$\alpha$ narrow-band images of 25 CHANG-ES galaxies were taken with the Apache Point Observatory (APO) $3.5$m Telescope and the Astrophysical Research Consortium Telescope Imaging Camera (ARCTIC), with detailed descriptions of the observation and data reduction procedure presented in \citet{Vargas19}. In the present paper, our aim is to  characterize the vertical extension of the eDIG, so we have excluded three galaxies with obviously distorted disks in their H$\alpha$ images (NGC~660, NGC~2992, NGC~4438). The vignetting correction and/or background subtraction of these H$\alpha$ images still have some uncertainties, which may affect quantitative measurements of the flux of large-scale features. Therefore, we adopt the H$\alpha$ images from \citet{Vargas19} without a flux calibration, in order to avoid any misleading results on flux measurements of large-scale features. The spatial analysis presented in this paper will not be affected, however. The surface brightness sensitivity of these H$\alpha$ images is in the range of $2.06-74.2\times 10^{-18} {\rm\ erg\ s}^{-1}{\rm\ cm}^{-2}{\rm\ arcsec}^{-2}$, with a median value of $8.81\times 10^{-18} {\rm\ erg\ s}^{-1}{\rm\ cm}^{-2}{\rm\ arcsec}^{-2}$.

We take the redshift independent distance ($d$), the optical diameter ($D_{25}$), and the inclination angle ($i$) from \citet{Irwin12a}; the stellar mass ($M_{\ast}$) and rotation velocity ($v_{\rm rot}$) from \citet{Li16}; the revised SFR (SFR$_{\rm H\alpha+22\mu m}$, based on a combination of the H$\alpha$ and WISE $22~{\rm \mu m}$ fluxes) and SFR surface density (SFR$_{\rm SD}$) from \citet{Vargas19}. The values of $M_{\ast}$ and SFR$_{\rm H\alpha+22\mu m}$ are calculated based on a scaled-down Salpeter initial mass function (IMF) \citep{Bell01} and a Kroupa IMF \citep{Murphy11}, respectively. All of the galaxies in our sample have high inclination angles ($i\geq 75^\circ$). The key parameters of the sample galaxies are summarized in Table \ref{gal_par}.

\begin{deluxetable*}{lcccccccccc}
\tablewidth{0pt}
\tabletypesize{\scriptsize}
\tablecaption{Galaxy Sample}
\startdata\\
Name & $d^{a}$ & $D_{25}^{b}$ & $M_{\ast}^{c}$ & SFR$_{\rm H\alpha +22\mu m}^{d}$ & SFR$_{\rm SD}^{d}$ & $v_{\rm rot}^{e}$ & $i^{f}$ & PA$^{g}$ & scale$^{h}$ \\
 & Mpc & arcmin & $\times 10^{10}~{\rm M_{\odot}}$ & ${\rm M_{\odot}~yr^{-1}}$ & $\times 10^{-3}~{\rm M_{\odot}~yr^{-1}~kpc^{-2}}$ & ${\rm km~s^{-1}}$ & deg & deg & ${\rm kpc~arcsec^{-1}}$ \\
\hline
NGC 2613 & 23.4 & 7.2 & 11.96$\pm$0.19 & 3.36$\pm$ 0.35 & 3.77$\pm$0.39 & 290.6 & 85 & 113 & 0.11 \\
NGC 2683 & 6.27 & 9.1 & 1.49$\pm$0.02 & 0.25$\pm$ 0.03 & 3.54$\pm$0.40 & 202.6 & 79 & 41.5 & 0.03 \\
NGC 2820 & 26.5 & 4.1 & 0.467$\pm$0.013 & 1.35$\pm$0.14 & 9.00$\pm$0.96 & 162.8 & 90 & 65 & 0.13 \\
NGC 3003 & 25.4 & 6 & 0.485$\pm$0.010 & 1.56$\pm$0.16 & 2.59$\pm$0.27 & 120.6 & 90 & 79 & 0.12 \\
NGC 3044 & 20.3 & 4.4 & 0.660$\pm$0.013 & 1.75$\pm$0.16 & 6.79$\pm$0.60 & 152.6 & 90 & 114 & 0.10 \\
NGC 3079$^*$ & 20.6 & 7.7 & 4.73$\pm$0.07 & 5.08$\pm$0.45 & 9.57$\pm$0.84 & 208.4 & 88 & 75 & 0.10 \\
NGC 3432$^*$ & 9.42 & 4.9 & 0.100$\pm$0.002 & 0.51$\pm$0.06 & 6.36$\pm$0.69 & 109.9 & 82 & 30 & 0.05 \\
NGC 3448 & 24.5 & 4.9 & 0.564$\pm$0.011 & 1.78$\pm$0.18 & 14.5$\pm$1.5 & 119.5 & 78 & 65 & 0.12 \\
NGC 3556$^*$ & 14.09 & 7.8 & 2.81$\pm$0.04 & 3.57$\pm$0.30 & 7.32$\pm$0.62 & 153.2 & 81 & 82.5 & 0.07 \\
NGC 3628 & 8.5 & 14.8 & 2.83$\pm$0.04 & 1.41$\pm$0.12 & 2.99$\pm$0.26 & 215.4 & 87 & 104 & 0.04 \\
NGC 3735 & 42.0 & 4 & 14.92$\pm$0.21 & 6.23$\pm$0.57 & 6.71$\pm$0.61 & 241.1 & 85 & 130 & 0.20 \\
NGC 3877 & 17.7 & 5.1 & 2.74$\pm$0.04 & 1.35$\pm$0.12 & 5.04$\pm$0.44 & 155.1 & 85 & 35 & 0.09 \\
NGC 4013 & 16.0 & 4.7 & 3.23$\pm$0.05 & 0.71$\pm$0.07 & 3.51$\pm$0.32 & 181.4 & 84 & 114 & 0.08 \\
NGC 4096$^*$ & 10.32 & 6.4 & 0.613$\pm$0.010 & 0.71$\pm$0.08 & 6.52$\pm$0.77 & 144.8 & 82 & 17.5 & 0.05 \\
NGC 4157 & 15.6 & 7 & 2.92$\pm$0.04 & 1.76$\pm$0.18 & 8.15$\pm$0.83 & 188.9 & 90 & 63.5 & 0.08 \\
NGC 4192$^*$ & 13.55 & 8.7 & 3.40$\pm$0.05 & 0.78$\pm$0.07 & 1.67$\pm$0.15 & 214.8 & 83 & 150 & 0.07 \\
NGC 4388 & 16.6 & 5.6 & 1.54$\pm$0.02 & 2.42$\pm$0.23 & 25.0$\pm$2.3 & 171.2 & 79 & 89.5 & 0.08 \\
NGC 4666$^*$ & 27.5 & 4.2 & 12.48$\pm$0.18 & 10.5$\pm$0.92 & 12.8$\pm$1.1 & 192.9 & 76 & 40 & 0.13 \\
NGC 4845 & 16.98 & 4.8 & 2.89$\pm$0.05 & 0.62$\pm$0.06 & 7.33$\pm$0.68 & 176.0 & 81 & 75 & 0.08 \\
NGC 5297 & 40.4 & 5.3 & 3.69$\pm$0.07 & 3.00$\pm$0.33 & 5.70$\pm$0.62 & 189.5 & 89 & 153 & 0.20 \\
NGC 5792 & 31.7 & 7.2 & 8.89$\pm$0.16 & 4.41$\pm$0.37 & 10.0$\pm$0.8 & 208.6 & 81 & 81.5 & 0.15 \\
UGC 10288 & 34.1 & 4.9 & 2.03$\pm$0.05 & 0.66$\pm$0.07 & 1.85$\pm$0.21 & 167.1 & 90 & 91 & 0.17 \\
\hline
\enddata
\label{gal_par}
\tablenotetext{*}{For these galaxies, when fitting the H$\alpha$ intensity profiles, we exclude the galactic disk area containing $90\%$ of the total emission on each side, instead of $70\%$ for other galaxies.}
\tablenotetext{a}{Redshift independent distance, from \citet{Irwin12a}.}
\tablenotetext{b}{Observed blue diameter at the 25$^{\rm th}$ mag arcsec$^{-2}$ isophote, from \citet{Irwin12a}.}
\tablenotetext{c}{Stellar mass calculated based on a scaled-down Salpeter IMF, from \citet{Li16}.}
\tablenotetext{d}{Star formation rate estimated by combining the H$\alpha$ and $22\mu\rm{m}$ data (SFR$_{\rm H\alpha +22\mu m}$) (adopting a Kroupa IMF) and the corresponding SFR surface density (SFR$_{\rm SD}$), from \citet{Vargas19}.}
\tablenotetext{e}{The maximum gas rotation velocity corrected for inclination, from \citet{Li16}.}
\tablenotetext{f}{Inclination angle, from \citet{Irwin12a}.}
\tablenotetext{g}{Position angle, obtained from NASA/IPAC Extragalactic Database (NED). The adopted values of PA for some galaxies are slightly adjusted, because the H$\alpha$ disk may not perfectly align with the stellar disk.}
\tablenotetext{h}{Physical scale of each galaxy, calculated from the adopted distance.}
\end{deluxetable*}

\subsection{Extraction and Calibration of the H$\alpha$ Vertical Intensity Profile}

We adopt the continuum-subtracted H$\alpha$ images from \citet{Vargas19} to extract the H$\alpha$ vertical intensity profile.
The first step is to detect and mask the optical bright point sources above the disk.
We use the {\tt sep} package \citep{Barbary16} and the python module of {\tt Source Extractor} \citep{Bertin96} to detect point sources in the broad-band images taken with a Sloan Digital Sky Survey (SDSS) \emph{r}-band filter.
These \emph{r}-band images are also adopted as the continuum filter images in \citet{Vargas19}
We then mask the positions of the detected sources on the H$\alpha$ images, which could be either foreground or background sources (stars or AGN), or local sources such as bright compact \ion{H}{2} regions.
In order to avoid over-masking the fine structures of the disk, we define an elliptical region covering the galactic disk within which we adopt a different source masking criterion.
The major axis of the elliptical region is $1.2 D_{\rm 25}$, while the minor axis is calculated in the same way as in \citet{Irwin12a},
\begin{equation}
    i=3~{\rm deg}+\cos^{-1}{\left( \sqrt{((b/a)^{2}-0.2^{2})/(1-0.2^{2})}\right)},
\end{equation}
where $b/a$ is the minor to major axis ratio.
Outside this elliptical region, we mask all the detected point sources, while inside it, we only mask the sources with a diameter less than 10 pixels (pixel size $0.228^{\prime\prime}$).
This strategy retains some extended H$\alpha$ structures, such as \ion{H}{2} regions, in the follow-up analysis. The masking process is conducted using the {\tt IRAF} tool {\tt imedit}  \citep{Tody86}.
We then fill in the masked holes using a linear interpolation of the surrounding pixels.
We also manually mask some strong sources that are not appropriately detected, such as some bright \ion{H}{2} regions and the luminous nucleus in NGC~4013.

The above procedure generally cleanly removes the contribution from bright compact sources.
However, the residual of the reflected or scattered light from bright compact sources may still affect the analysis of the low surface brightness extended emissions, especially when narrow-band filters are used \citep{Karabal17, Boselli18}.
We slightly modified the fitting models when some significant reflected or scattered feature is present.
For example, such an effect may cause the significantly elevated background level on one side of NGC~5297 (see the online only figures in the appendix).
We therefore only fit the H$\alpha$ profile of this galaxy with a single component exponential model, which may cause some biases.
This galaxy is separated from our main sample in the statistical analysis.
We present H$\alpha$ images and vertical intensity profiles in the Appendix, so readers could appreciate whether individual cases such features could potentially affect the results or not.

Before extracting the vertical profile, a world coordinate system (WCS) is assigned using {\tt IRAF}.
We combine the H$\alpha$ images of the galaxies with more than one observation field using {\tt Montage} \citep{Berriman03}.
We then rotate the H$\alpha$ images by the position angles (PA) of the galaxies.
We extract the H$\alpha$ vertical intensity profile of each galaxy along its minor axis.
The horizontal range used to calculate the average value of the intensity is $D_{25}/2$, centered at the mid-plane across the galactic center.
As an example, we present the H$\alpha$ vertical intensity profile and the area used to extract it for the case of NGC~3003 in Fig.~\ref{fig:N3003_result}.

The flat fielding of the H$\alpha$ images using the twilight flat seems imperfect \citep{Vargas19}, with a significant residual vignetting effect.
Therefore, we fit the background beside the galactic thick disk with a second-order polynomial model to account for the curvature of the background, using the curve fitting code {\tt LMFIT}\footnote{\url{https://lmfit.github.io/lmfit-py/}}.
This fitting accounts for both the sky background and a further correction of the flat field.
We choose different vertical ranges in the fitting for individual galaxies, excluding prominent H$\alpha$ emitting features mainly from the galactic disk.
The results are in general insensitive to the choice of this fitting range.
The background fitting results of individual galaxies are presented in the \hyperref[Appendix]{Appendix}, with one example shown in Fig.~\ref{fig:N3003_result}.
We then subtract the best-fit background and obtain the flattened intrinsic H$\alpha$ vertical profile of each galaxy.
We notice that in some cases the central region of the vertical profile (close to the disk) may have the background slightly over-subtracted due to the curvature of the second-order polynomial model (e.g., Fig.~\ref{fig:N3003_fit}).
This may affect the follow-up exponential fit of the H$\alpha$ intensity profile, especially the compact component describing the emission significantly affected by the galactic thin disk.
This is another reason for not including this component in the scientific analysis (\S\ref{subsec:HaScaleHeight}).
Furthermore, there may also be some additional uncertainties in continuum subtraction in this area, due to the color gradient of the underlying stellar population (e.g., \citealt{Spector12}).
However, as we are mainly interested in the extended eDIG distributions, we expect a simple mask of the disk area in the H$\alpha$ vertical intensity profile should be sufficient to minimize this effect.

\subsection{Exponential Fit to the H$\alpha$ Vertical Profile} \label{subsec:HaScaleHeight}

\begin{figure*}
    \centering
    \subfigure[]{
        \includegraphics[width=0.401\textwidth]{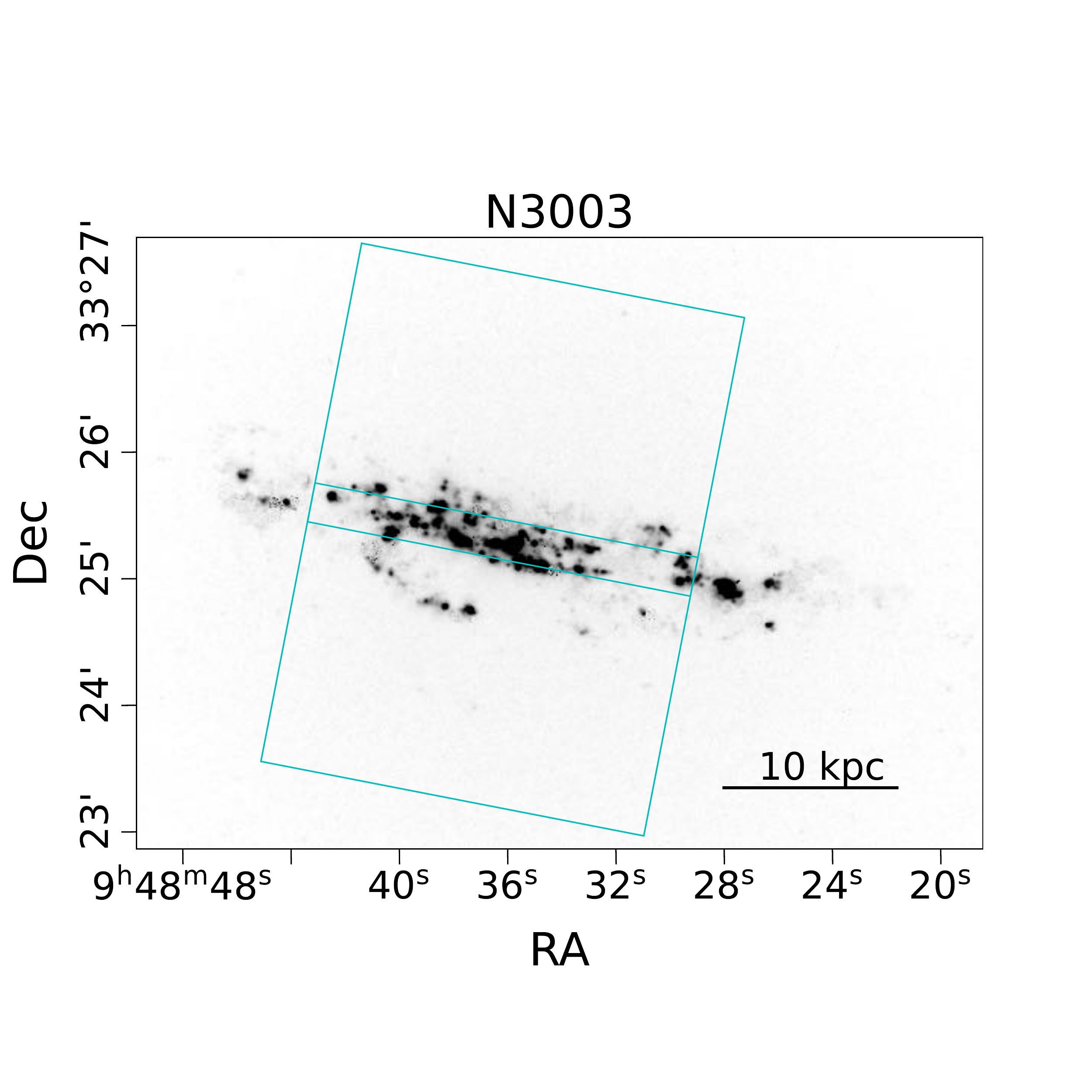}
        \label{fig:N3003_img}
        }
    \subfigure[]{
        \includegraphics[width=0.561\textwidth]{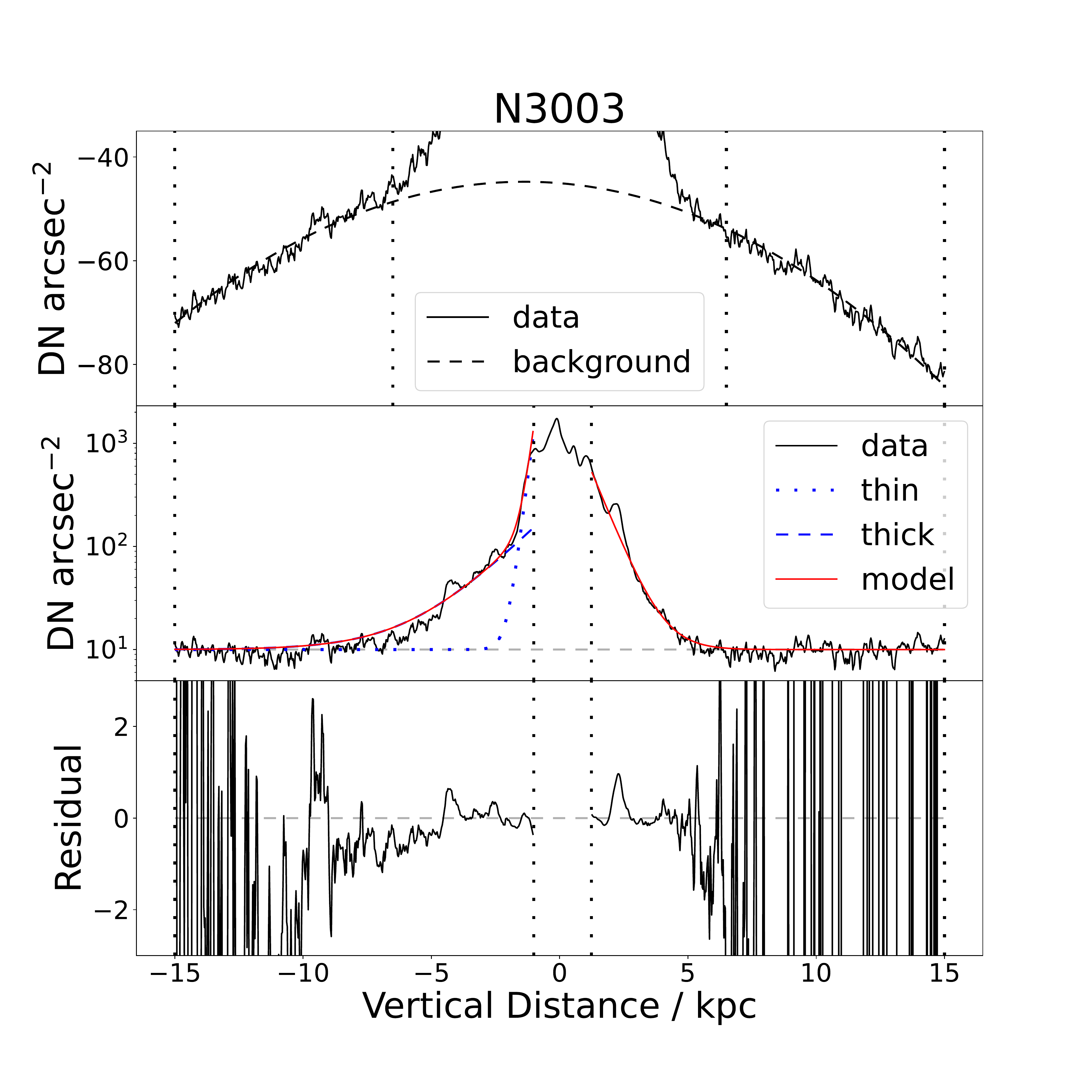}
        \label{fig:N3003_fit}
        }
    \caption{(a) H$\alpha$ image of NGC~3003 observed with the APO 3.5m telescope.
    The cyan boxes illustrate the exponential fitting region.
    The profile is averaged along the major axis.
    The central region along the disk is excluded from the fitting due to high extinction.
    (b) Exponential fitting of the vertical H$\alpha$ intensity profile (with digital number arcsec$^{-2}$ unit)
    \emph{Top:} background analysis.
    The solid black curve is the zoomed-in H$\alpha$ profile to better show the background.
    The background is fitted with a second-order polynomial model (the dashed black curve).
    The vertical dotted lines indicate the fitting region boundary.
    \emph{Middle:} Exponential fit to the background-subtracted H$\alpha$ intensity profile (black solid curve).
    The red solid curve is the best-fit exponential model, with the two components (if available) shown as the blue dotted (thin component) and dashed (thick component) curves.
    The horizontal grey dashed line indicates the zero level with a $10$ kpc offset.
    \emph{Bottom:} residual of the best-fit defined as $(data - model) / model$.
    \label{fig:N3003_result}}
\end{figure*}

Based on previous studies of the eDIG \citep{Rand96, Rand97, Boettcher16, Boettcher19, Reach20}, we need multiple components to describe the distribution of the extraplanar H$\alpha$ emission.
In particular, we fit the profiles above and below the galactic disk separately, to account for the asymmetry of the eDIG.
We need to mask the disk since it could be highly affected by emission from the \ion{H}{2} regions and extinction by the dust lane.
Therefore, we block a region typically containing $70\%$ of the total H$\alpha$ emission on each side of the disk.
For some of the galaxies in the sample, which have unusually extended disk emission or absorption features
(most of these are less inclined than the other galaxies, as listed in Table \ref{gal_par}), we increase this threshold to $90\%$ to ensure that the eDIG component we are interested in will be well characterized by the exponential model(s).
In order to clarify the above choices of blocked disk regions ($70$ or $90\%$),
we plot them both on the images of individual galaxies and on the vertical intensity profiles (Fig.~\ref{fig:N3003_result}; the \hyperref[Appendix]{Appendix}).
In many cases, the selected fitting area still includes some disk structures such as the spiral arms (e.g., Fig.~\ref{fig:N3003_img}).
These structures could affect the fitting of the H$\alpha$ intensity profiles, especially the compact exponential component.
In order to further reduce the influence of the bright compact structures on the analysis of the diffuse extraplanar emission, we perform a Levenberg-Marquardt least-squares fit to minimize the scalar value converted from a negative log-likelihood function in a Cauchy distribution, which is $-\sum{\log(1/({\pi(1+r^{2})}))}$, where $r$ is the residual array.
This method reduces the weight of data points that have a large departure from the fitting model in the fitting process.
Those outlier points may be caused by, e.g., spiral arms or off-disk \ion{H}{2} regions that were not masked out. 

We fit the background-subtracted H$\alpha$ vertical intensity profiles with a double-exponential function using {\tt LMFIT},
\begin{equation}\label{eq:double_e}
    I_{\rm H\alpha}(z) = I_{\rm H\alpha ,1}(0)e^{-|z|/h_{z,1}} + I_{\rm H\alpha ,2}(0)e^{-|z|/h_{z,2}},
\end{equation}
where $z$ is the projected distance to the mid-plane of the galaxy, $I_{\rm H\alpha (,1,2)}(0)$ 
is the peak intensity of the H$\alpha$ emission (of each component) at $z=0$, and $h_{z,1(2)}$ is the scale height of each component.
In some cases when the second exponential component is not necessary, we adopt only a single exponential function instead.
There is no background component in the function because the background described with a polynomial function has already been subtracted before the fitting.
We set $h_{z,2}>h_{z,1}$, so the second component represents the diffuse thick envelope which will be discussed in the following sections, while the first component typically represents the residual of the galactic thin disk.
The best-fit values of $h_{z}$ of each galaxy are listed in Table~\ref{fit_res}.
The subscripts {\emph 1} and {\emph 2} denote the different components, while {\emph n} and {\emph p} denote different sides of the mid-plane.
As an example, we present the best-fit result of NGC~3003 in Fig.~\ref{fig:N3003_fit}, where we adopt double and single exponential functions on the two sides, respectively.
The figures for other galaxies are presented in the \hyperref[Appendix]{Appendix}.

\begin{deluxetable*}{lcccccccccc}
\tablewidth{0pt}
\tabletypesize{\scriptsize}
\tablecaption{Exponential fit results of the vertical H$\alpha$ intensity profiles and corresponding multi-wavelength parameters}
\startdata\\
Name & $h_{z,n,1}$ & $h_{z,n,2}$ & $h_{z,p,1}$ & $h_{z,p,2}$ & $h_{\rm H\alpha}$ & $h_{\rm HI}^{a}$ & $h_{\rm X}^{b}$ & $h_{\rm Cband}^{c}$ & $h_{\rm Lband}^{c}$ \\
 & kpc & kpc & kpc & kpc & kpc & kpc & kpc & kpc & kpc \\
\hline
NGC 2613 & $1.029\pm 0.129$ & - & $1.107\pm 0.024$ & - & - & - & - & - & - \\
NGC 2683 & $0.232\pm 0.024$ & $0.682\pm 0.089$ & $0.281\pm 0.019$ & $0.714\pm 0.076$ & $0.698\pm 0.059$ & $0.76\pm 0.28$ & - & - & - \\
NGC 2820 & $0.240\pm 0.005$ & $0.837\pm 0.034$ & $0.552\pm 0.030$ & - & $0.837\pm 0.034$ & - & - & $1.16\pm 0.07$ & $1.48\pm 0.19$ \\
NGC 3003 & $0.219\pm 0.033$ & $1.761\pm 0.229$ & $0.713\pm 0.025$ & - & $1.761\pm 0.229$ & $3.48\pm 2.85$ & - & $1.28\pm 0.25$ & $3.66\pm 1.19$ \\
NGC 3044 & $0.249\pm 0.008$ & $1.304\pm 0.148$ & $0.353\pm 0.014$ & $1.224\pm 0.213$ & $1.264\pm 0.130$ & $0.59\pm 0.20$ & - & $1.03\pm 0.11$ & $1.33\pm 0.16$ \\
NGC 3079 & $0.158\pm 0.011$ & $1.062\pm 0.017$ & $0.247\pm 0.024$ & $1.409\pm 0.105$ & $1.235\pm 0.053$ & $1.52\pm 1.02$ & $4.01\pm 0.21$ & $1.54\pm 0.04$ & $1.18\pm 0.19$ \\
NGC 3432 & $0.122\pm 0.002$ & $1.050\pm 0.038$ & $0.168\pm 0.028$ & $0.611\pm 0.084$ & $0.830\pm 0.046$ & - & - & $0.69\pm 0.21$ & $0.93\pm 0.25$ \\
NGC 3448 & $0.291\pm 0.030$ & $1.332\pm 0.289$ & $0.184\pm 0.008$ & $1.011\pm 0.057$ & $1.171\pm 0.147$ & - & - & - & - \\
NGC 3556 & $0.321\pm 0.007$ & $1.404\pm 0.092$ & $0.590\pm 0.009$ & - & $1.404\pm 0.092$ & $1.19\pm 0.35$ & $1.29\pm 0.07$ & - & - \\
NGC 3628 & $0.545\pm 0.002$ & - & $0.603\pm 0.005$ & - & - & - & $3.78\pm 0.69$ & - & - \\
NGC 3735 & $1.046\pm 0.040$ & - & $0.337\pm 0.040$ & $1.710\pm 0.185$ & $1.710\pm 0.185$ & - & - & $1.43\pm 0.23$ & $1.66\pm 0.45$ \\
NGC 3877 & $0.269\pm 0.022$ & $1.284\pm 0.461$ & $0.187\pm 0.003$ & $1.152\pm 0.034$ & $1.218\pm 0.231$ & $1.33\pm 0.22$ & $0.90\pm 0.07$ & $0.86\pm 0.16$ & $1.08\pm 0.17$ \\
NGC 4013 & $0.481\pm 0.033$ & $1.299\pm 0.125$ & $0.315\pm 0.006$ & $1.333\pm 0.037$ & $1.316\pm 0.065$ & - & $1.02\pm 0.15$ & $0.64\pm 0.05$ & $1.03\pm 0.14$ \\
NGC 4096 & $0.216\pm 0.017$ & $0.700\pm 0.163$ & $0.309\pm 0.007$ & - & $0.700\pm 0.163$ & $1.05\pm 0.25$ & - & - & - \\
NGC 4157 & $0.439\pm 0.006$ & - & $0.576\pm 0.009$ & - & - & $0.75\pm 0.28$ & - & $0.99\pm 0.10$ & $1.09\pm 0.07$ \\
NGC 4192 & $0.887\pm 0.047$ & - & $0.826\pm 0.037$ & - & - & - & - & - & - \\
NGC 4388 & $0.134\pm 0.002$ & $0.521\pm 0.006$ & $0.311\pm 0.011$ & $1.171\pm 0.103$ & $0.846\pm 0.052$ & - & $2.00\pm 0.15$ & - & - \\
NGC 4666 & $0.367\pm 0.006$ & $2.609\pm 0.031$ & $0.304\pm 0.005$ & $2.376\pm 0.033$ & $2.492\pm 0.023$ & - & $1.96\pm 0.32$ & - & - \\
NGC 4845 & $0.202\pm 0.026$ & $0.541\pm 0.045$ & $0.251\pm 0.006$ & $0.882\pm 0.024$ & $0.711\pm 0.026$ & - & - & - & - \\
NGC 5297 & $0.411\pm 0.000$ & - & $0.701\pm 0.051$ & - & - & - & - & - & - \\
NGC 5792 & $0.683\pm 0.009$ & - & $0.904\pm 0.107$ & - & - & - & - & - & - \\
UGC 10288 & $0.212\pm 0.037$ & $0.948\pm 0.030$ & $0.346\pm 0.041$ & $1.225\pm 0.053$ & $1.087\pm 0.030$ & $1.17\pm 0.07$ & - & $0.73\pm 0.02$ & $0.84\pm 0.08$ \\
\hline
\enddata
\label{fit_res}
\tablecomments{
The subscripts ``n'' and ``p'' denote the negative and positive sides beside the galactic midplane, while ``1'' and ``2'' denote the compact and extended components, respectively.
$h_{\rm H\alpha}$ is the average value of the negative and positive sides of the extended component, and is used for statistical analysis.}
\tablenotetext{a}{The average scale height of \ion{H}{1} $21$~cm emission, from \citet{Zheng22a}.}
\tablenotetext{b}{The average scale height of $0.5-1.5$~keV emission, from \citet{Li13a}.}
\tablenotetext{c}{The scale height of the emission in C and L bands, from \citet{Krause18}.}
\end{deluxetable*}

\section{Results and Discussions}\label{sec:Discussion}

In our exponential fit of the 22 sample galaxies, 16 have two components at least on one side.
The remaining six galaxies can be well fitted with only one exponential component.
We notice that the scale heights of these six galaxies ($0.41 - 1.11$~kpc) are somewhat 
comparable to the typical value of the extended component of other galaxies (median value $1.13\pm 0.14$~kpc).
The lack of the thin disk component may be caused by the strong extinction by the dust lane, which was filtered in the fitting.
Alternatively, the extended component of many moderately inclined galaxies may be highly affected by the disk structures, instead of the real extraplanar gas.
Regardless, for consistency, we will only include the extended component ($h_{z,2}$) in the following statistical analysis.
We will also separate the highly and moderately inclined galaxies in the plots for comparison.
For six galaxies where the extended component is detected only on one side (e.g., Fig.~\ref{fig:N3003_fit}), we directly adopt the measured $h_{z,2}$ on this side as the value for this galaxy.
The H$\alpha$ emission on the other side may be largely blocked by the galactic disk, especially when its inclination angle is relatively small.
For galaxies with the extended component detected on both sides, we adopt the average value of $h_{z,n,2}$ and $h_{z,p,2}$ in the following analysis.
The adopted $h_{z,2}$ values of the 16 galaxies calculated this way are in the range of $0.70-2.49$~kpc.
The median value of $h_{z,2}$ is $1.13\pm 0.14$~kpc, consistent with what has been found in previous works (typically $\sim 1-2$~kpc; e.g., \citealt{Rossa03}).

\begin{figure}
    \centering
    \hspace{-0.3cm}
    \includegraphics[width=0.48\textwidth]{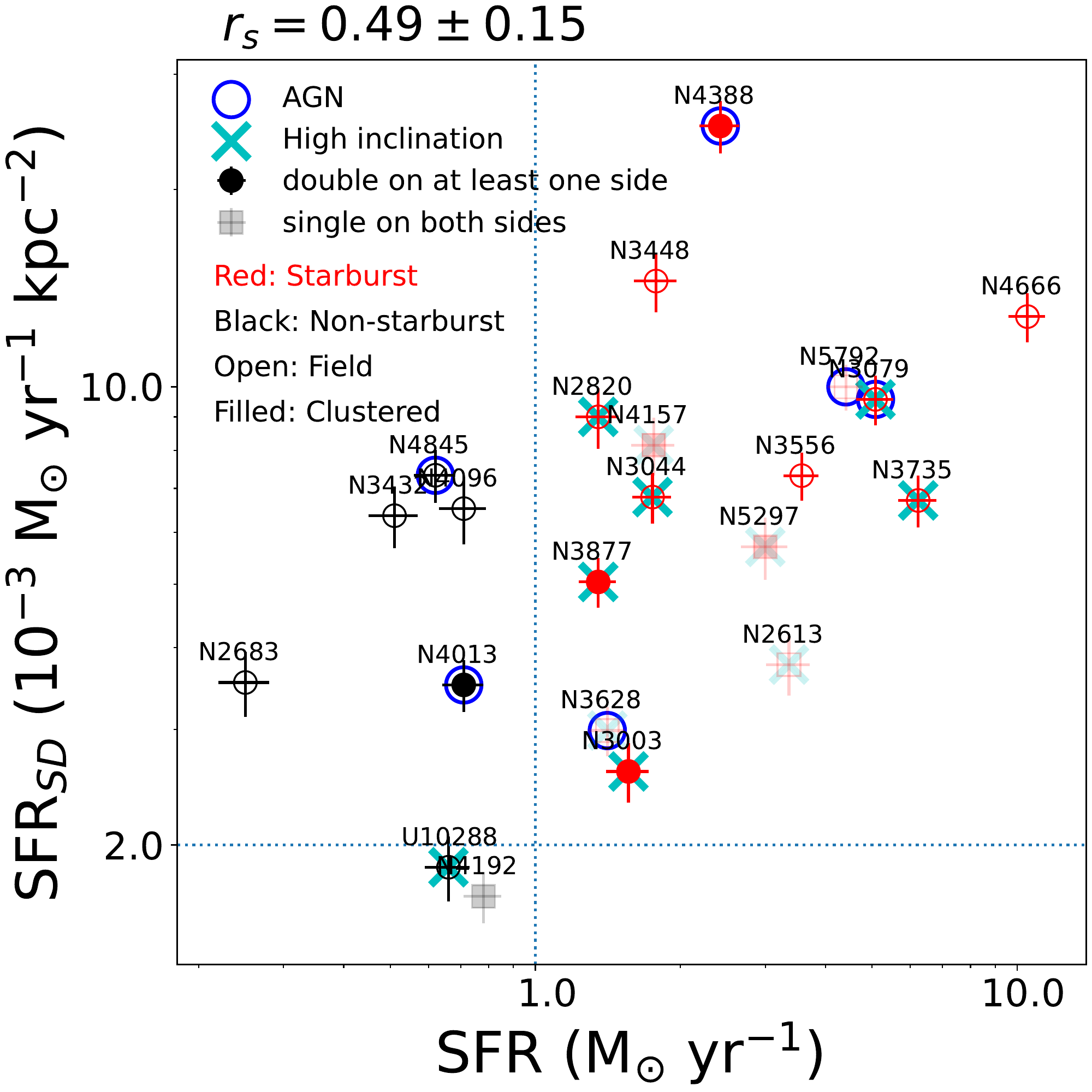}
    \caption{The SFR and the SFR surface density $\rm SFR_{SD}$ of all 22 sample galaxies.
    Different symbols represent different sub-samples: starburst (red) versus non-starburst (black), field galaxy (open) versus clustered galaxy (filled).
    The galaxies fitted with just one exponential component on both sides are plotted in square symbols with faded color.
    Highly inclined galaxies ($i\geqslant 85^{\circ}$) are marked by green X-shape symbols.
    The blue circles represent galaxies hosting an identified radio bright AGN.
    The two blue dotted lines (${\rm SFR}=1~{\rm M_{\odot}~yr^{-1}}$ and ${\rm SFR_{SD}}=0.002~{\rm M_{\odot}~yr^{-1}~kpc^{-2}}$) roughly separate starburst and non-starburst galaxies.
    See \citet{Li16} for details.
    \label{fig:SFRSDvsSFR}}
\end{figure}

Using a similar approach adopted in some other CHANG-ES papers \citep[e.g.,][]{Li16, Vargas19}, we divide the sample into starburst/non-starburst, and field/clustered subsamples.
We define the division of starburst/non-starburst galaxies based on their total SFR and ${\rm SFR_{SD}}$.
A starburst galaxy is defined to have ${\rm SFR}>1\ {\rm M_{\odot}\ yr^{-1}}$ and ${\rm SFR_{SD}}>0.002\ {\rm M_{\odot}\ yr^{-1}\ kpc^{-2}}$.
We define field/clustered galaxies using the local galaxy number density ($\rho$; \citealt{Irwin12a}), which is the density of galaxies brighter than -16~mag in the vicinity of the entry \citep{Tully88}.
Galaxies with $\rho <0.6\rm~Mpc^{-3}$ are classified as being in the field, while those with $\rho >0.6\rm~Mpc^{-3}$ are classified as clustered.
In the following analysis, we will plot these different galaxy subsamples with different colors or symbols (starburst: red; non-starburst: black; field: open symbols; clustered: filled symbols; see e.g., Fig. \ref{fig:SFRSDvsSFR}).
Furthermore, we mark the galaxies that have an identified active galactic nucleus (AGN) with a large blue circle (see e.g., \citealt{Li16,Irwin19}).

We use the Spearman’s rank order coefficient ($r_s$) to describe the goodness of a correlation (e.g., \citealt{Li13b}).
In the statistical analysis, we estimate the errors of the fitted or statistical parameters by applying the same analysis to 1000 bootstrap-with-replacement sampled data within their measurement errors.
Both the measurement error and the scatter of the sample are taken into account by simultaneously re-sampling both the distribution of the measured parameter values (assuming Gaussian distribution characterized with the errors) and the sample galaxies in the bootstrapping.
The same approach has been adopted and detailed in our previous works (e.g., \citealt{Li13b,Li16}).

\subsection{Contamination caused by the disk inclination}

\begin{figure*}
    \centering
    \subfigure[]{
        \includegraphics[width=0.49\textwidth]{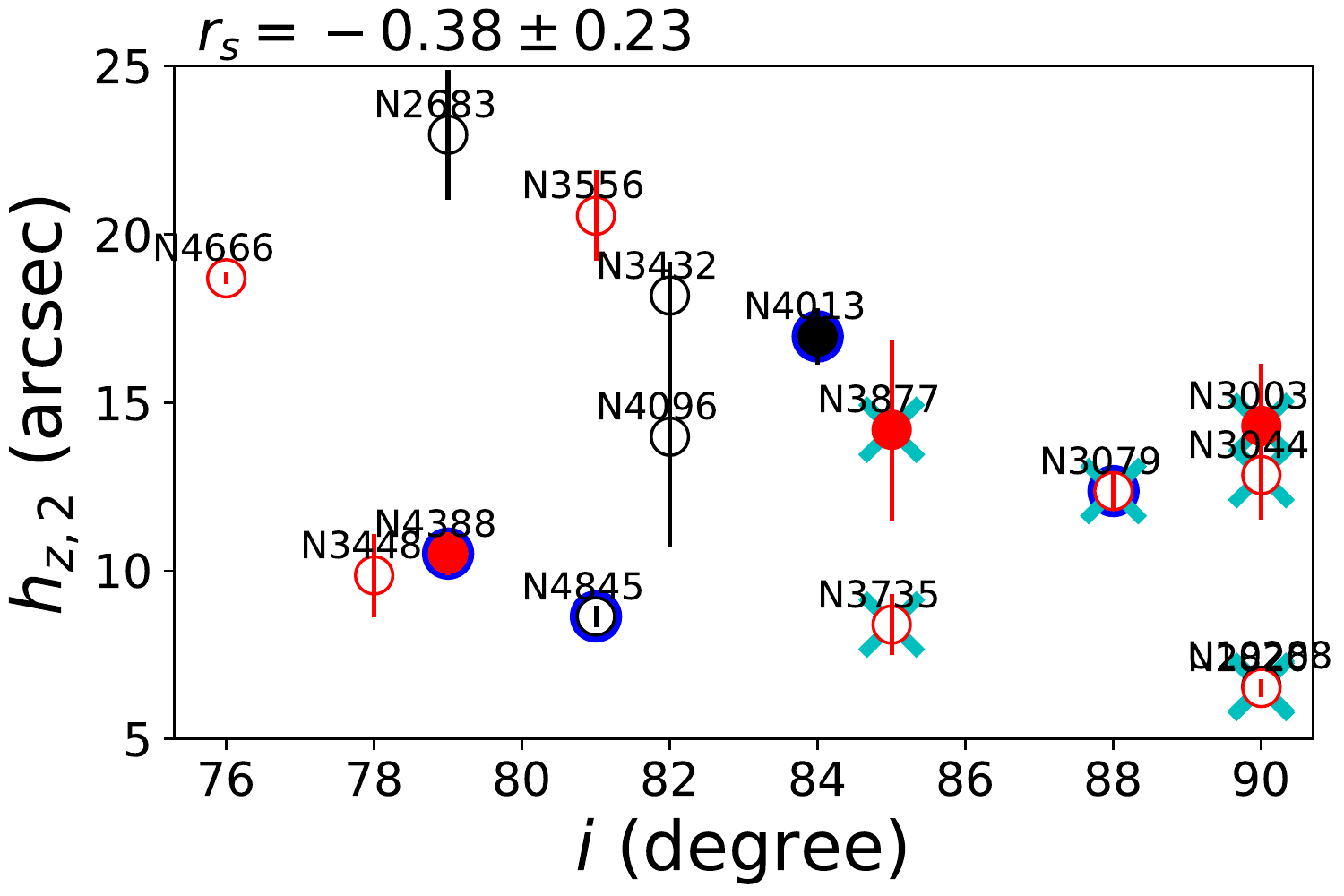}
        \label{fig:hvsi}
        }\hspace{-0.3cm}
    \subfigure[]{
        \includegraphics[width=0.49\textwidth]{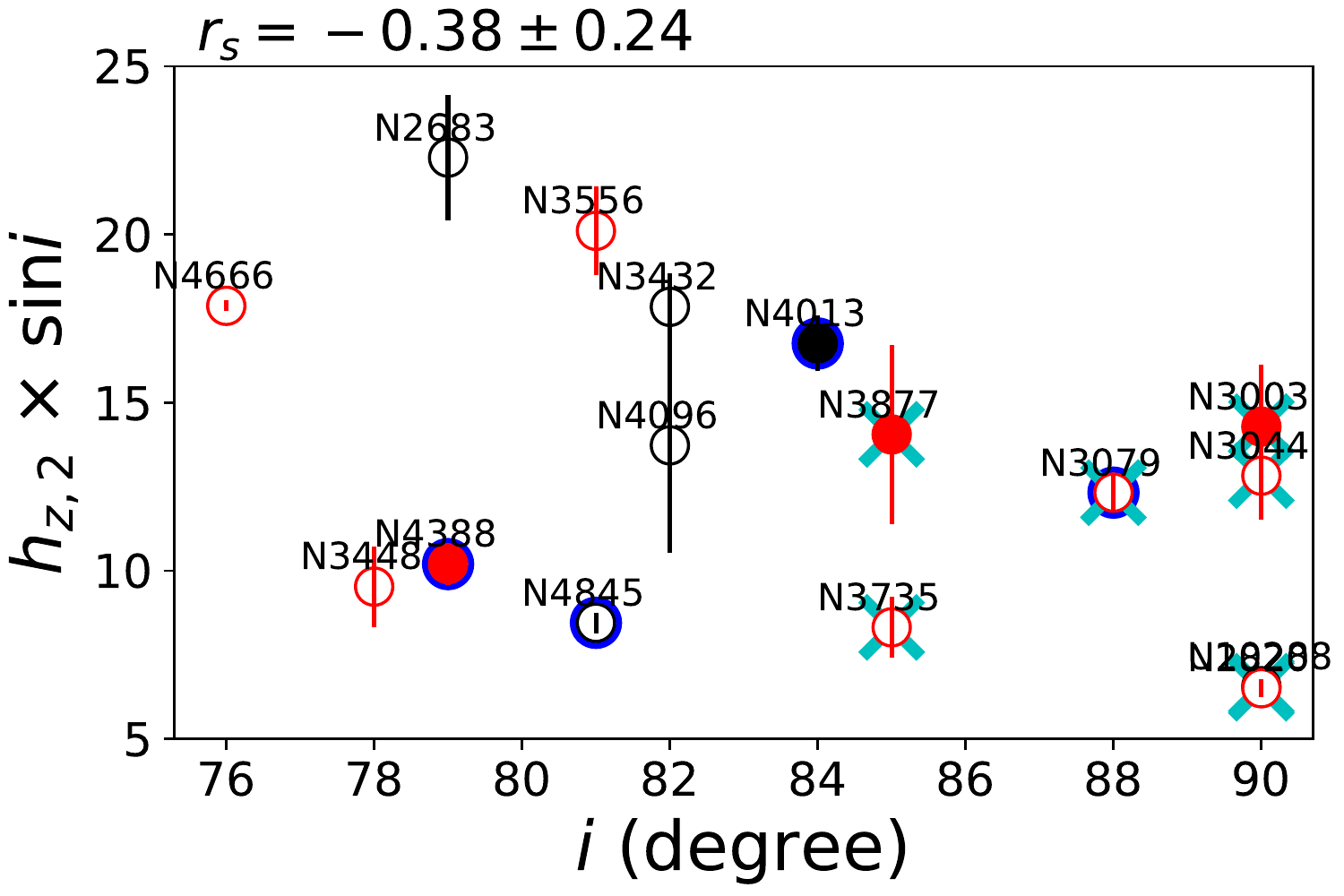}
        \label{fig:hsinivsi}
        }
    \subfigure[]{
        \includegraphics[width=0.49\textwidth]{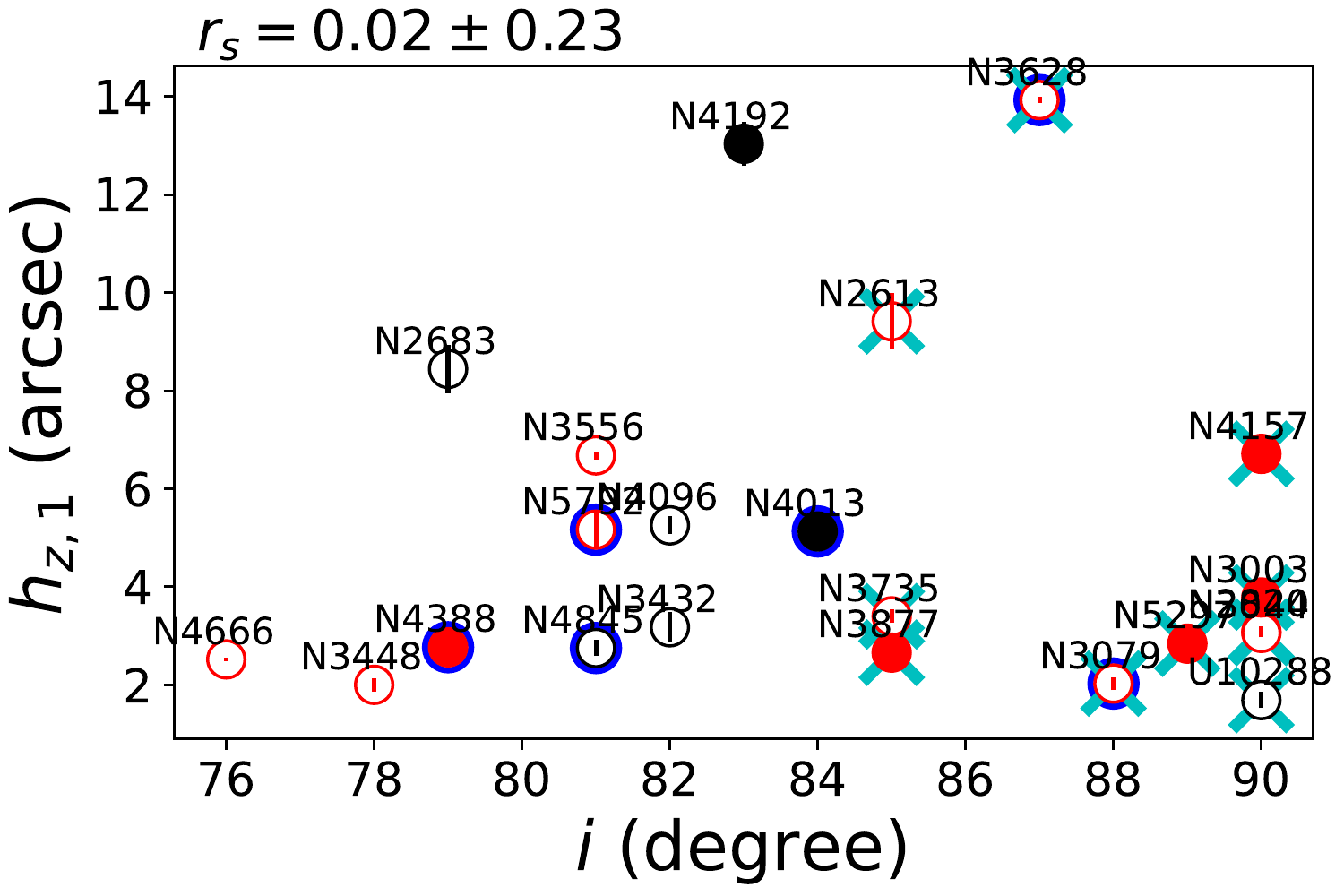}
        \label{fig:hdiskvsi}
        }\hspace{-0.3cm}
    \subfigure[]{
        \includegraphics[width=0.49\textwidth]{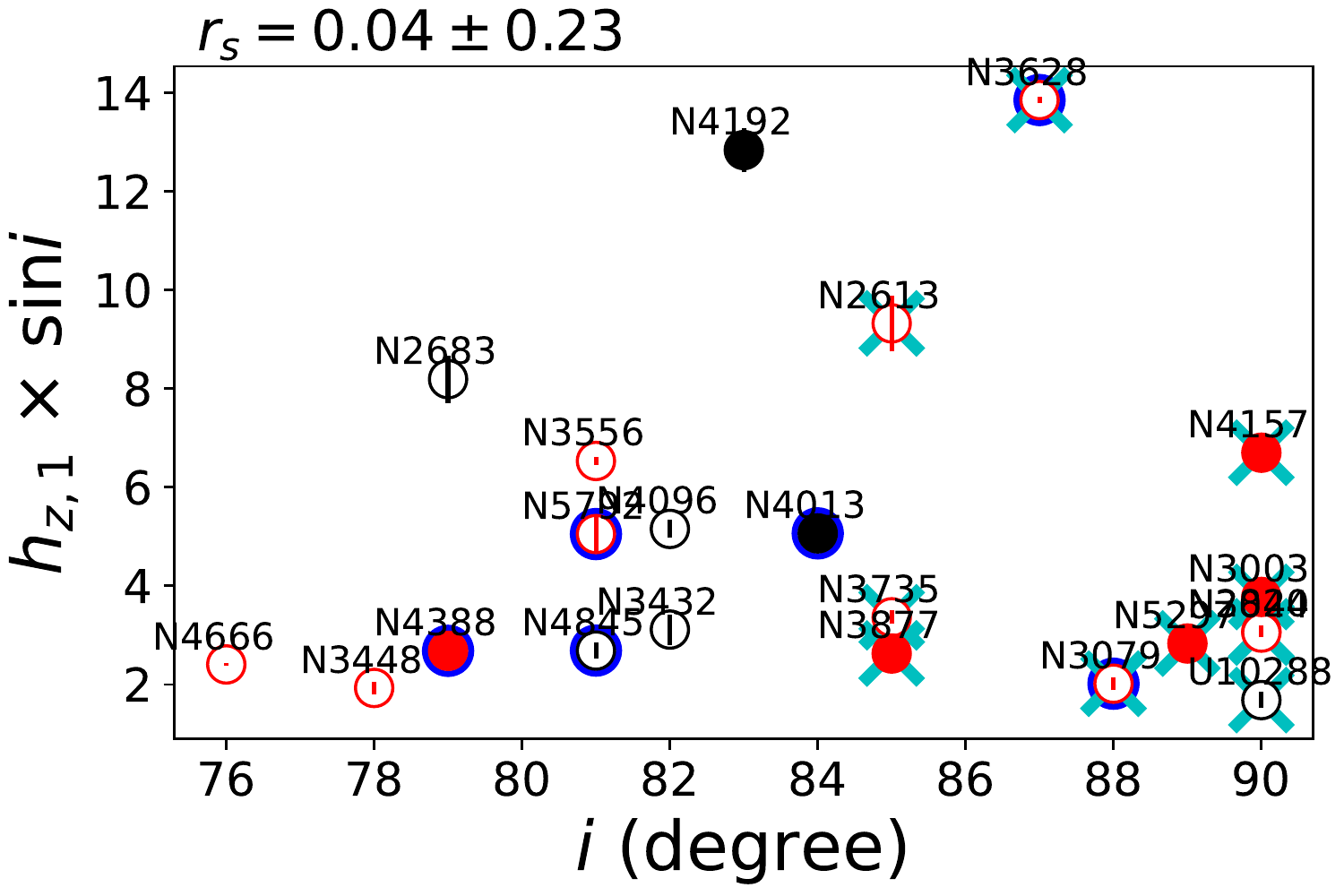}
        \label{fig:hdisksinivsi}
        }
    \caption{Comparison of the H$\alpha$ scale heights in (a) component 2 (thick) and (c) component 1 (thin) with the inclination angles ($i$).
    Panels (b) and (d) show scale heights that have been modified by multiplying with $\sin i$ in order to check the effect of inclination.
    }
\end{figure*}

The emission and/or absorption features from the galactic disk may affect our measurement of the H$\alpha$ scale height, especially in moderately inclined galaxies.
We first check the dependence of the measured H$\alpha$ scale height of the extended component $h_{z,2}$ (with or without a simple geometric correction of the projection) on the inclination angle $i$ (Fig.~\ref{fig:hvsi}, \ref{fig:hsinivsi}).
We find no significant correlations between $i$ and $h_{z,2}$ ($r_s=-0.38\pm 0.23$), or $h_{z,2}$ after a simple geometric correction by multiplying by $\sin{i}$ ($r_s=-0.38\pm 0.24$).
Because of the lack of correlation between the measured H$\alpha$ scale height and the inclination angle, and also because our sample galaxies all have high inclination angles ($i=76^\circ -90^\circ$), we do not expect a significant systematic bias caused by the extended range of inclination angle of the sample galaxies.
Nevertheless, we highlight the highly inclined galaxies ($i\geqslant85^\circ$) with a cyan cross in all the plots, in order to examine the potential difference between them and the moderately inclined ones. 

Similar results are also shown for the compact component representing the galactic disk (component 1 in Eq.~\ref{eq:double_e}; \S\ref{subsec:HaScaleHeight}).
Here we find a similar lack of correlation between the H$\alpha$ scale height and the inclination angle (Fig.~\ref{fig:hdiskvsi}, \ref{fig:hdisksinivsi}).
This component is in principle more significantly affected by some bright highly structured H$\alpha$ features embedded in or around the disk, such as the \ion{H}{2} regions, filaments, bubbles, etc (e.g., \citealt{Vargas19,Li19,Li22}).
As discussed above, involving this compact component in the statistical analysis could cause confusion in the interpretation of the results. Therefore we focus on the extended eDIG component in the following analysis (the average scale height of the two sides of the extended component is quoted as $h_{\rm H\alpha}$ hereafter; Table~\ref{fit_res}).

\subsection{Dependence of H$\alpha$ scale height on global galaxy parameters}\label{ssec:Discussion_parameters}

\begin{figure*}
    \centering
    \hspace{-0.5cm}
    \subfigure[]{
        \includegraphics[width=0.33\textwidth]{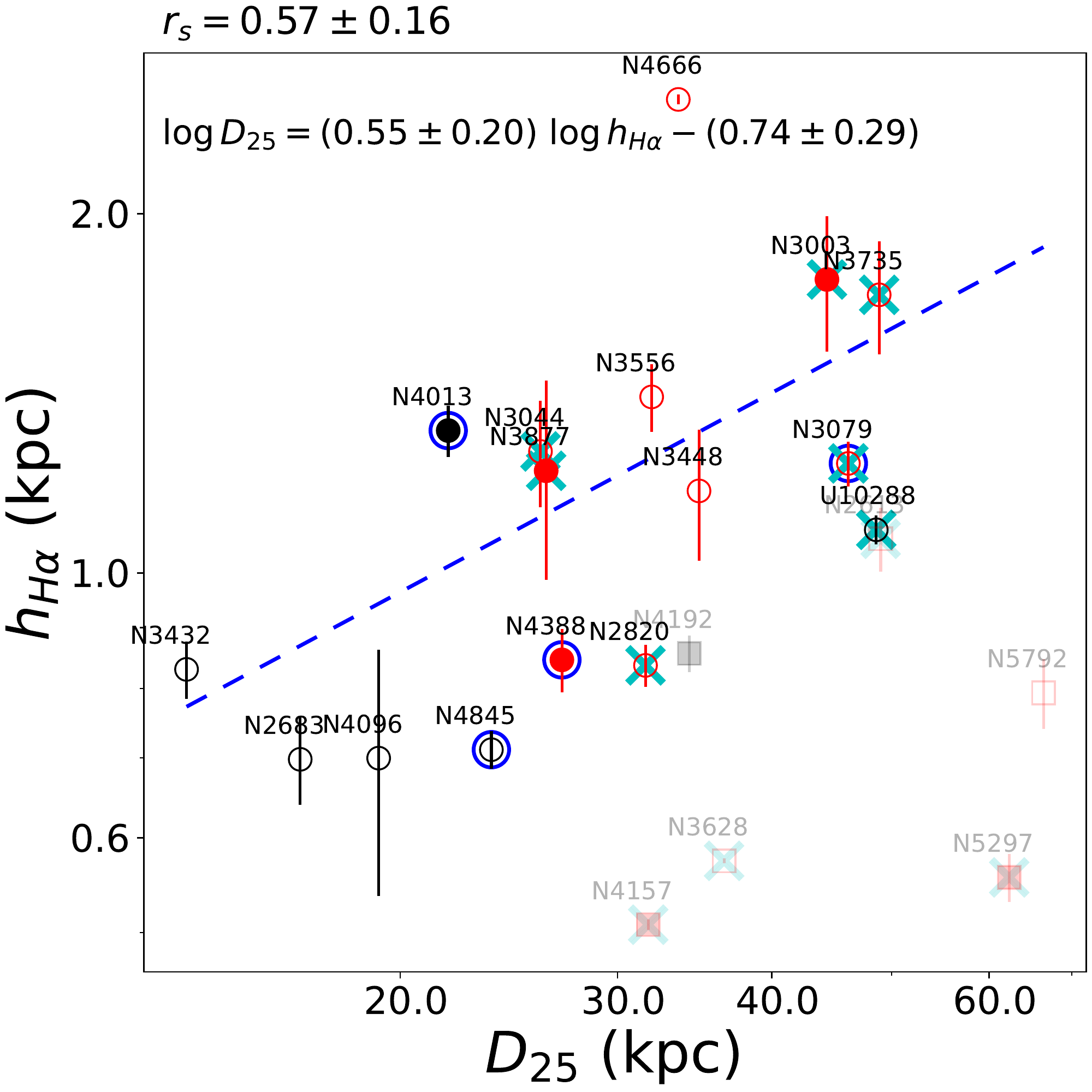}
        \label{fig:hvsd25}
        }\hspace{-0.3cm}
    \subfigure[]{
        \includegraphics[width=0.33\textwidth]{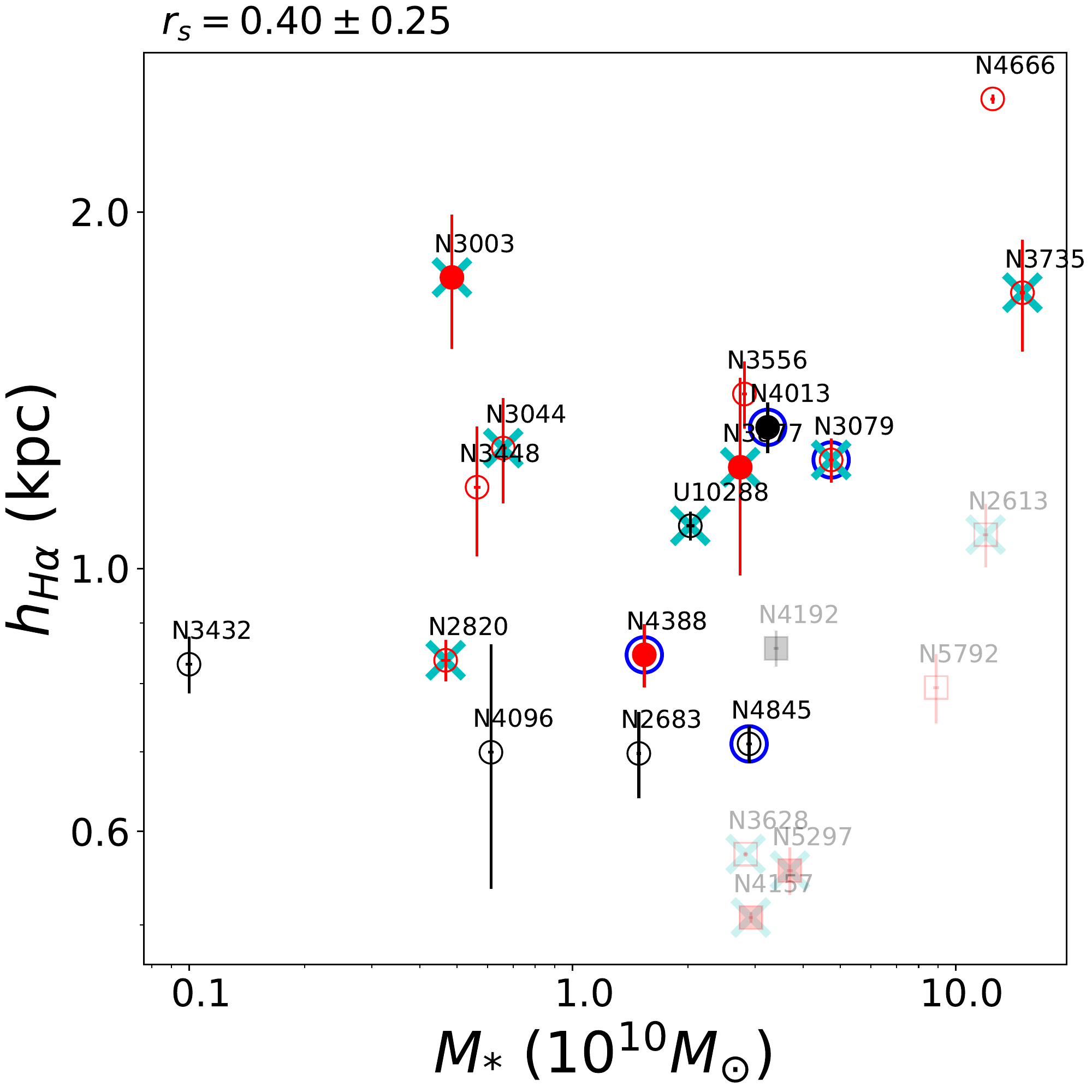}
        \label{fig:hvsMstar}
        }\hspace{-0.3cm}
    \subfigure[]{
        \includegraphics[width=0.33\textwidth]{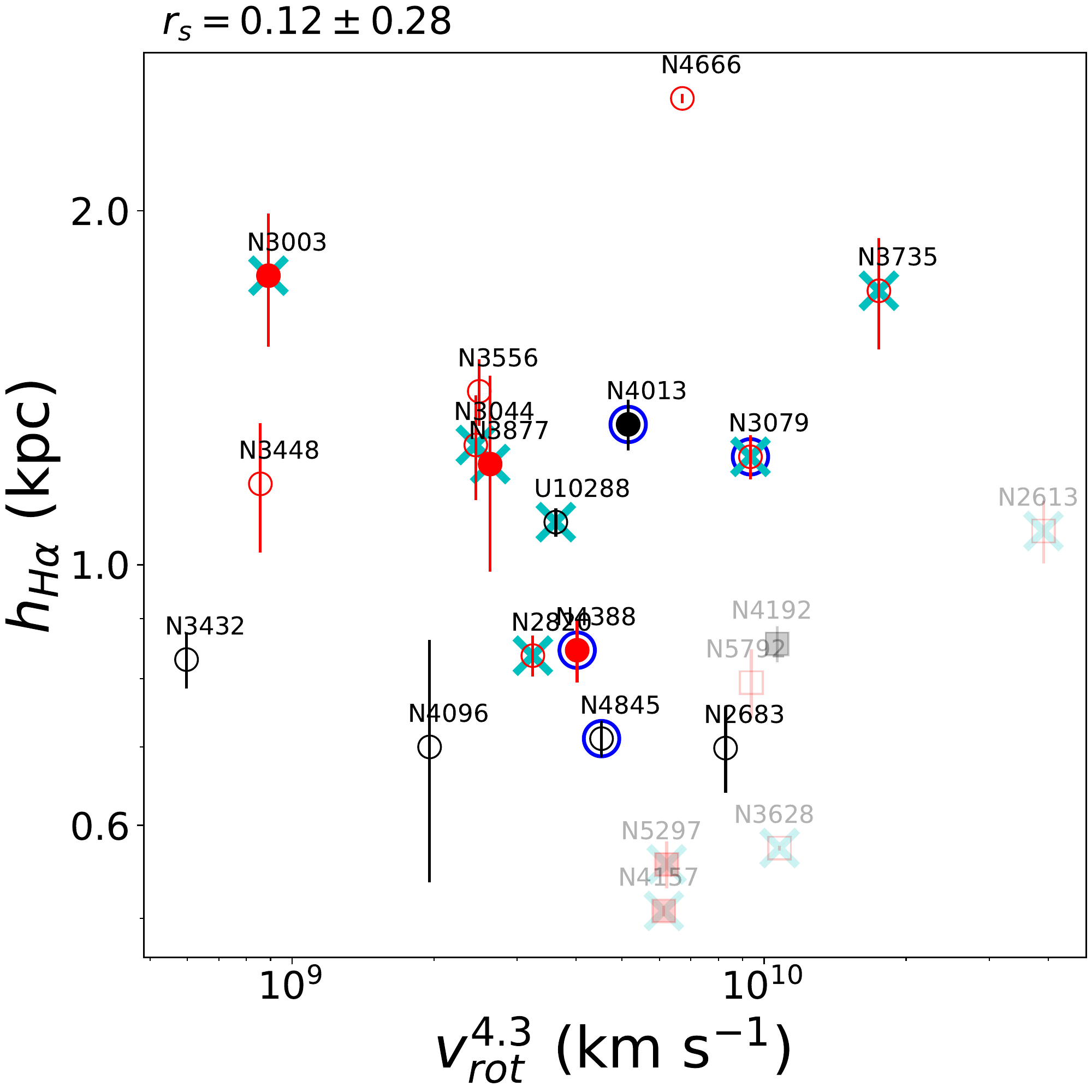}
        \label{fig:hvsvrot}
        }
    \caption{The measured scale heights of extended components in H$\alpha$ emission plotted against the $D_{25}$, $M_{\ast}$, and $v_{\rm rot}^{4.3}$.
    Different symbols represent different sub-samples of galaxies as described in Fig.~\ref{fig:SFRSDvsSFR}.
    The Spearman's rank order correlation coefficient ($r_{s}$) with its $1~\sigma$ error of each relation is noted at the top left of each panel.
    The blue dashed line in panel (a) denotes the  best-fitting power law model, which is presented at the top of the panel.
    The galaxies fitted with just one exponential component on both sides (square symbols) are plotted in faded color to facilitate readers.
    \label{fig:h-parameters}}
\end{figure*}

We plot the measured H$\alpha$ scale height against various galaxy parameters (SFR, $D_{25}$, $M_{\ast}$, $v_{\rm rot}^{4.3}$) in Fig. \ref{fig:h-parameters}.
As described in \S\ref{subsec:Sample}, the SFR is estimated based on a combination of the extinction-corrected H$\alpha$ emission and the \emph{WISE} $22\ \mu$m emission.
This revised SFR is systematically higher than the SFR based only on the mid-IR luminosity \citep{Vargas19}.
$D_{25}$ is the observed blue diameter at the 25$^{\rm th}$ mag arcsec$^{-2}$ isophote and is used to represent the size of the galaxies. $M_{\ast}$ is the photometric stellar mass estimated using the \emph{Two Micron All-Sky Survey}\footnote{\url{https://www.ipac.caltech.edu/project/2mass}} (\emph{2MASS}) \emph{K}-band magnitude and a color-dependent mass-to-light ratio \citep{Bell01}.
$v_{\rm rot}$ is the inclination-corrected rotation velocity obtained from the HyperLeda\footnote{\url{http://leda.univ-lyon1.fr}} database.
The index 4.3 in $v_{\rm rot}^{4.3}$ is obtained from the baryonic Tully-Fisher relation \citep{Bell01}, so $v_{\rm rot}^{4.3}$ is proportional to the baryonic mass of the galaxy measured in a dynamical way.

We find a tight positive correlation ($r_{\rm s} = 0.66\pm 0.16$) between $h_{\rm H\alpha}$ and SFR (Fig.~\ref{fig:hvsSFR}).
The best-fit $h_{\rm H\alpha}-{\rm SFR}$ relation indicates a significant sub-linear slope:
\begin{equation} \label{eq:h-SFR}
    h_{\rm H\alpha}/{\rm kpc} = 1.02\pm 0.09~({\rm SFR}/{\rm M_\odot~yr^{-1}})^{0.29\pm 0.08}.
\end{equation}
Many previous works indicate that bright and widespread extraplanar H$\alpha$ emission is preferentially detected around galaxies with high SFR or high SFR surface density (e.g., \citealt{Rand96,Rossa03}).
However, the connection between the extraplanar H$\alpha$ features and the disk star formation process may not be the direct reason for the apparent tight $h_{\rm H\alpha}$-SFR correlation, where $h_{\rm H\alpha}$ is mostly determined by the diffuse eDIG component instead of the fine structures.
Alternatively, the correlation may be a result of the global scaling relations of the galaxies, which means larger galaxies tend to have both higher global parameters (e.g., mass, SFR, etc.) and larger extensions of the multi-phase gas (e.g., \citealt{Wang16, Jiang19}).
We will further discuss how to disentangle the global scaling relation from the real physical connection between the eDIG and the SF activity in \S\ref{ssec:SFRimpact}.

As a result of the global scaling relations, we also find a fairly good correlation between $h_{\rm H\alpha}$ and some other global galaxy parameters such as $D_{25}$ ($r_{\rm s} = 0.57\pm 0.16$; Fig.~\ref{fig:hvsd25}).
The slope of the $h_{\rm H\alpha}$-$D_{25}$ relation is  sub-linear ($0.55\pm0.20$).
On the other hand, the correlations between $h_{\rm H\alpha}$ and the mass indicators become significantly poorer ($r_{\rm s} = 0.40\pm 0.25$ with $M_{\ast}$; $r_{\rm s} = 0.12\pm 0.28$ with $v_{rot}^{4.3}$; Fig.~\ref{fig:hvsMstar}, \ref{fig:hvsvrot}).
\citet{Zheng22a,Zheng22b} found that the scale height of the \ion{H}{1} $21$~cm emission decreases with the increase of both the total mass surface density and the baryon mass surface density of the galaxies. Such anti-correlations could be explained as a gravitational confinement effect.
This effect is also indicated by the sub-linear slope of the $h_{\rm H\alpha}$-$D_{25}$ relation, which suggests that eDIG of bigger galaxies tends to be more elongated.
However, we do not see any significant correlation between $h_{\rm H\alpha}$ and the mass indicators.
This may be a result of various other effects, such as feedback, which has an opposite effect to the gravitational confinement.

\subsection{How does star formation feedback impact the ionized gas distribution?} \label{ssec:SFRimpact}

Star formation and the following feedback of energy and metal-enriched materials are key processes linking the galaxies and their environments.
The existence of extraplanar H$\alpha$ emission around nearby galaxies is believed to be largely related to the active star formation processes in the disk (e.g., \citealt{Rand96, Rossa00}).
However, as discussed in \S\ref{ssec:Discussion_parameters}, the apparent correlation between $h_{\rm H\alpha}$ and the global galaxy parameters including the SFR is not necessarily caused by the real connection between the star formation processes and the eDIG.

In a sample of 38 nearby edge-on galaxies, \citet{Jo18} reported even tighter correlations between $h_{\rm H\alpha}$ and the SFR or the SFR surface density ($\Sigma _{\rm SFR}$) than reported in the present paper, which is largely caused by the extension of their sample to lower SFR or $\Sigma _{\rm SFR}$ (Fig.~\ref{fig:h-SFR_Jo}).
In addition, they also reported a weak correlation between the normalized H$\alpha$ scale height ($h_{\rm H\alpha}/D_{25}$) and $\Sigma _{\rm SFR}$, which is quite marginal in our sample (Fig.~\ref{fig:hdd25vssigmaSFR_Jo}).
We caution that the SFR and its surface density are measured in a different way in the present paper than in \citet{Jo18}. Our $\rm SFR_{SD}$ is calculated from the revised SFR ($\rm SFR_{H\alpha +22\mu m}$, based on a combination of the H$\alpha$ and mid-IR data; \citealt{Vargas19}) and the mid-IR-based star formation radius (measured from the \emph{WISE} $22{\rm~\mu m}$ image; \citealt{Wiegert15}):
\begin{equation}\label{eq:SFRSD}
    {\rm SFR_{SD}}={\rm SFR_{H\alpha +22\mu m}}/(\pi R_{22}^{2}),
\end{equation}
where $R_{22}$ is the $22~\mu$m radii from \citet{Wiegert15}.
In contrast, \citet{Jo18} calculated $\Sigma _{\rm SFR}$ from the far-IR-based SFR and the optical diameter $D_{25}$:
\begin{equation}
    \Sigma _{\rm SFR,FIR}={\rm SFR_{FIR}}/(\pi D_{25}^{2}).
\end{equation}
In the figures presented in this paper, we have replaced the diameter $D_{25}$ with the corresponding radius $R_{25}$ in the above equation and re-calculated $\Sigma _{\rm SFR,FIR}$ for \citet{Jo18}'s sample.
For a more consistent comparison, we also calculate $\rm SFR_{FIR}$ and $\Sigma _{\rm SFR,FIR}$ for our sample galaxies using the same definition as adopted in \citet{Jo18}.
We quote the FIR luminosity $L_{\rm FIR}$ of our sample from \citet{Irwin12a} and calculate $\rm SFR_{FIR}$ using the relation from \citet{Kennicutt98}.
We also caution that although the FIR luminosity in \citet{Irwin12a} and \citet{Jo18} are both calculated from the \emph{IRAS} $60$ and $100~\mu$m fluxes ($f_{60}$ and $f_{100}$; using the relation in \citealt{Sanders96}), the fluxes are indeed quoted from different references.
\citet{Jo18} quoted the values from \citet{Moshir90} and \citet{Irwin12a} quoted them from \citet{Sanders03}, except for several galaxies that used the most recent entry listed in NED, usually from the IRAS Faint Source Catalog (FSC).

We compare our sample to \citet{Jo18}'s sample in Fig.~\ref{fig:h-SFR_Jo}. Our sample is in general well consistent with that of \citet{Jo18} on the global $h_{\rm H\alpha}-{\rm SFR_{FIR}}$ relation, and most of our $\sim L^\star$ galaxies are located in the high SFR range because they are generally more massive (Fig.~\ref{fig:hvsSFR_Jo}). The SFR surface density $\Sigma _{\rm SFR,FIR}$ is a normalized entity and for this reason it is a better tracer of the intrinsic activity of star formation in galaxies. In Fig.~\ref{fig:hvssigmaSFR_Jo} and~\ref{fig:hdd25vssigmaSFR_Jo}, we compare our sample to that of \citet{Jo18} by plotting $h_{\rm H\alpha}$ and $h_{\rm H\alpha}/D_{\rm 25}$ against $\Sigma _{\rm SFR,FIR}$. It is clear that the two samples are consistent with each other on the relations, and show significant positive correlations. This indicates real positive effects of star formation on the vertical extension of the eDIG, as also suggested by many previous works (e.g., \citealt{Collins00,Rossa00,Haffner09}).

\begin{figure*}
    \centering
    \hspace{-0.5cm}
    \subfigure[]{
        \includegraphics[width=0.33\textwidth]{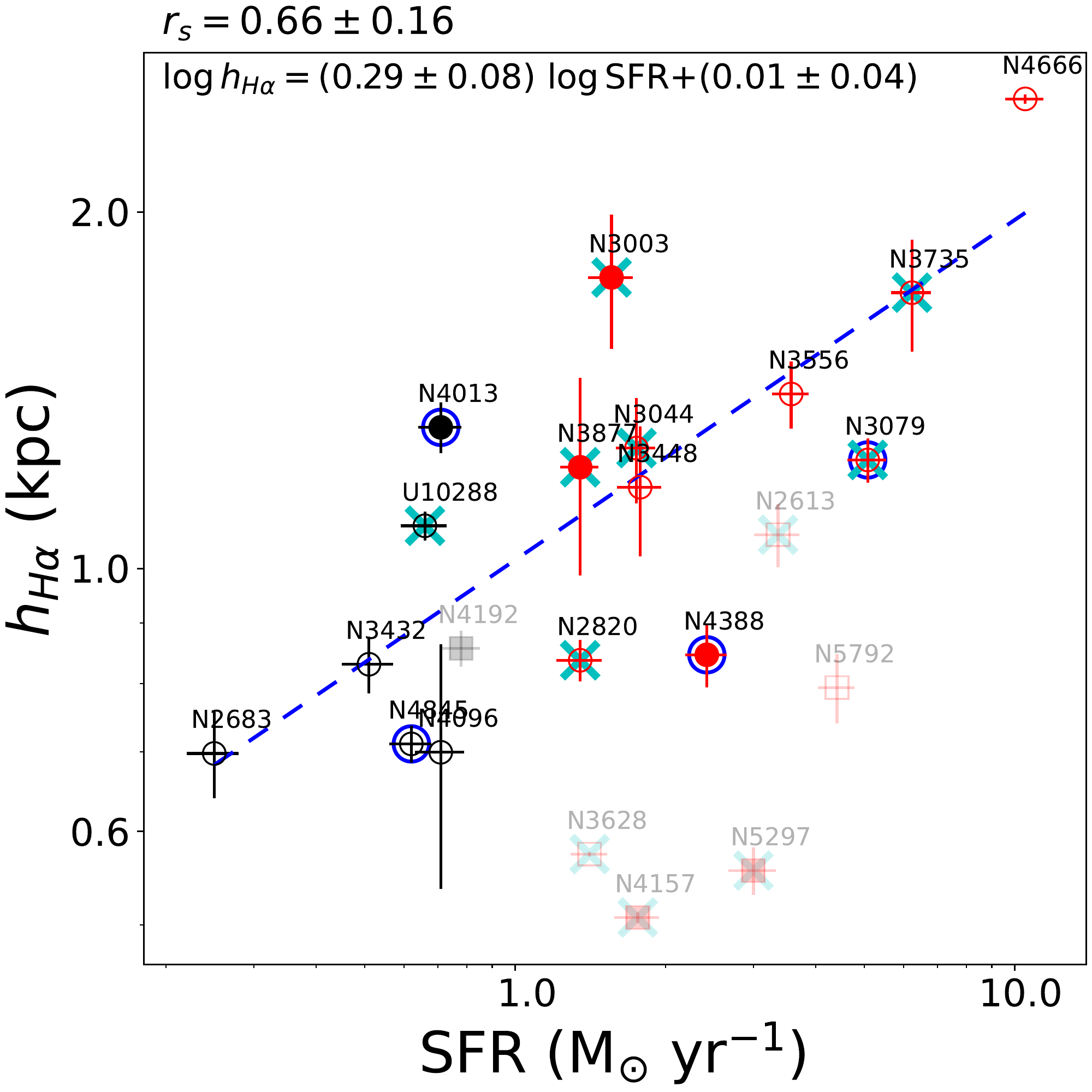}
        \label{fig:hvsSFR}
        }\hspace{-0.3cm}
    \subfigure[]{
        \includegraphics[width=0.33\textwidth]{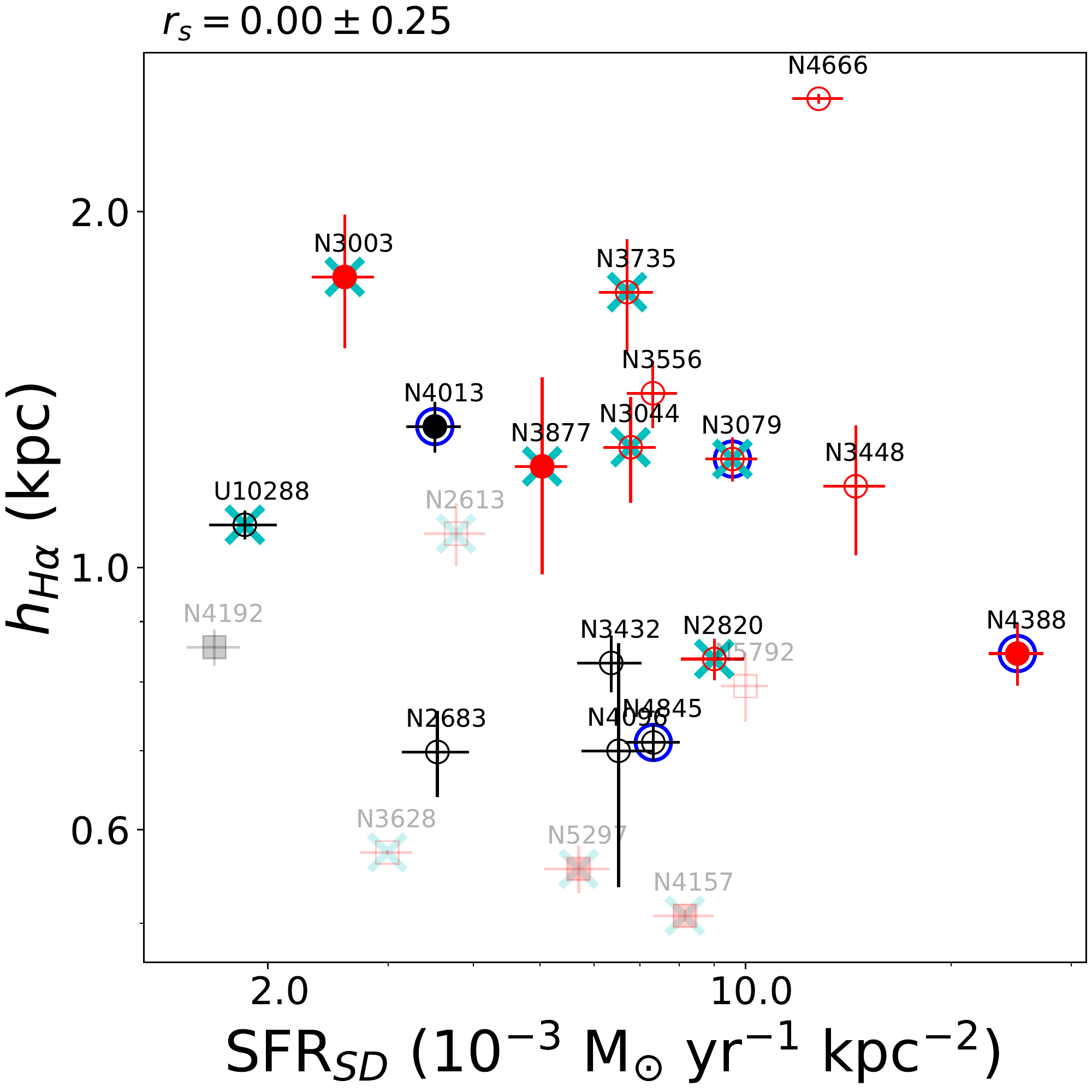}
        \label{fig:hvsSFRSD}
        }\hspace{-0.3cm}
    \subfigure[]{
        \includegraphics[width=0.33\textwidth]{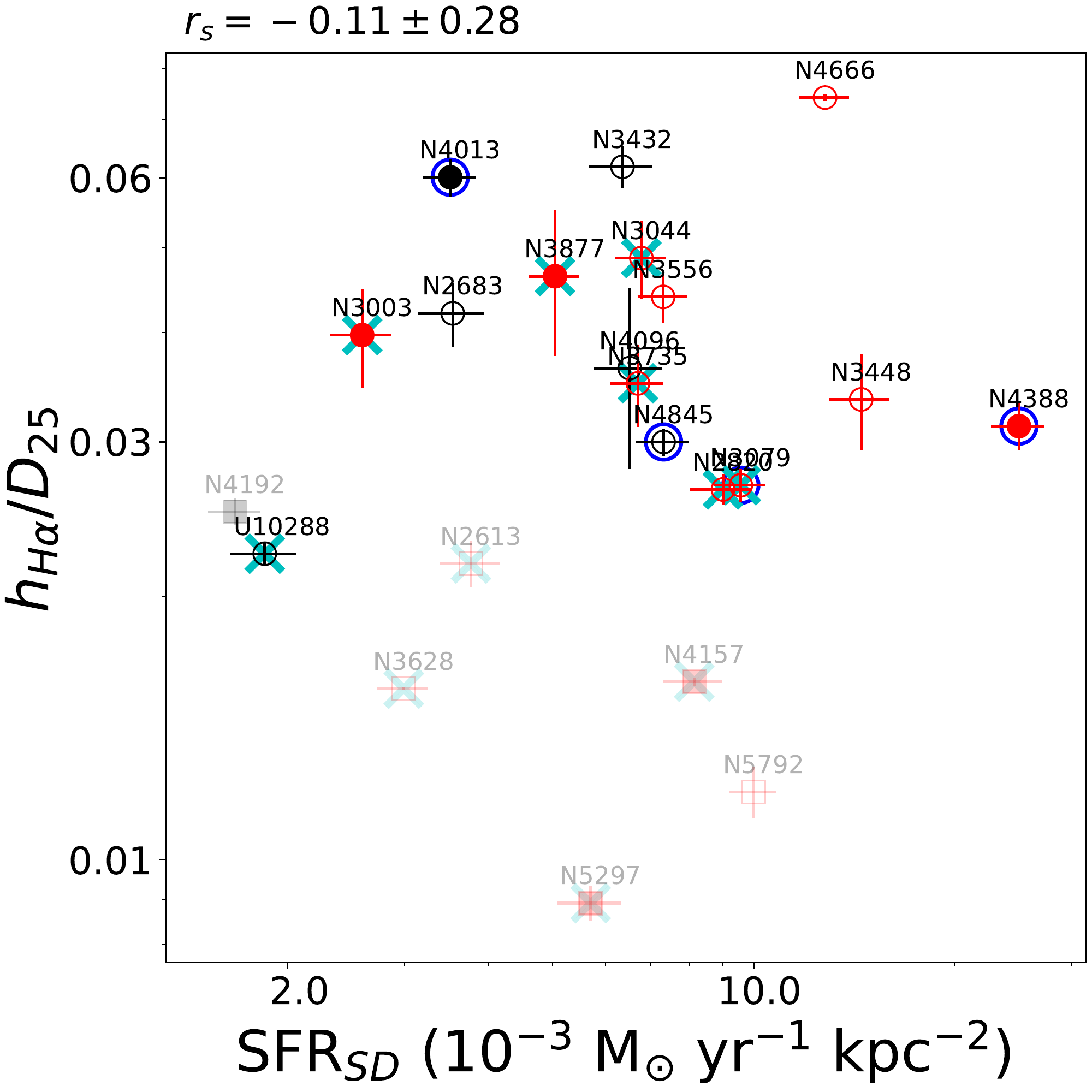}
        \label{fig:hdd25vsSFRSD}
        }
    \caption{Comparison of the scale heights of the H$\alpha$ emission with (a) ${\rm SFR}$, and (b) $\rm SFR_{SD}$.
    (c) Comparison of the normalized scale heights of the H$\alpha$ emission with $\rm SFR_{SD}$.
    The $r_{\rm s}$ with its $1\sigma$ error of each relation is noted at the top left of each panel.
    The blue dashed line in panel (a) denotes the best-fitting log-log linear line and the relation is presented at the top of the panel.
    \label{fig:h-SFR}}
\end{figure*}

\begin{figure*}
    \centering
    \hspace{-0.5cm}
    \subfigure[]{
        \includegraphics[width=0.33\textwidth]{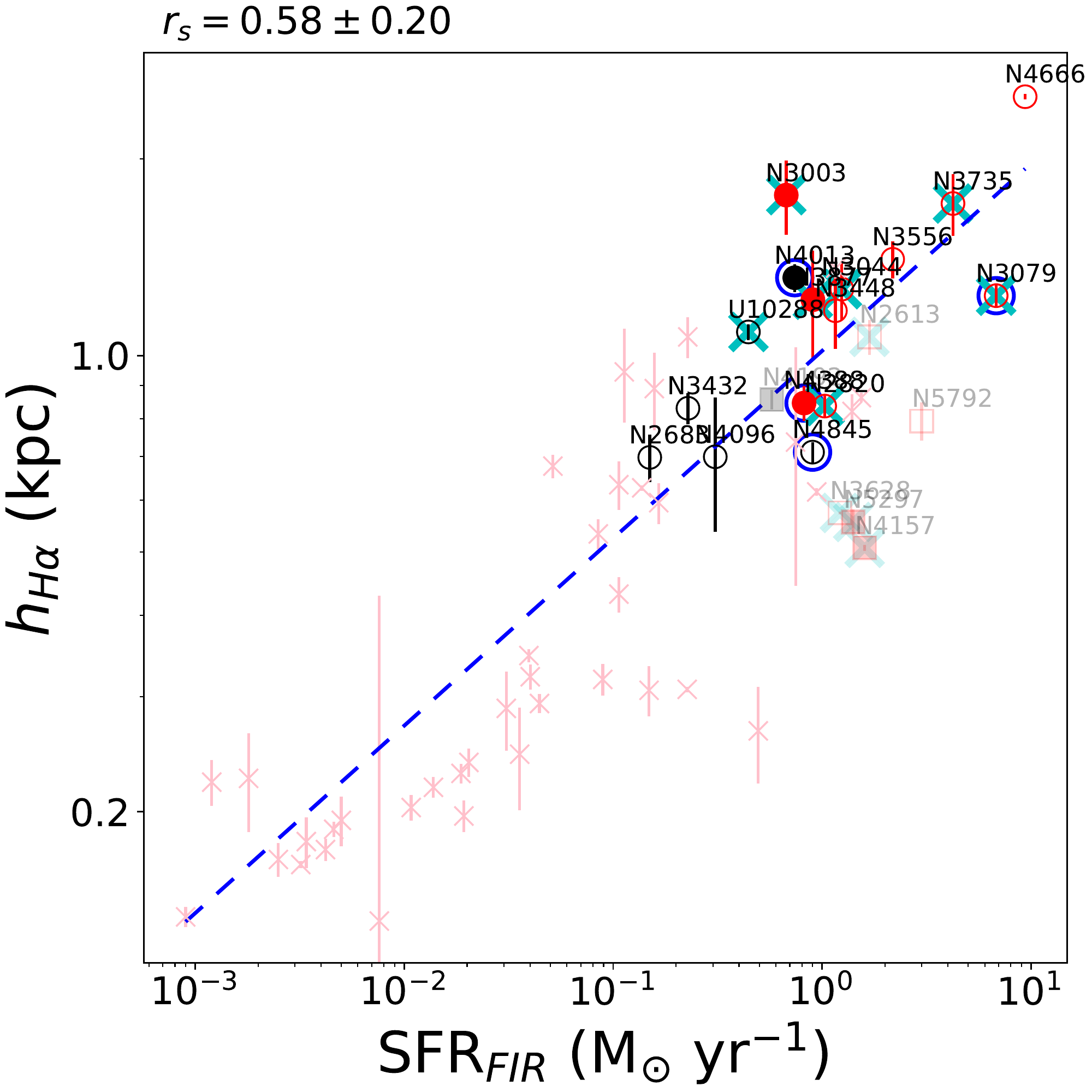}
        \label{fig:hvsSFR_Jo}
        }\hspace{-0.3cm}
    \subfigure[]{
        \includegraphics[width=0.33\textwidth]{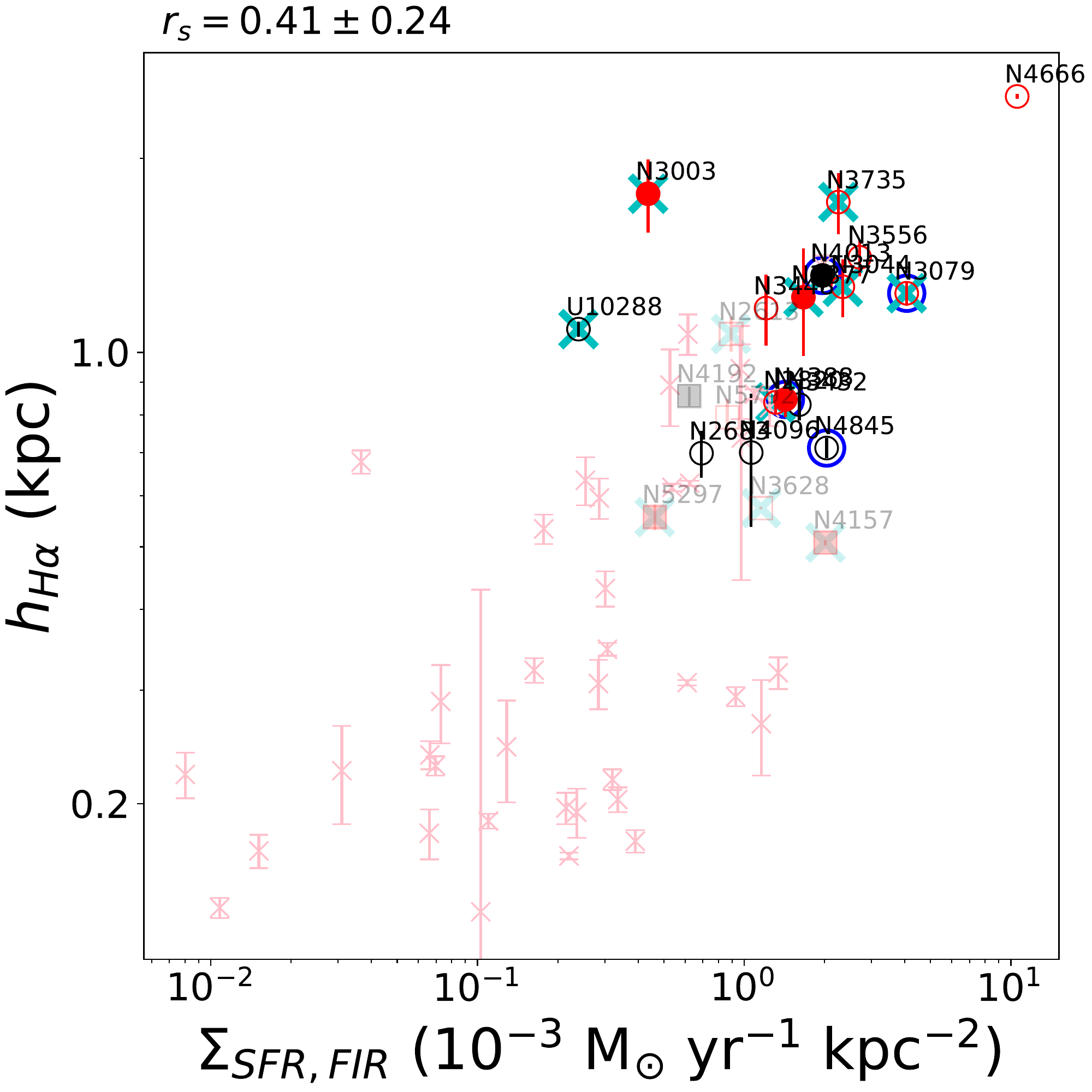}
        \label{fig:hvssigmaSFR_Jo}
        }\hspace{-0.3cm}
    \subfigure[]{
        \includegraphics[width=0.33\textwidth]{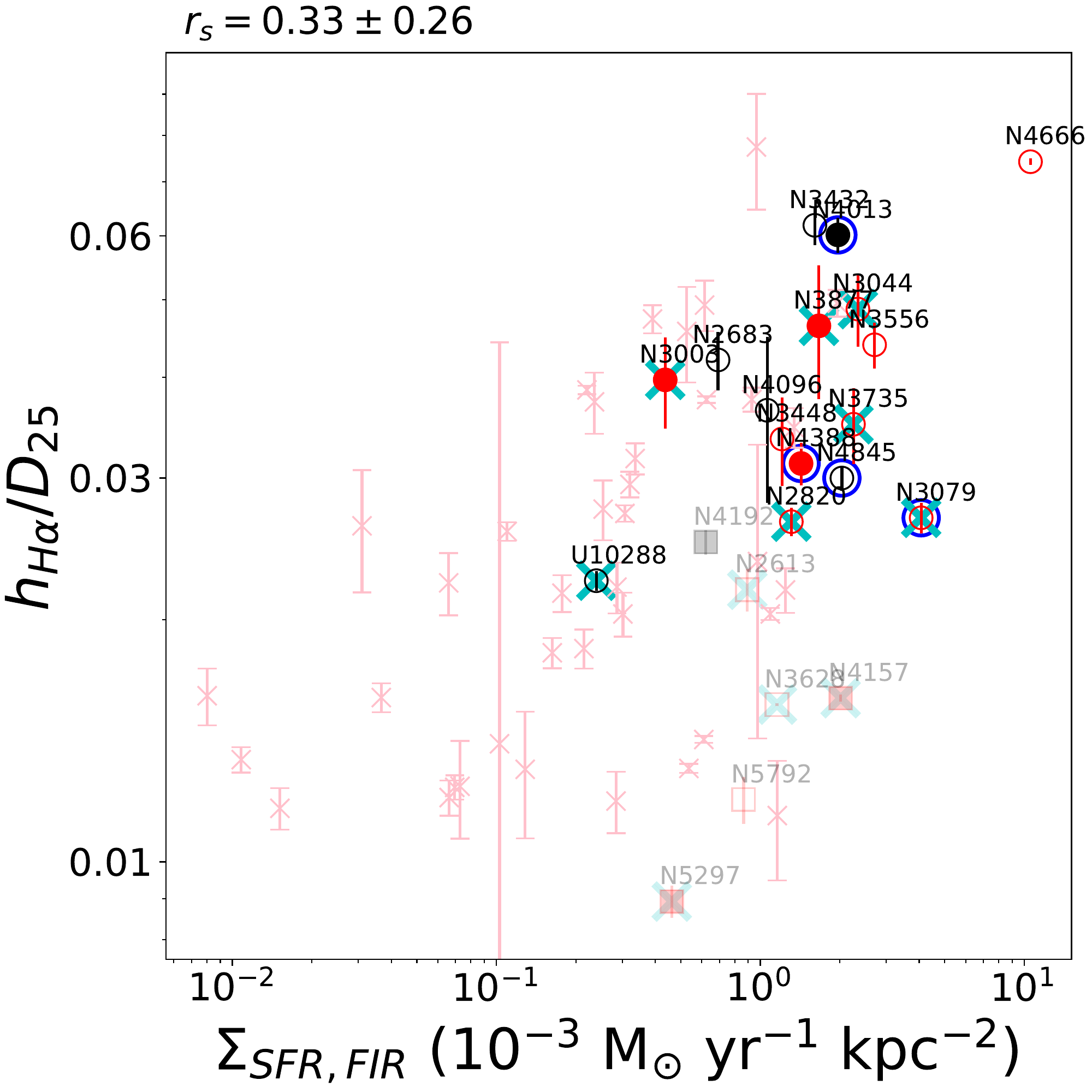}
        \label{fig:hdd25vssigmaSFR_Jo}
        }
    \caption{Comparison of the scale heights of the H$\alpha$ emission with (a) ${\rm SFR_{FIR}}$, and (b) $\Sigma _{\rm SFR,FIR}$.
    (c) Comparison of the normalized scale heights of the H$\alpha$ emission with $\Sigma _{\rm SFR,FIR}$.
    The points from \citealt{Jo18} are denoted by pink x-shaped symbols, and these data are not involved in any calculation and fitting in this paper.
    The measurement and definition of the SFR and SFR surface density are the same for the different samples: ${\rm SFR_{FIR}}$ is estimated by FIR luminosity and $\Sigma _{\rm SFR,FIR}={\rm SFR_{FIR}}/\pi R_{25}^{2}$, see the context for details
    The $r_{\rm s}$ with its $1\sigma$ error of each relation for our sample is noted at the top left of each panel.
    The blue dashed line in panel (a) is as same as that in Fig.~\ref{fig:hvsSFR} and shown here for comparison.
    \label{fig:h-SFR_Jo}}
\end{figure*}

\begin{figure}
    \centering
    \hspace{-0.3cm}
    \includegraphics[width=0.48\textwidth]{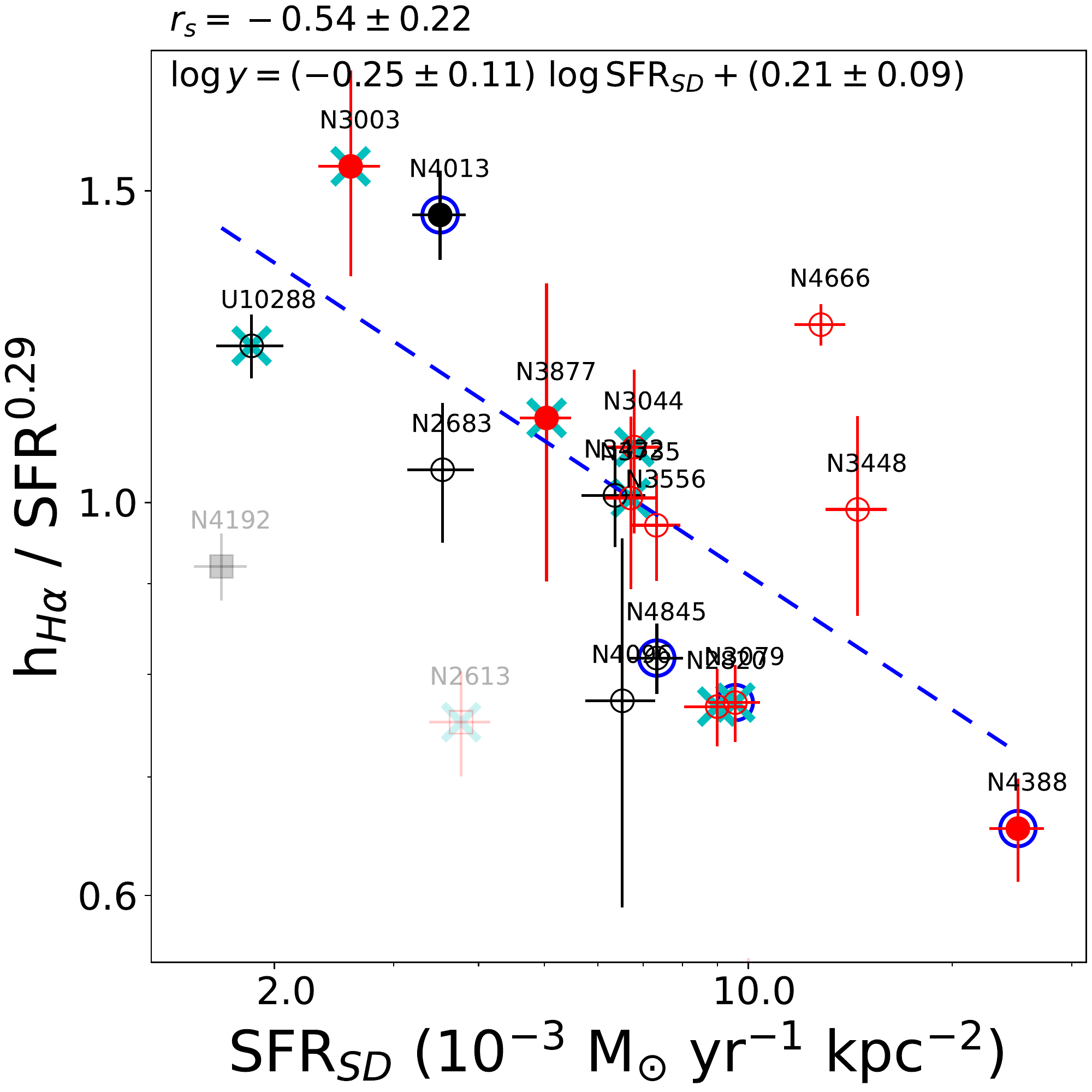}
    \caption{The offsets in the $h_{\rm H\alpha}-{\rm SFR}$ relation plotted against the SFR surface density ($\rm SFR_{SD}$).
    The $r_{\rm s}$ with its $1\sigma$ error is noted at the top left. The blue dashed line denotes the best-fitting log-log linear lines and the relation is presented at the top right.
    \label{fig:hdSFRvsSFRSD}}
\end{figure}

We also notice a fairly strong anti-correlation between $h_{\rm H\alpha}/{\rm SFR}^{0.29}$ and $\rm SFR_{SD}$ ($r_{\rm s}=-0.54\pm 0.22$; Fig.~\ref{fig:hdSFRvsSFRSD}).
Here the index 0.29 is the best-fit power-law index of the $h_{\rm H\alpha}-{\rm SFR}$ relation (Fig.~\ref{fig:hvsSFR}), the strongest correlation between $h_{\rm H\alpha}$ and the global galaxy parameters.
Therefore, $h_{\rm H\alpha}/{\rm SFR}^{0.29}$ describes the offset of different galaxies from the best-fit $h_{\rm H\alpha }-{\rm SFR}$ relation, or the residual of the primary correlation which may be affected by the secondary effects.
The clear anti-correlation between $h_{\rm H\alpha}/\rm{SFR}^{0.29}$ and $\rm SFR_{SD}$, together with the sublinear $h_{\rm H\alpha }-{\rm SFR}$ relation, suggests that the size of the eDIG is disproportionately scaled to the star formation activity, or a more actively star forming galaxy will tend to extend the eDIG less efficiently.

The eDIG could be ionized in various ways. The most common one is by photo-ionization caused by leaking Lyman continuum (LyC) photons from \ion{H}{2} regions embedded in the galactic disk (e.g., \citealt{Rozas96,Zurita00,Tacchella21}).
Some other processes may also contribute, including, but not limited to, shock-ionization (e.g., \citealt{Chevalier85}), turbulence (e.g., \citealt{Slavin93, Binette09}), in situ ionization by decaying massive neutrinos (e.g., \citealt{Sciama90}), and/or extra ionizing photons from the UV background (e.g., \citealt{Madau15}).
In case of photo-ionization, the vertical distribution of the H$\alpha$ emission is determined mainly by two factors: the vertical distribution of the neutral gas and the vertical distribution of the ionizing photons.
The disk star formation activity could strongly affect the intensity of the ionizing photons, but have less effect on the global distribution of the neutral gas reservoir.

When the eDIG is mostly ionized by some external sources such as the uniform UV background, its extension is mostly affected by the distribution of the gas instead of the ionizing photons, so we do not expect a tight correlation between $h_{\rm H\alpha}$ and the SFR surface density. But a positive $h_{\rm H\alpha}$-SFR correlation is still expected as bigger galaxies tend to have both larger SFR and more extended gas reservoirs. On the other hand, if internal ionizing sources, such as LyC photons or shock, dominate the ionization of the eDIG, $h_{\rm H\alpha}$ will be highly affected by the star formation activity. The former case (external ionizing source dominates) appears at large vertical distances from the disk or in quiescent galaxies, while the latter case (internal ionizing source dominates) appears close to an actively star forming disk. The vertical H$\alpha$ scale height is a characterization of the overall distribution of the eDIG, which is not only affected by its outermost extension. When the SFR surface density increases, not only the outermost extension of the eDIG increases, but also the H$\alpha$ emission tends to be more concentrated toward the galactic plane due to its high photon intensity and gas density. The latter effect could even reduce the vertical scale height. The combined effect is just a slight increase of $h_{\rm H\alpha}$, as indicated by the significant sublinear $h_{\rm H\alpha}$-SFR relation (Figs.~\ref{fig:hvsSFR}, \ref{fig:hvsSFR_Jo}). Furthermore, the possible negative effect of intense star formation on $h_{\rm H\alpha}$ is also indicated by the anti-correlation between $h_{\rm H\alpha}/\rm{SFR}^{0.29}$ and $\rm SFR_{SD}$ (Fig.~\ref{fig:hdSFRvsSFRSD}), which means galaxies with a high $\rm SFR_{SD}$ disproportionately increase $h_{\rm H\alpha}$ compared to less active galaxies. \citet{Li13b} discovered a similar effect in their X-ray studies of a sample of edge-on galaxies, where the X-ray radiation efficiency, or the offset of the first order $L_{\rm X}-\dot{E}_{\rm SN}$ relation ($L_{\rm X}$ is the X-ray luminosity of the galactic halo, while $\dot{E}_{\rm SN}$ is the total SN energy injection rate), also shows an anti-correlation with the SN rate surface density (roughly proportional to the SFR surface density in late-type galaxies). The observed relations between the eDIG extension and the star formation properties are caused by a combination of many complicated processes, and so appear to have a large scatter. Detailed theoretical modeling, in comparison with deeper optical emission line observations, are needed to unveil the impact of star formation feedback on the eDIG (e.g., \citealt{Tacchella21}). 


\subsection{Vertical extension of different CGM phases}

\begin{figure*}
    \centering
    \subfigure[]{
        \includegraphics[width=0.48\textwidth]{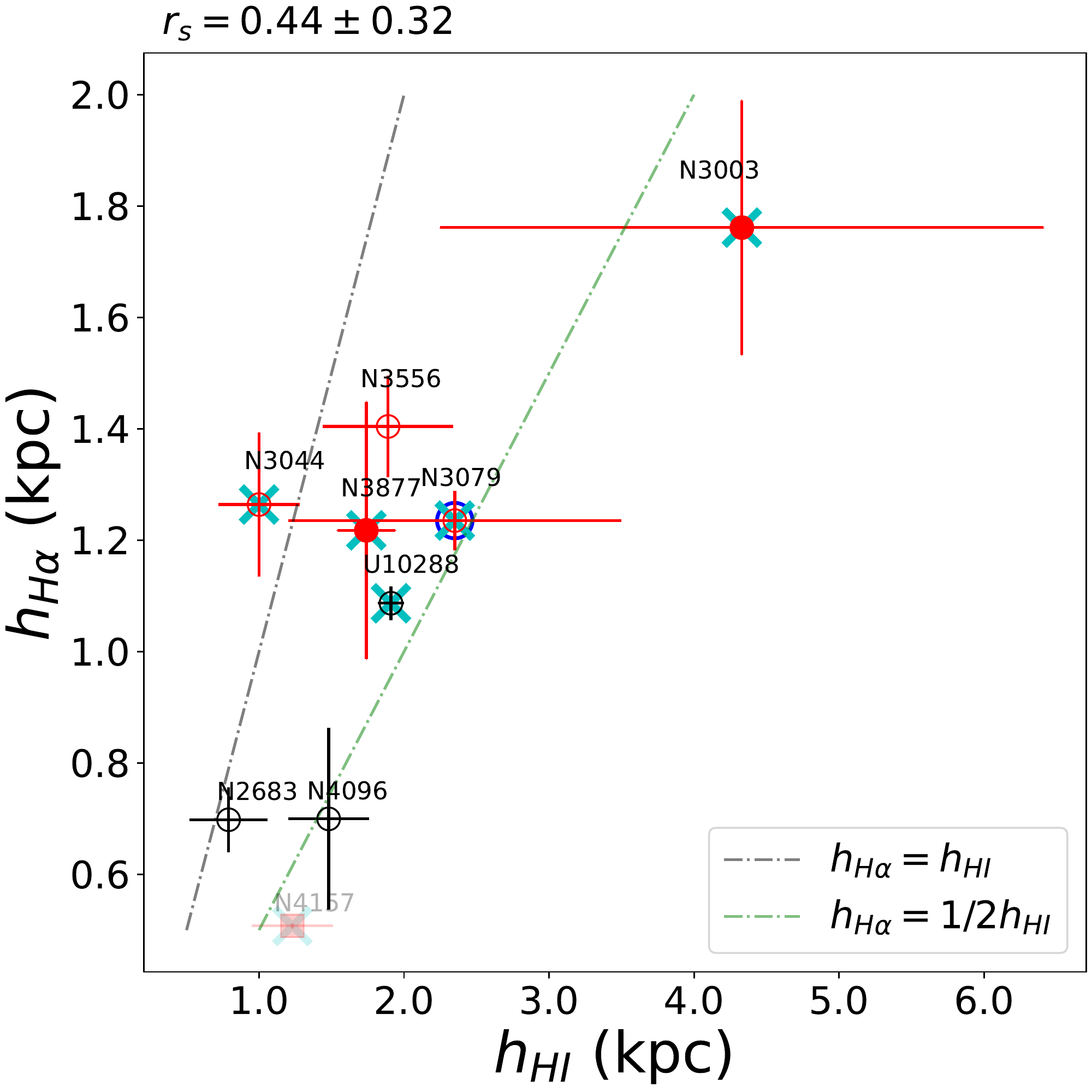}
        \label{fig:hvshHI}
        }
    \subfigure[]{
        \includegraphics[width=0.48\textwidth]{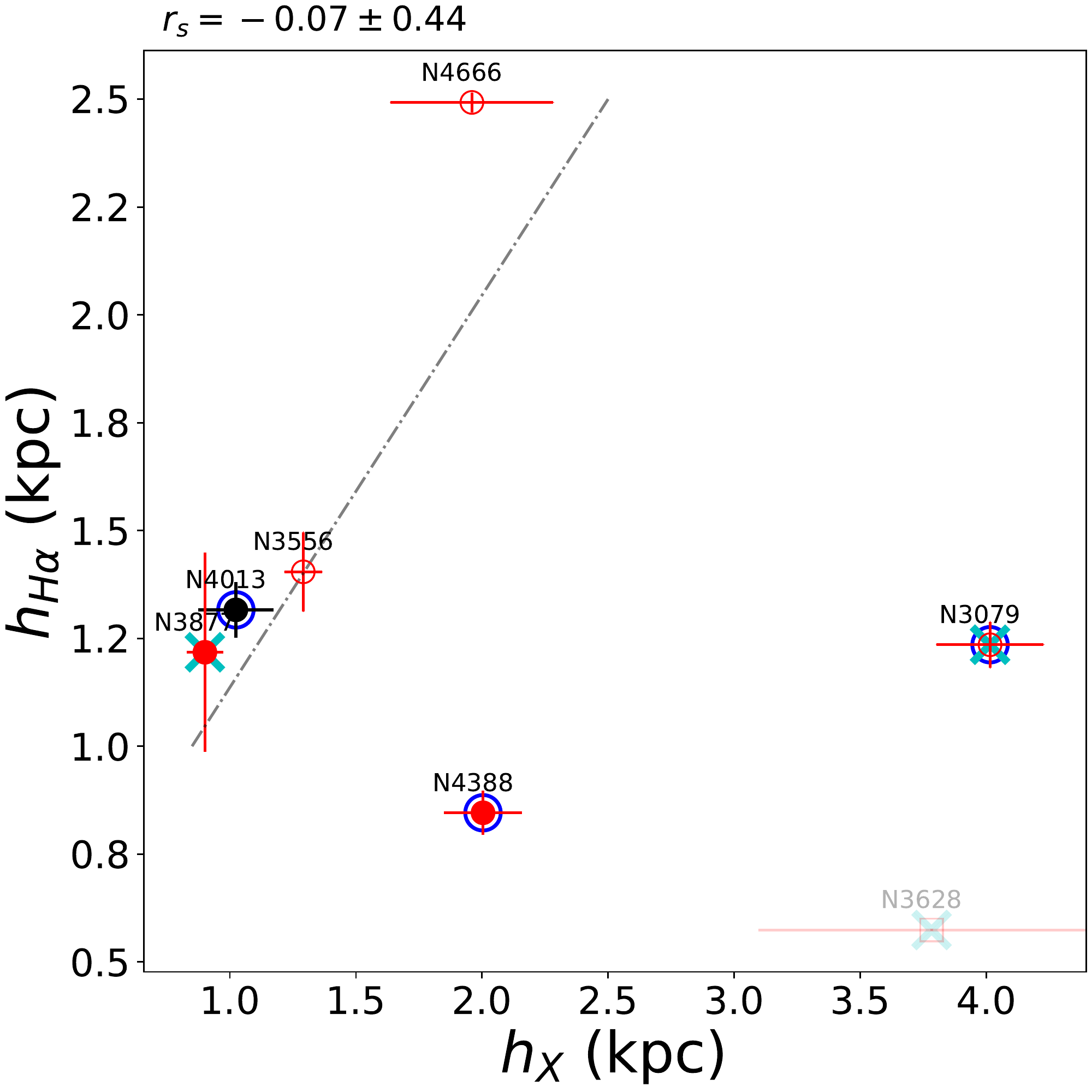}
        \label{fig:hvshX}
        }
    \subfigure[]{
        \includegraphics[width=0.48\textwidth]{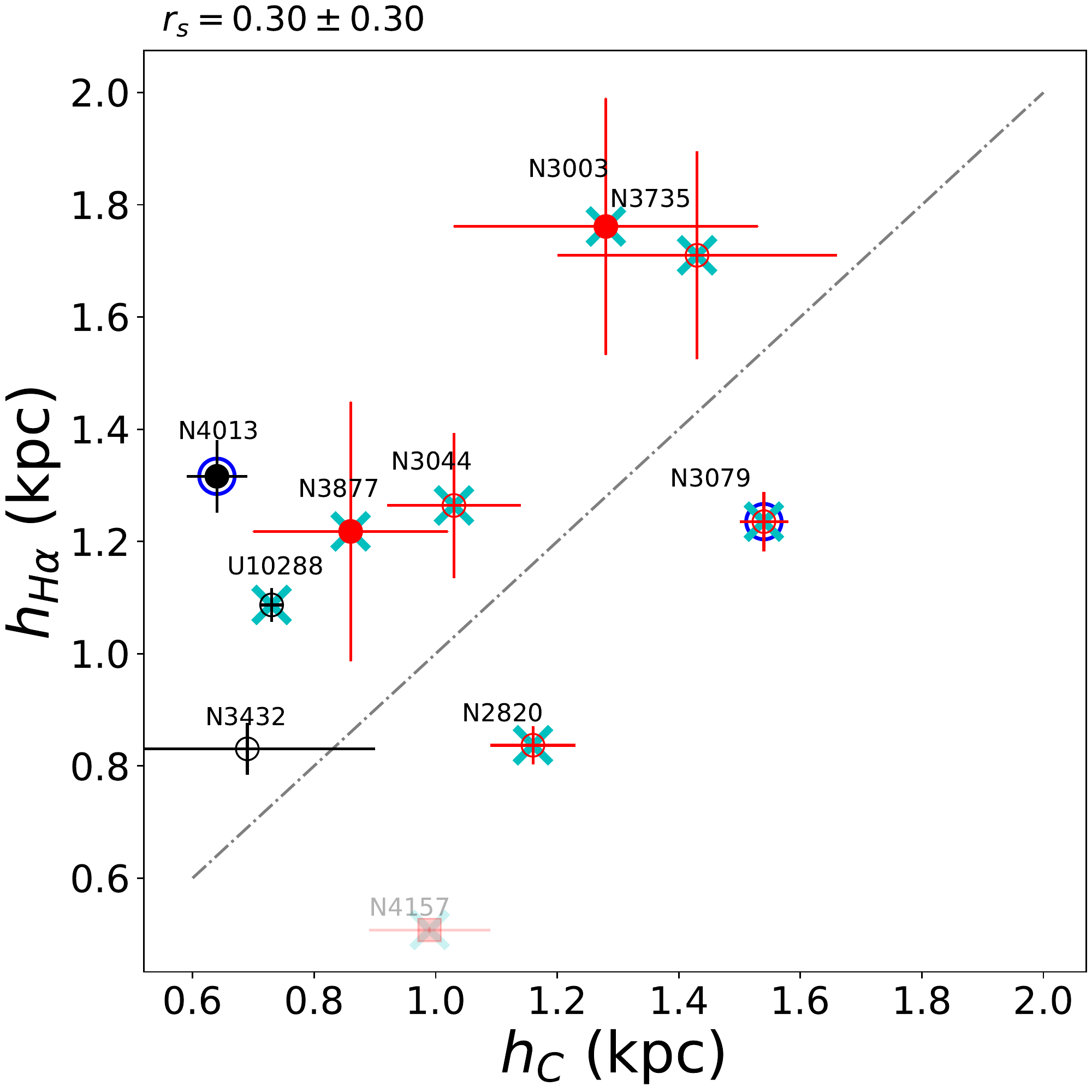}
        \label{fig:hvshC}
        }
    \subfigure[]{
        \includegraphics[width=0.48\textwidth]{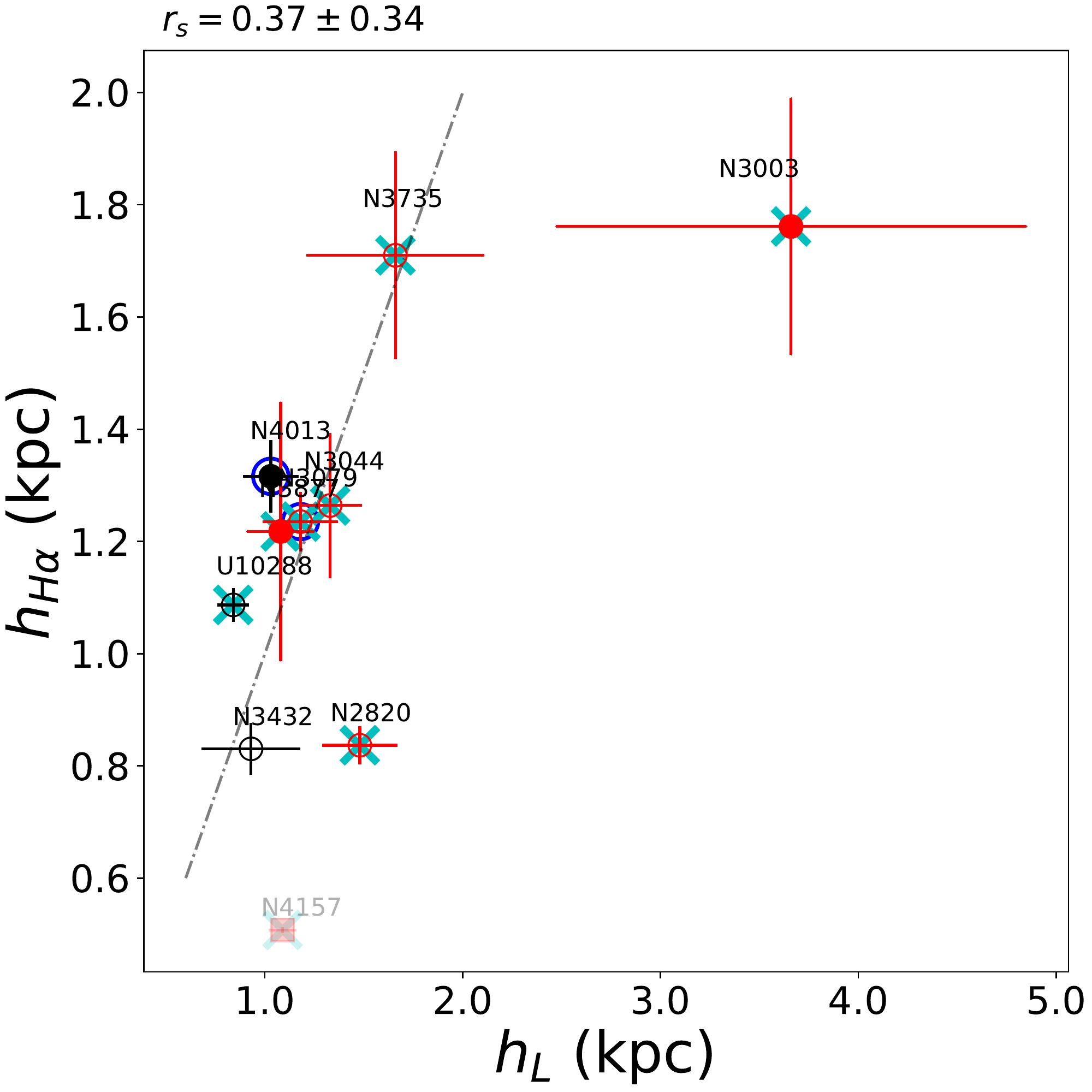}
        \label{fig:hvshL}
        }
    \caption{Comparison of the scale heights of the H$\alpha$ emission with those of (a) \ion{H}{1} $21$ cm, (b) X-ray, (c) radio C-band, and (d) L-band emission.
    The black dot-dashed lines indicate one-to-one correspondence lines.
    The green dot-dashed line in panel (a) indicates the line where $h_{\rm H\alpha}=1/2h_{\rm HI}$.
    The value of $r_{s}$ with its $1\sigma$ error of each relation is noted at the top left of each panel.
    \label{fig:h-otherh}}
\end{figure*}

In this section, we compare our H$\alpha$ scale height of the eDIG to the vertical scale heights of a few other circum-galactic medium (CGM) phases obtained from multi-wavelength observations of the CHANG-ES galaxies: the \ion{H}{1} $21$~cm line tracing the neutral hydrogen ($h_{\rm HI}$; \citealt{Zheng22a,Zheng22b}), the soft X-ray emission tracing the hot gas ($h_{\rm X}$; \citealt{Li13a}), and the radio continuum emission at C-band (centered at 6~GHz) and L-band (centered at 1.5~GHz) tracing the synchrotron emission of cosmic ray (CR) electrons in the galactic scale magnetic field ($h_{\rm C}$ and $h_{\rm L}$; \citealt{Krause18}). The results are presented in Fig.~\ref{fig:h-otherh}, where the black dot-dashed line in each panel indicates where the scale heights of the two phases included in the comparison are equal to each other. We caution that due to the limited number of objects with reliable multi-wavelength measurements, the large uncertainty in the measured scale heights, and the low statistical significance of the relations (see below), the comparisons in this section are largely initial and far from conclusive. But they still show some potentially interesting trends which may worth discussions and some further observational confirmations in the future.

We first compare our H$\alpha$ scale height to the \ion{H}{1} $21$~cm line scale height obtained from \citet{Zheng22a,Zheng22b} (Fig.~\ref{fig:hvshHI}).
\citet{Zheng22a,Zheng22b} studied the \ion{H}{1} $21$~cm line emission from 19 galaxies based on CHANG-ES L-band data observed with the \emph{VLA} in its C-configuration.
These wide-band observations were designed to detect the radio continuum halo, so they are not optimized for studying narrow emission lines, especially their dynamics. Nevertheless, \citet{Zheng22a,Zheng22b} could still extract the \ion{H}{1} $21$~cm line vertical profile and measure its global scale height.

Only eight galaxies have reliable measurements of the scale heights in both our H$\alpha$ sample and \citet{Zheng22a,Zheng22b}'s \ion{H}{1} sample. Most of the galaxies have $h_{\rm H\alpha}<h_{\rm HI}$ (Fig.~\ref{fig:hvshHI}). The median value of $h_{\rm H\alpha}/h_{\rm HI}$ is $0.7\pm 0.1$. A smaller $h_{\rm H\alpha}$ than $h_{\rm HI}$ has also been revealed in some previous works on nearby edge-on galaxies, often indicating an extended \ion{H}{1} envelope (e.g., \citealt{Reach20}). Since the H$\alpha$ emission is $\propto n_{\rm e}^2$, while the \ion{H}{1} emission is $\propto n_{\rm HI}$, we also plot a line of $h_{\rm H\alpha}=1/2h_{\rm HI}$ in Fig.~\ref{fig:hvshHI}, which marks where the H$\alpha$ emission exactly follows the distribution of the atomic gas. Most of the galaxies have $h_{\rm H\alpha}$ slightly larger than $1/2 h_{\rm HI}$. Therefore, the eDIG may be slightly more vertically extended than the atomic CGM. This could possibly be consistent with the existence of some extended sources of ionizing photons, either from the UV background or from the escaped LyC photons from the galactic disk. We caution that the H$\alpha$ and \ion{H}{1} scale heights adopted in the present paper are both insensitive to the large scale structures (partially caused by the high detection limit of the observations; \citealt{Zheng22a}), so the comparison here is probably only sensitive to the compact gas components close to the disk. For example, because the radiative cooling curve peaks at the typical temperature of the warm ionized gas \citep{Sutherland93}, the radiative cooling timescale of the eDIG is often small compared to that of the hot or cold gases or the dynamical timescale of the galaxies. The warm gas in the eDIG thus tends to cool fast and form some filamentary structures. These structures have been detected in optical emission lines on different scales from sub-kpc to $\sim 10^{2}{\rm~kpc}$ (e.g., \citealt{Li08,Li19,Li22,Kenney08,Rupke19,HodgesKluck20}). Because of their low filling factors, our vertical intensity profiles are generally insensitive to these fine structures. Therefore, the vertical extension of the eDIG quoted here only includes the diffuse component and could bias the comparison.

NGC~3044 is the only galaxy in our sample whose H$\alpha$ scale height is larger than the \ion{H}{1} scale height (Fig.~\ref{fig:hvshHI}).
This galaxy has a relatively high dynamical-to-stellar mass ratio ($M_{\rm TF}/M_*\sim 4.12\pm 0.08$, where $M_{\rm TF}$ is the mass calculated with the rotation velocity $v_{\rm rot}$ and the Tully-Fisher relation, while $M_*$ is the photometric stellar mass, both from \citealt{Li16}).
This is significantly higher than the average value of all the CHANG-ES galaxies ($M_{\rm TF}/M_*\sim 2.44\pm 0.01$).
The high $M_{\rm TF}/M_*$ value could be a result of the deeper gravitational potential which may help to confine the cold atomic gas close to the mid-plane of the galaxy \citep{Richards18}.
Another galaxy, NGC~3003, shows an unusually large \ion{H}{1} scale height, together with a very extended eDIG.
The extended and distorted multi-phase CGM could be explained by the tidal interaction with the neighboring galaxy NGC~3021 \citep{Hoopes99, Karthick14}.

X-ray emission from the halos of disk galaxies are most commonly explained as the hot gas produced by stellar feedback (e.g., \citealt{Li13b,Li14}). The eDIG layer or some fine structures in it are found to be spatially correlated with coherent soft X-ray emission features (e.g., \citealt{Strickland04a, Tullmann06a, Tullmann06b, Li08,Li19,Li22}). \citet{Li13a} studied the large-scale diffuse soft X-ray emission around 53 nearby and highly inclined disk galaxies observed by \emph{Chandra}, and measured the exponential scale height of the point-source-removed $0.5-1.5$ keV emission. Although there are only six galaxies with reliable measurements in both H$\alpha$ and X-ray, it seems that most galaxies tend to have an X-ray scale height smaller than the H$\alpha$ scale height, except for two outliers: NGC~4388 and NGC~3079 (Fig. \ref{fig:hvshX}). These two galaxies are well known for their extremely large H$\alpha$ and X-ray filaments extending to a few tens of kpc above the disk (e.g., \citealt{Cecil01, Yoshida02, Iwasawa03, Strickland04a, Kenney08, Li19, HodgesKluck20}). However, the H$\alpha$ filaments always appear much thinner than the coherent X-ray filaments, because they are intrinsically denser and less volume filling, and also because the H$\alpha$ images typically have higher angular resolutions than the X-ray images. Therefore, these large scale filaments may not contribute significantly in our H$\alpha$ scale height measurements, but they could largely affect the X-ray scale height. If we do not consider these two outliers, the typically smaller scale height in X-ray than in H$\alpha$ may suggest that X-ray emission detected with typically shallow exposures is actually from the thick disk in the close vicinity of the galaxies instead of the extended CGM, representing the interplay between hot and cool gases \citep{Bogdan13,Anderson16,Li17,Li18}.

\citet{Krause18} measured the radio continuum scale heights in C-band (centered at $6$~GHz) and L-band (centered at $1.5$~GHz) of 13~CHANG-ES galaxies, by applying a two-component exponential function.
They found no clear correlations between the radio scale heights and the SFR or SFR surface density.
The average values of the radio scale heights of the extended halo component are $h_{\rm C}=1.1\pm 0.3$ kpc in C-band and $h_{\rm L}=1.4\pm 0.7$ kpc in L-band, which are comparable to the median H$\alpha$ scale height obtained in the present paper ($1.13\pm 0.14$ kpc).
As shown in Fig.~\ref{fig:hvshC} and \ref{fig:hvshL}, except for the significantly larger L-band scale height of NGC~3003 (and possibly the slightly larger radio scale heights of NGC~2820 in both bands), most of the galaxies have a smaller scale height in C-band, while a comparable scale height in L-band, than the H$\alpha$ scale height.
The scale height of the C-band emission is smaller than those of the L-band and H$\alpha$.
This could be explained by the stronger synchrotron cooling at the higher CR electron energy or a larger fraction of thermal radio emission at higher frequencies (e.g., \citealt{Vargas18}).
The most significant outlier, NGC~3003, as mentioned above, is a galaxy undergoing interaction with a neighboring galaxy, which may result in unusually large radio features, at least on low resolution images.
Furthermore, there exists a background radio source projected close to the disk of this galaxy, which may also bias the radio scale height measurements \citep{Wiegert15}.
This galaxy is also a significant outlier in many statistical analyses of radio scale heights \citep{Krause18}.

The comparable scale heights in H$\alpha$ and L-band might not be by chance.
The synchrotron emission intensity can be described as:
\begin{equation}\label{equi:Isyn}
    I_{\rm syn}\propto n_{\rm CRE} B^{1+\alpha_{\rm syn}},
\end{equation}
where $n_{\rm CRE}$ is the number density of the CR electrons responsible for the synchrotron emission, $B$ is the magnetic field strength, and $\alpha_{\rm syn}$ is the synchrotron spectral index.
Therefore, the corresponding scale heights of the synchrotron emission, CR electrons, and the magnetic field, have the following relationship:
\begin{equation}\label{equi:ScaleHeight1}
    1/h_{\rm syn}=1/h_{\rm CRE}+(1+\alpha_{\rm syn})/h_{\rm B}.
\end{equation}
If we assume energy equipartition between the CR electrons and the magnetic field, i.e., $n_{\rm CRE}\propto B^2$, we can obtain $h_{\rm CRE}=h_{\rm B}/2$.
Substituting this into Eq.~\ref{equi:ScaleHeight1}, we can obtain the relation between the synchrotron emission scale height and the CR electron or magnetic field scale height:
\begin{equation}\label{equi:ScaleHeight2}
\begin{aligned}
    &h_{\rm syn}=h_{\rm B}/(3+\alpha_{\rm syn}),\\
    &h_{\rm syn}=2h_{\rm CRE}/(3+\alpha_{\rm syn}).
\end{aligned}
\end{equation}
$\alpha_{\rm syn}$ depends on the star formation properties. 
For most of the CHANG-ES galaxies without a radio-bright AGN, the typical value of $\alpha_{\rm syn}$ is $\sim1$ \citep{Li16}.
Therefore, we obtain $h_{\rm syn}=h_{\rm B}/4=h_{\rm CRE}/2$.
On the other hand, as the H$\alpha$ emission $I_{\rm H\alpha}\propto n_{e}^2$ (here $n_{e}$ is the thermal electron density), we can obtain $h_{\rm H\alpha}=h_{\rm e}/2$.
From Fig.~\ref{fig:hvshL}, we see $h_{\rm H\alpha}\approx h_{\rm L}$, so $h_{\rm e}\approx h_{\rm CRE}$.

The thermal and non-thermal electrons are produced by different mechanisms, but they have comparable scale heights.
This indicates that they may be linked in some ways.
A possible explanation is that both of them are transported outwards by an accelerating galactic wind.
They suffer from the same adiabatic expansion, so their vertical distributions are similar.
The similar scale heights of the thermal and non-thermal electrons could even indicate energy equipartition between the thermal gas and the CR electrons, as also suggested in magnetohydrodynamics (MHD) simulations at a relatively small distance from the galactic plane (e.g., $z\lesssim1\rm~kpc$; \citealt{Girichidis18}).
However, we caution that at larger distances, the CR pressure typically decreases slower than the thermal pressure and could play a more important role (e.g., \citealt{Everett08,Girichidis18,Hopkins20}).

\section{Summary} \label{sec:Summary}

In this paper, we present measurements and a statistical analysis of the eDIG of 22 CHANG-ES galaxies, using on H$\alpha$ images taken with the APO~3.5m telescope.
We characterize the vertical extension of the H$\alpha$ emission with a two-component exponential model.
While the compact component in the two-component model is thought to be the galactic thin disk, we assume that the extended component represents the extended H$\alpha$ envelope or the thick disk surrounding the galaxies.
The latter is used in the follow-up statistical analysis.

The median H$\alpha$ scale height of the thick disk component for the 16 galaxies where it is detected is $\langle {h}_{\rm H\alpha}\rangle =1.13\pm 0.14~{\rm kpc}$.
We also compare $h_{\rm H\alpha}$ of our sample galaxies to those from a bigger sample, including dwarf galaxies.
We examined the relation between $h_{\rm H\alpha}$ and some global galaxy parameters, and find a tight sub-linear (with a logarithm slope of $\alpha=0.29$) correlation between $h_{\rm H\alpha}$ and the SFR.
We also find an anti-correlation between the offsets from the best-fit SFR-$h_{\rm H\alpha}$ relation (characterized by $h_{\rm H\alpha}$/SFR$^\alpha$, where $\alpha=0.29$) and the SFR surface density, $\rm SFR_{SD}$.
This secondary effect indicates that galaxies with more intense star formation tend to have disproportionately extended eDIG compared to those with less intense star formation.
The vertical extension of the eDIG is affected by both the spatial distribution of neutral gas and ionizing photons, which need to be further studied with numerical simulations and deeper H$\alpha$ observations.

Based on the multi-wavelength data collected by the CHANG-ES consortium, we further compare the vertical extensions of the eDIG and the other phases of the CGM.
When comparing to the \ion{H}{1} 21-cm line scale height measured from the CHANG-ES \emph{VLA} data, we find that most of the galaxies have $h_{\rm H\alpha}/h_{\rm HI}=0.5-1$, with a median value of $\langle h_{\rm H\alpha}/h_{\rm HI}\rangle = 0.7\pm 0.1$.
This means that the thick disk component of the eDIG is slightly more extended than the cold neutral gas envelope, not accounting for the very extended component of both phases.
Most galaxies have an X-ray scale height smaller than the H$\alpha$ scale height, suggesting that the majority of the X-ray emission detected in shallow observations is actually from the thick disk rather than the extended CGM.

The H$\alpha$ scale height is in general comparable to the L-band radio continuum scale height, both slightly larger than the scale height at higher radio frequencies (C-band).
This indicates that the thermal and non-thermal electrons have similar spatial distributions, at least close to the galactic disk, a natural result if both of them are transported outwards by a galactic wind.
This explanation further indicates that the thermal gas, the CRs, and the magnetic field are probably close to energy equipartition, which has important implications in understanding the galactic outflow.
The smaller scale height in C-band could be explained by a stronger synchrotron cooling or a larger thermal fraction at higher frequencies.

\acknowledgments

This work is supported by the National Key R\&D Program of China No. 2017YFA0402600, and NSFC grants No. 11890692 and 12133008. We acknowledge the science research grants from the China Manned Space Project with NO. CMS-CSST-2021-A04.
H.L. is supported by NASA through the NASA Hubble Fellowship grant HST-HF2-51438.001-A awarded by the Space Telescope Science Institute, which is operated by the Association of Universities for Research in Astronomy, Incorporated, under NASA contract NAS5-26555. RJD acknowledges support by Deutsche Forschungsgemeinschaft Sonderforschungsbereich 1491. 

Data Availability

The H$\alpha$ images and vertical intensity profiles of individual galaxies are available as supplementary materials. Figures of them will be put online only. The original data could be shared by reasonable request to the corresponding author.

\begin{appendix} \label{Appendix}

Images and H$\alpha$ vertical intensity profiles of individual galaxies are included in the appendix as online only supplementary materials. These figures of individual galaxies are similar as the example shown in Fig.~\ref{fig:N3003_result}.

\begin{figure}[!h]
    \centering
    \subfigure[]{
        \includegraphics[width=0.401\textwidth]{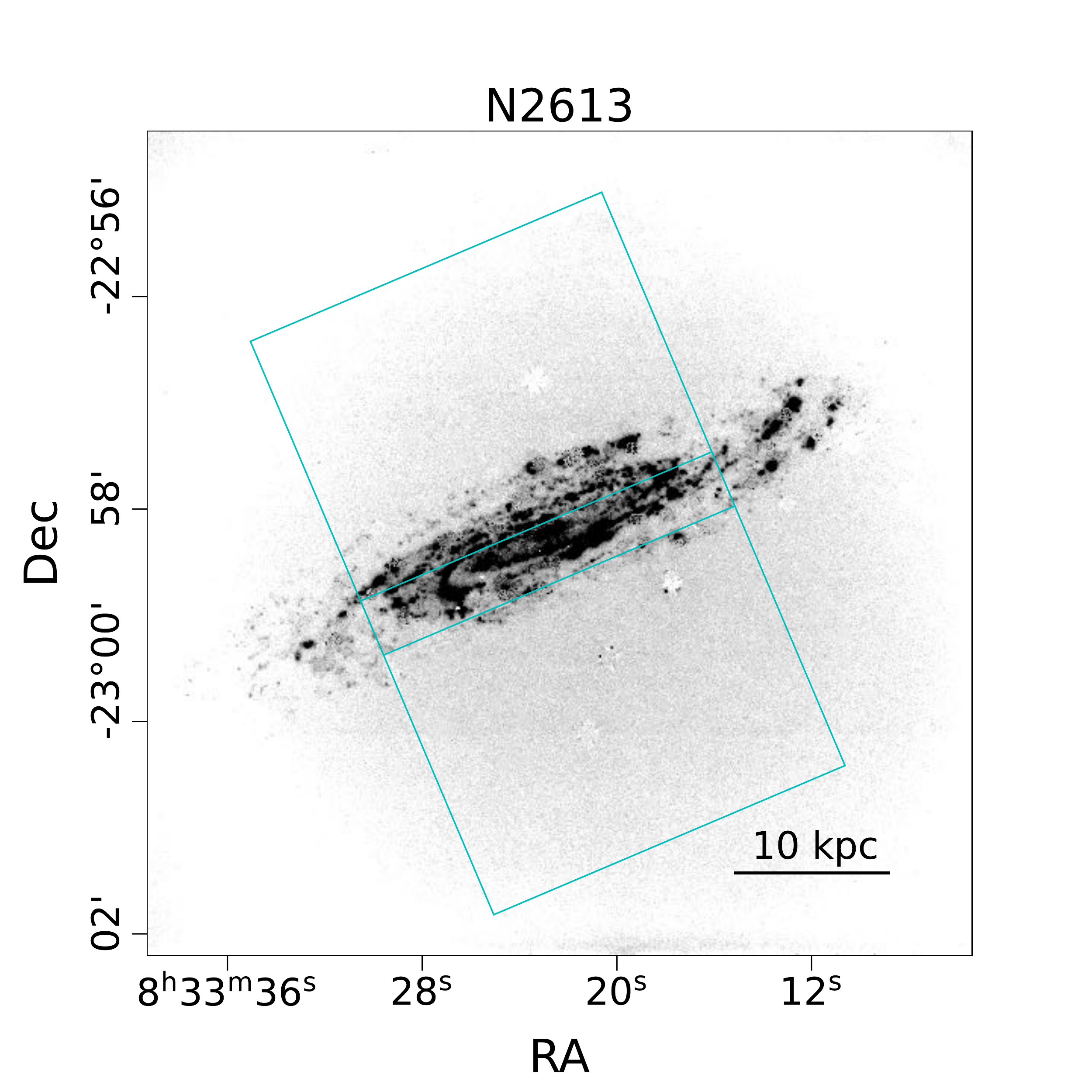}
        }
    \subfigure[]{
        \includegraphics[width=0.561\textwidth]{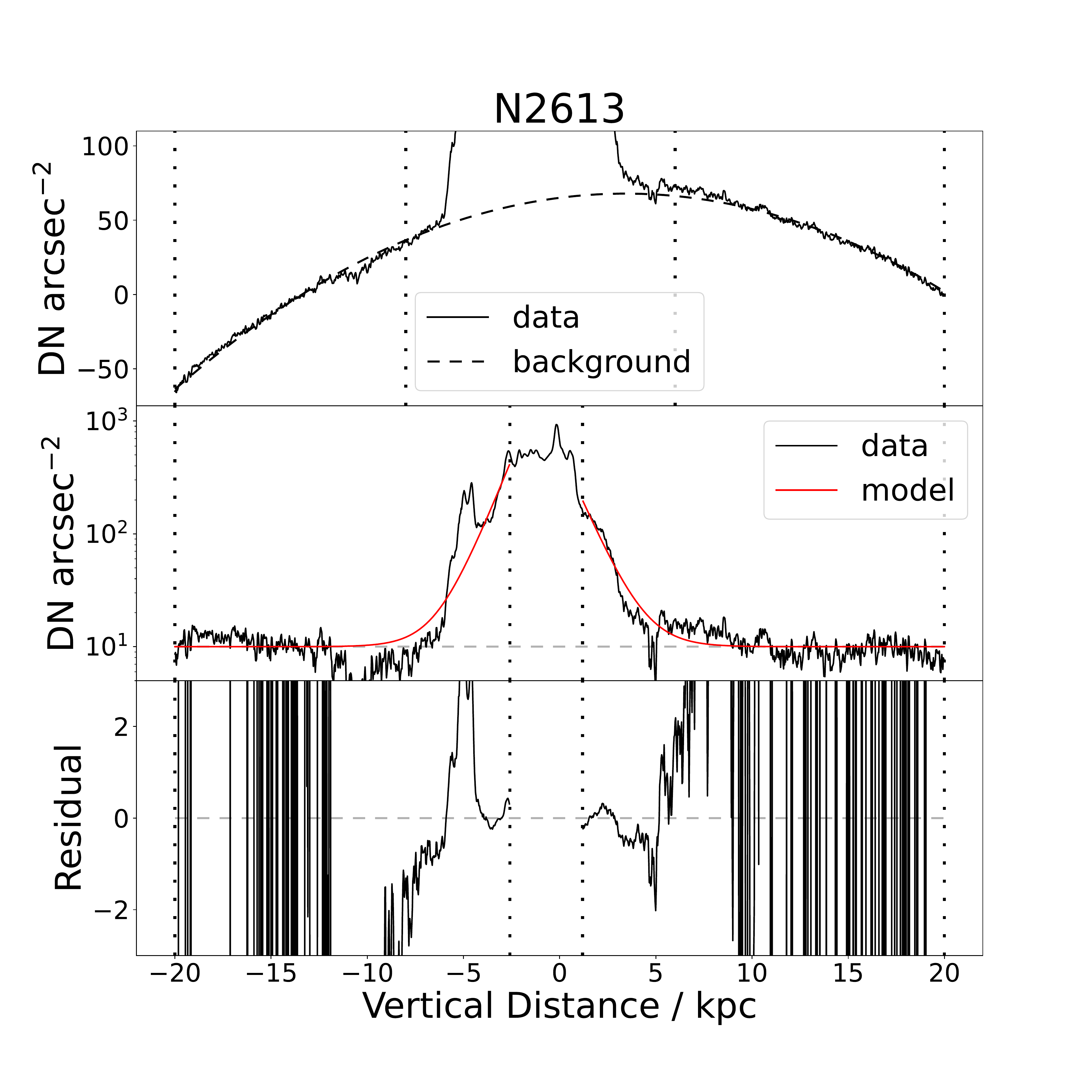}
        }
    \caption{Same as Fig. \ref{fig:N3003_fit}.
    }
\end{figure}
\begin{figure}[!h]
    \centering
    \subfigure[]{
        \includegraphics[width=0.401\textwidth]{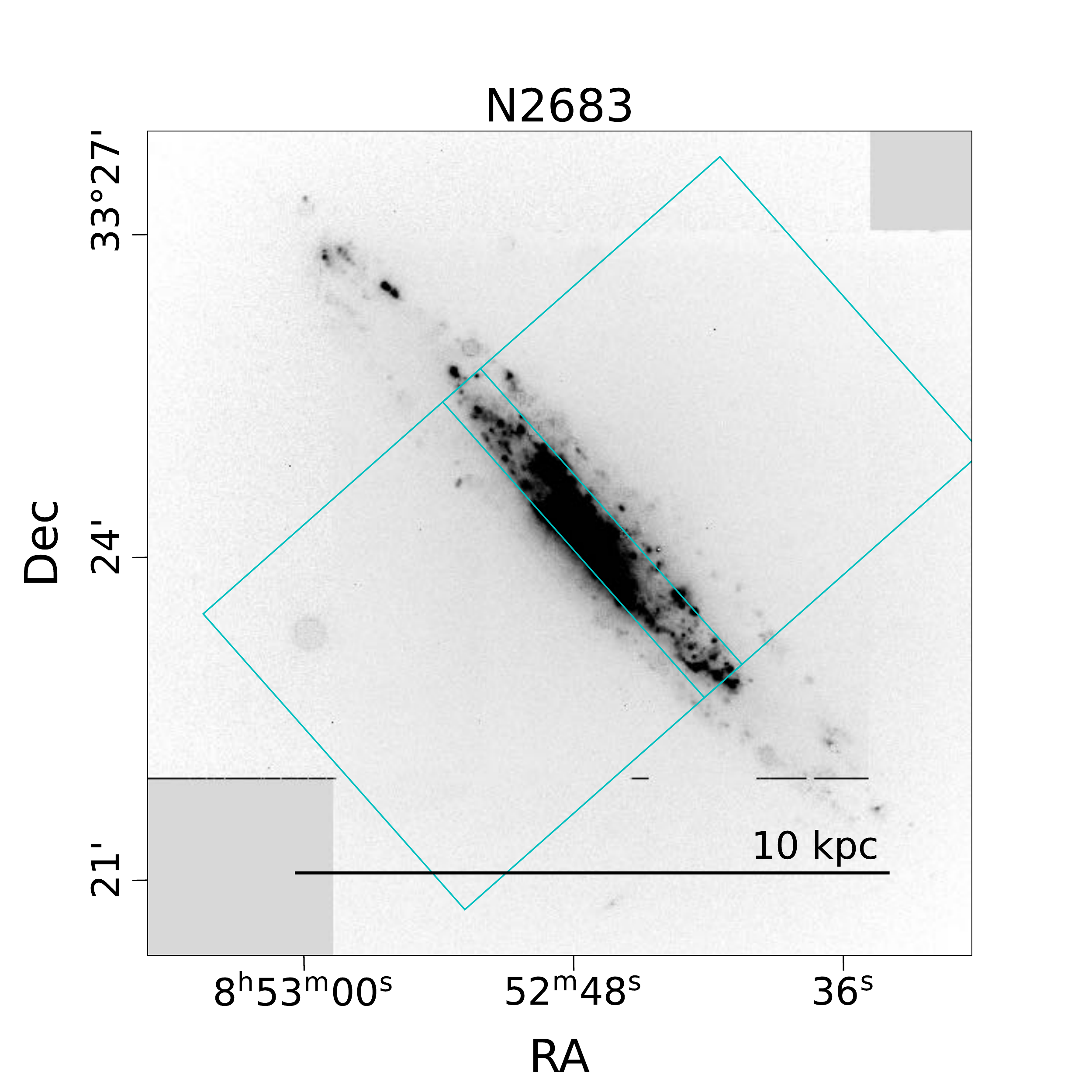}
        }
    \subfigure[]{
        \includegraphics[width=0.561\textwidth]{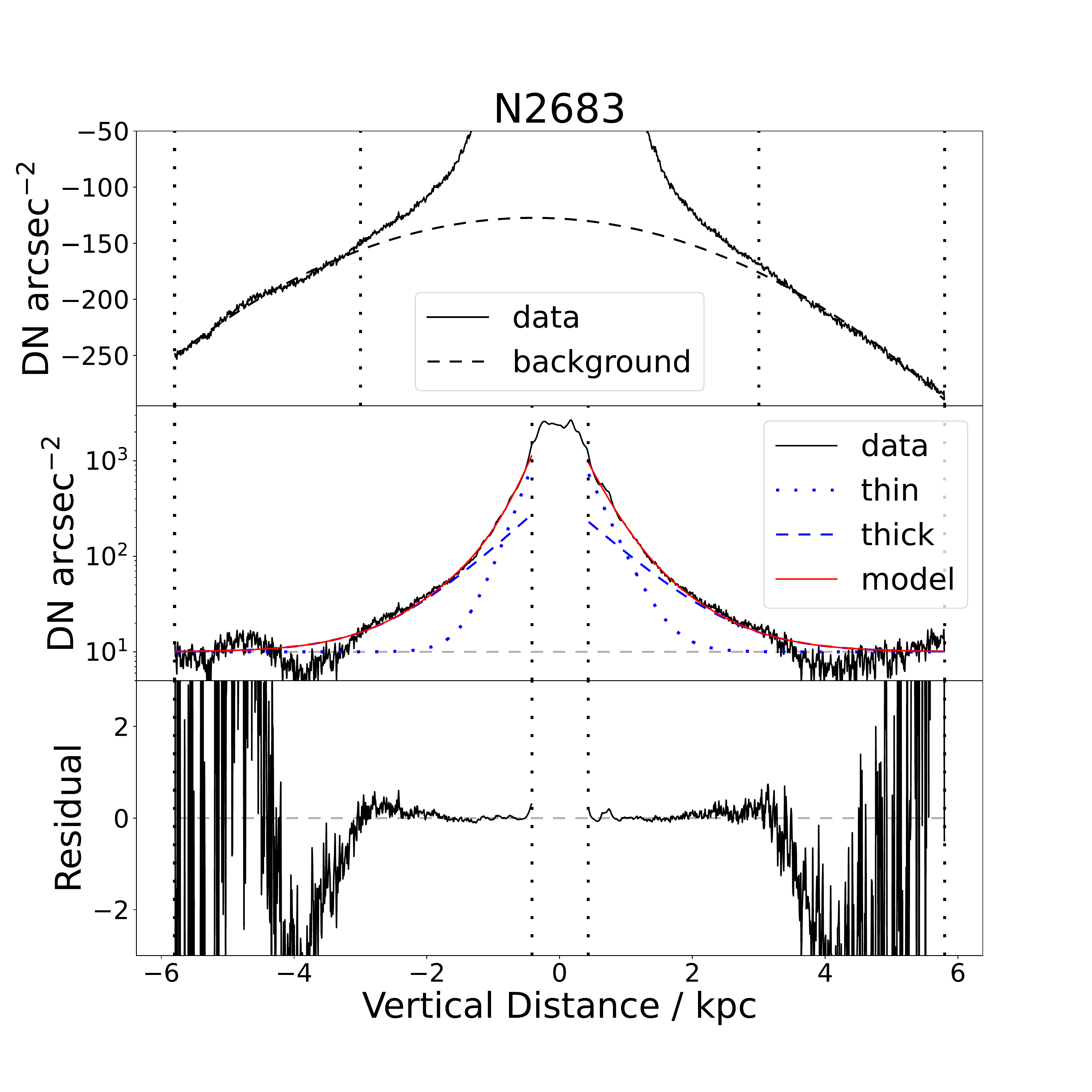}
        }
    \caption{
    }
\end{figure}
\begin{figure}[!h]
    \centering
    \subfigure[]{
        \includegraphics[width=0.401\textwidth]{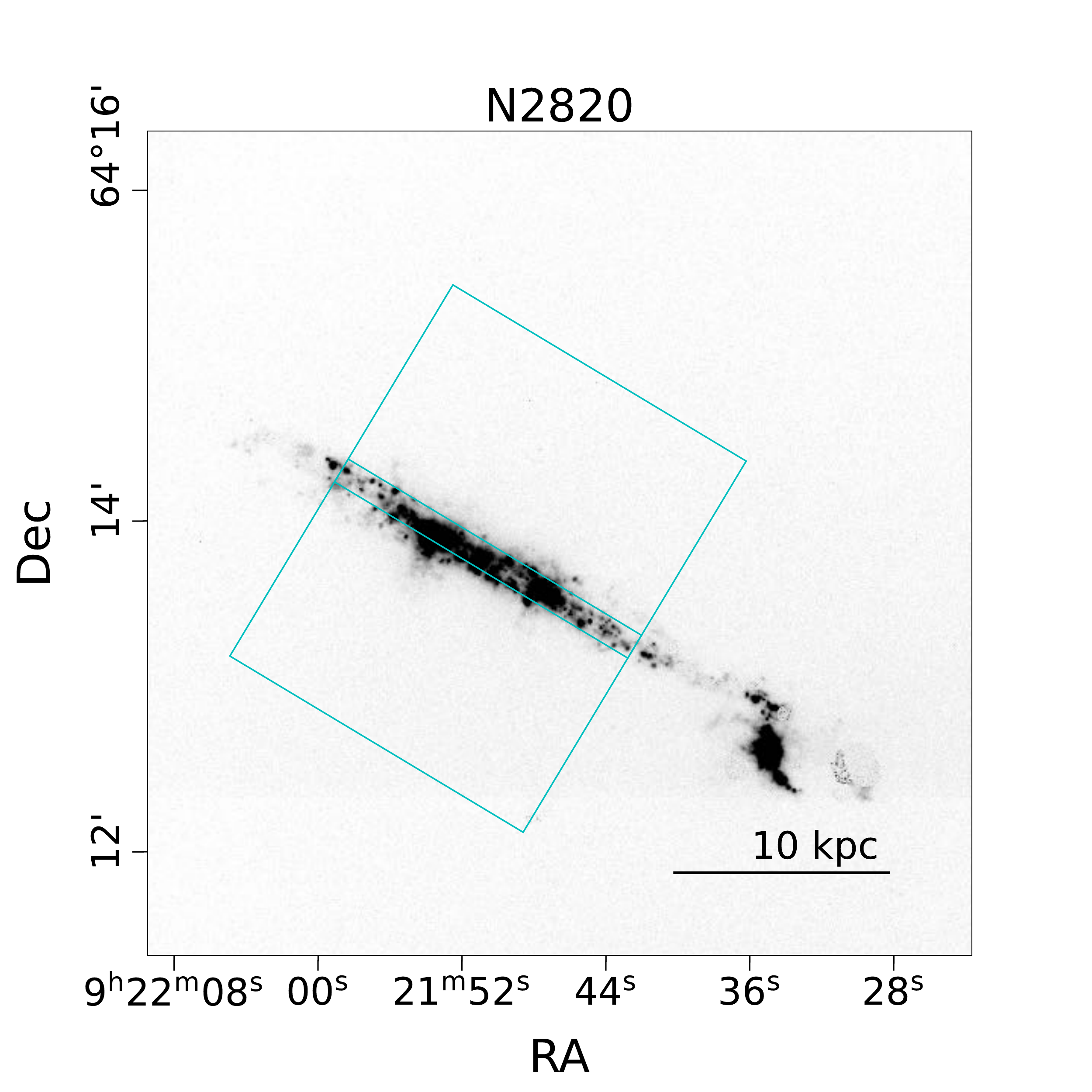}
        }
    \subfigure[]{
        \includegraphics[width=0.561\textwidth]{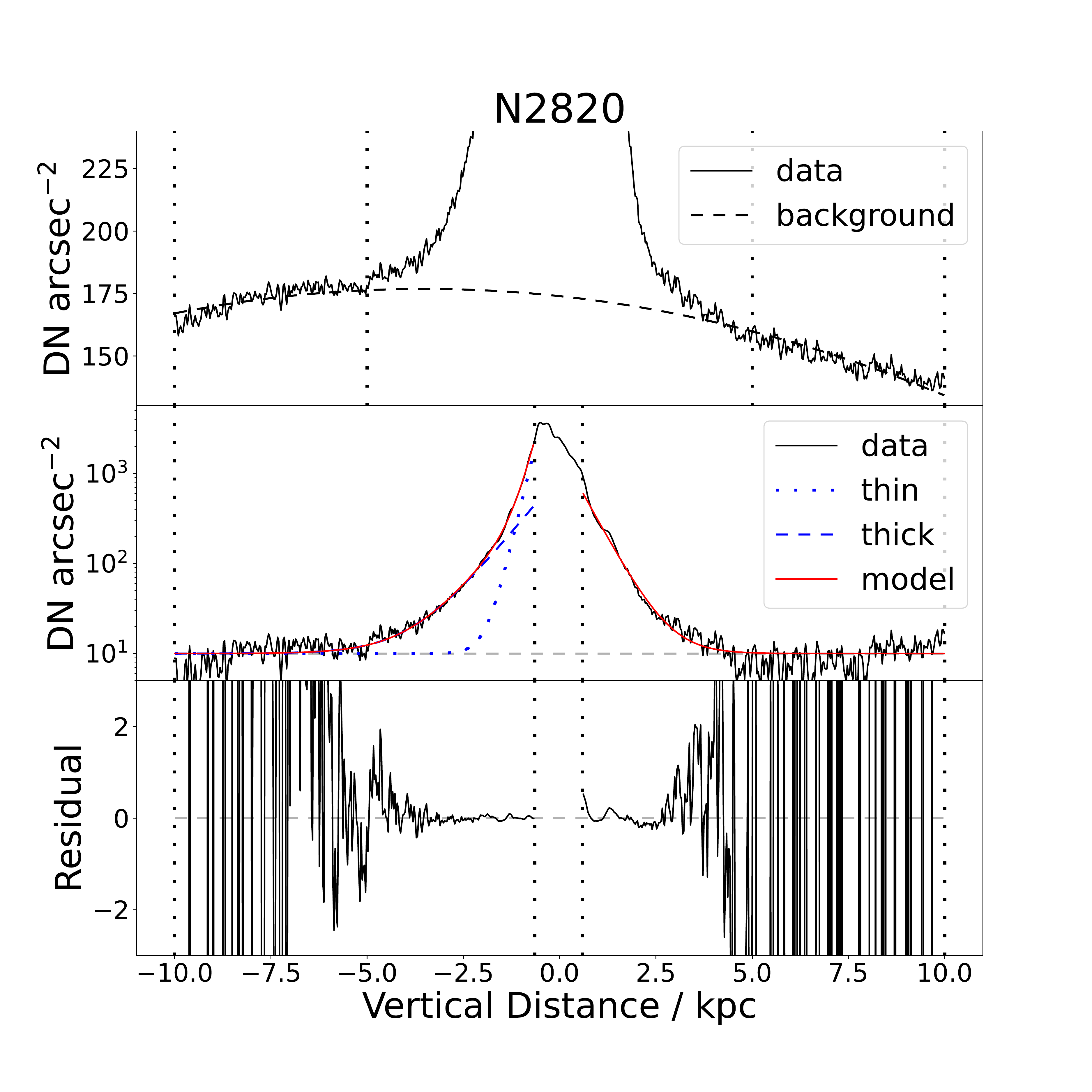}
        }
    \caption{
    }
\end{figure}
\begin{figure}[!h]
    \centering
    \subfigure[]{
        \includegraphics[width=0.401\textwidth]{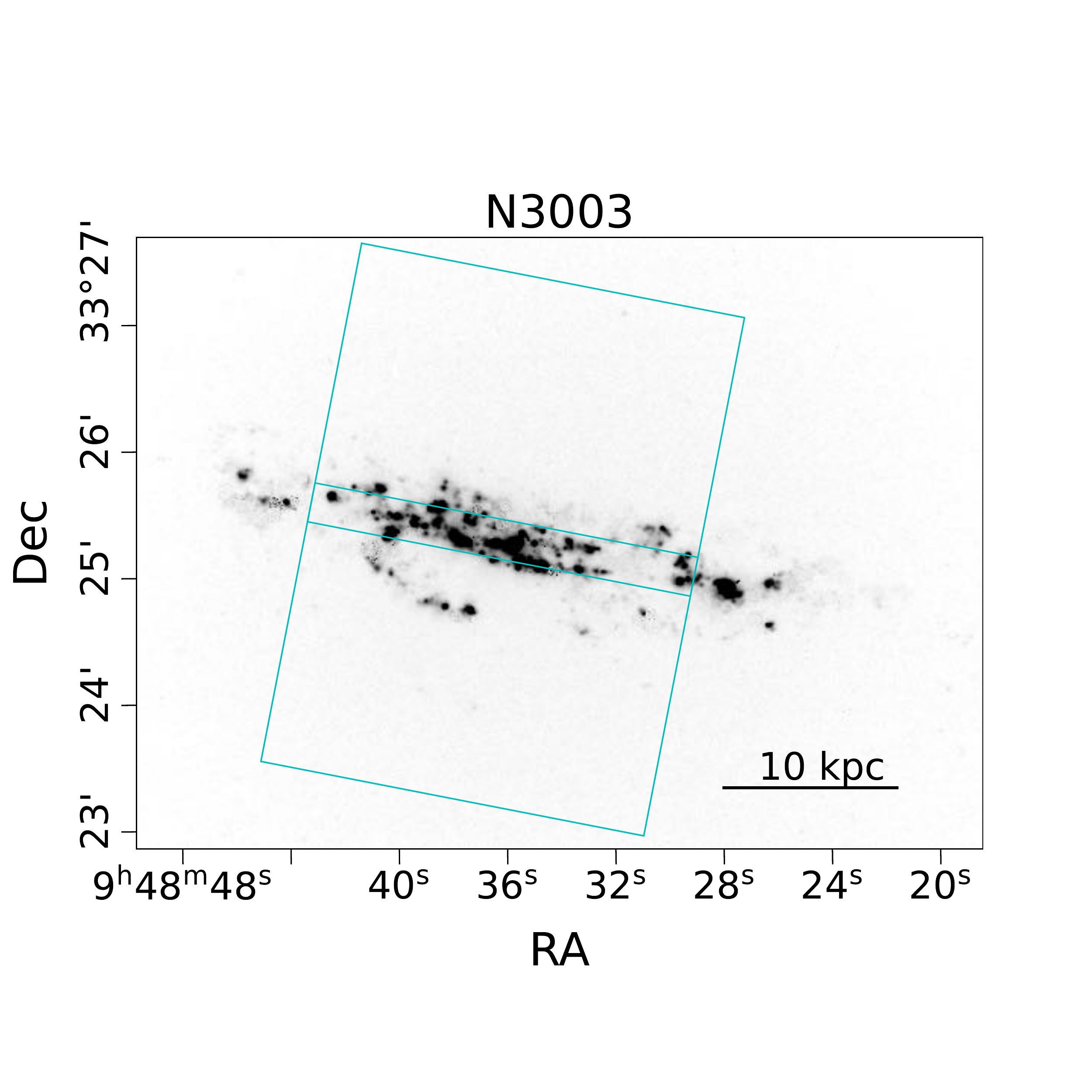}
        }
    \subfigure[]{
        \includegraphics[width=0.561\textwidth]{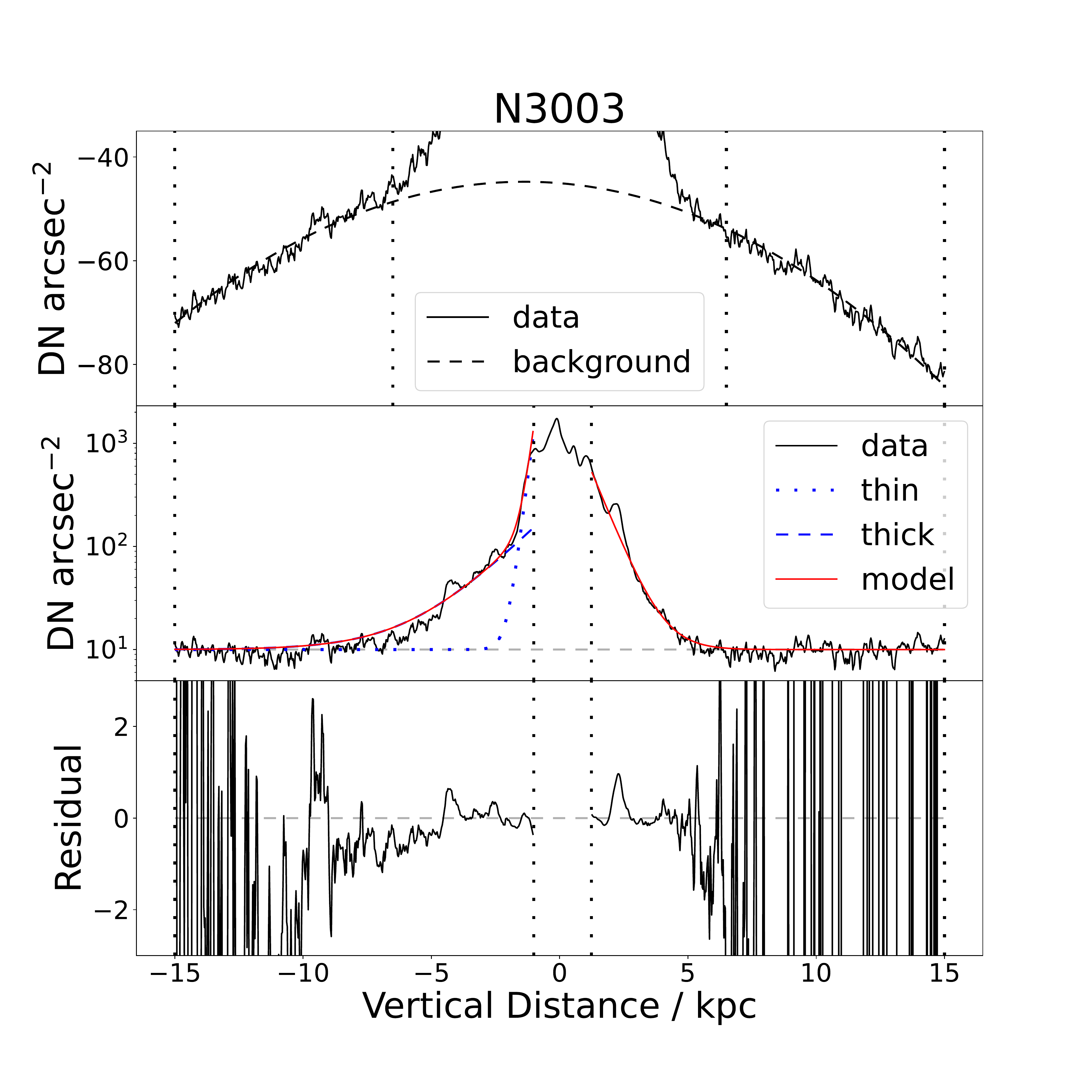}
        }
    \caption{
    }
\end{figure}
\begin{figure}[!h]
    \centering
    \subfigure[]{
        \includegraphics[width=0.401\textwidth]{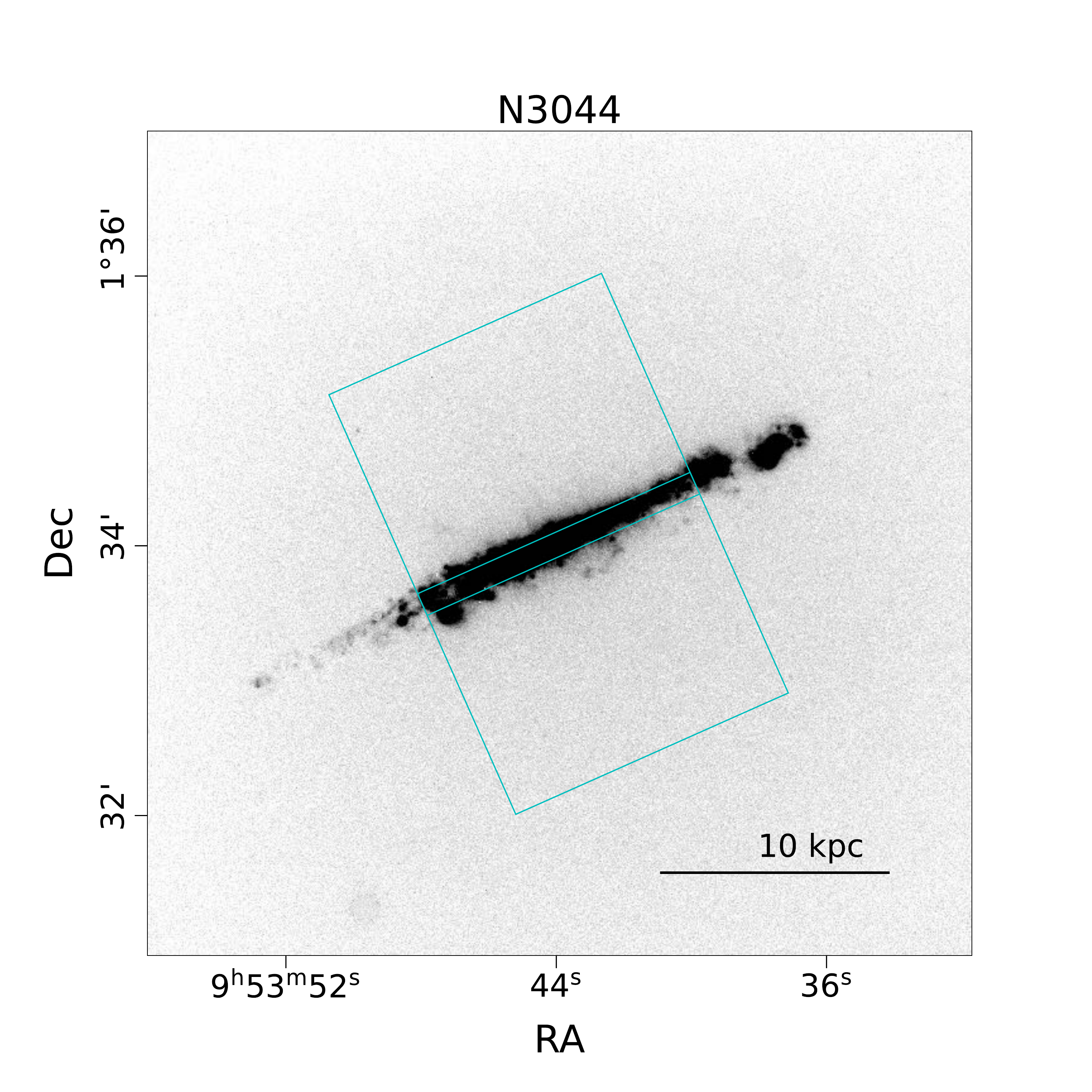}
        }
    \subfigure[]{
        \includegraphics[width=0.561\textwidth]{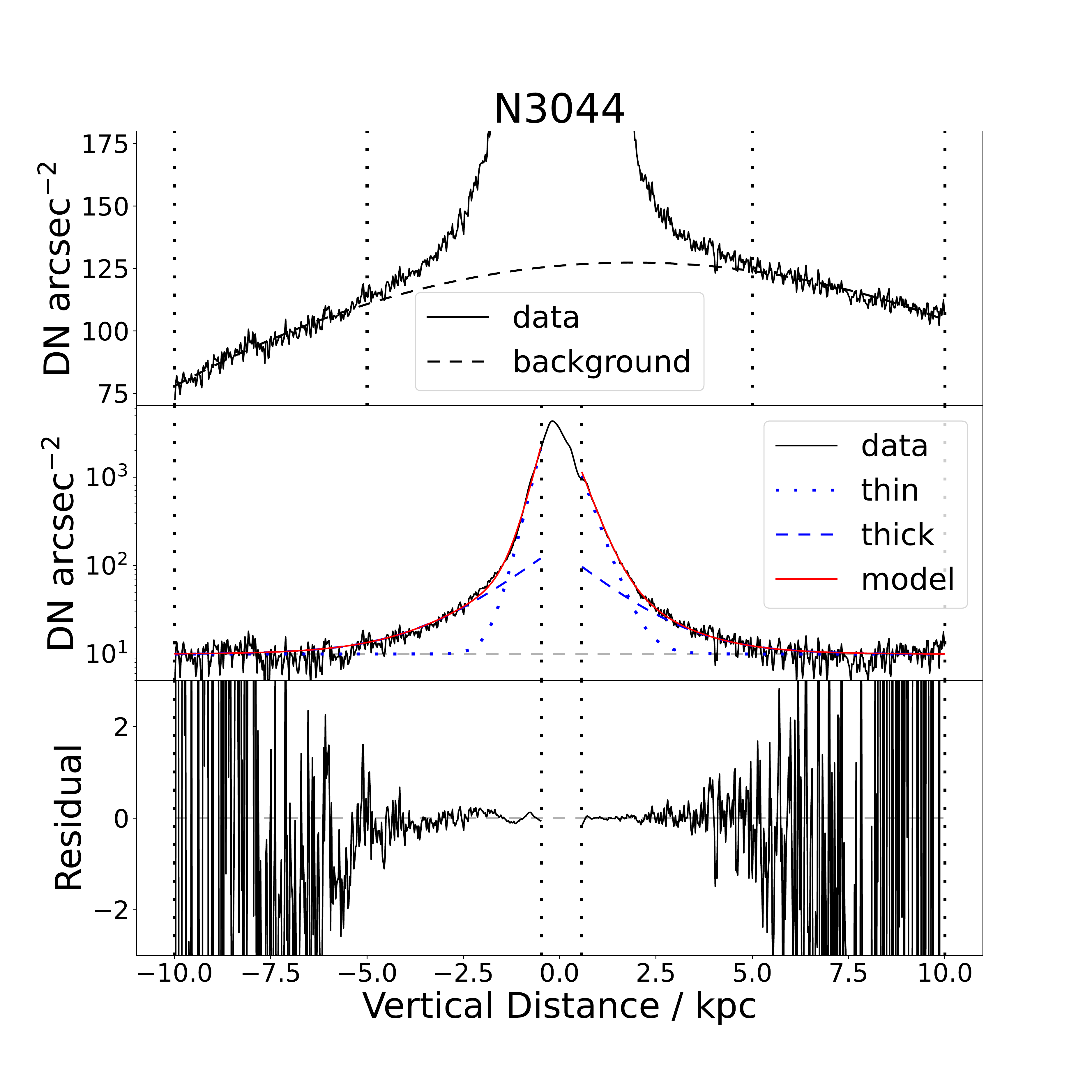}
        }
    \caption{
    }
\end{figure}
\begin{figure}[!h]
    \centering
    \subfigure[]{
        \includegraphics[width=0.401\textwidth]{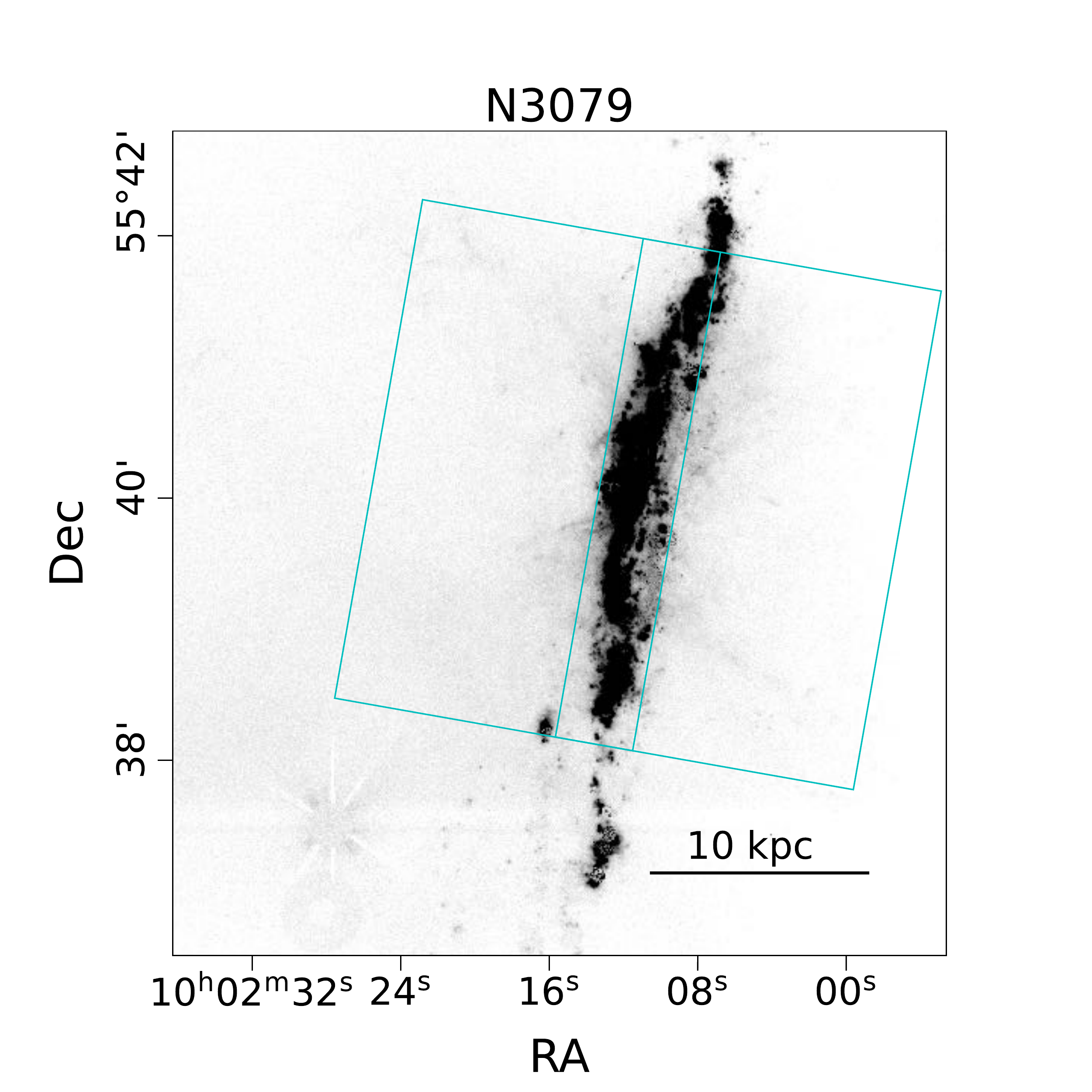}
        }
    \subfigure[]{
        \includegraphics[width=0.561\textwidth]{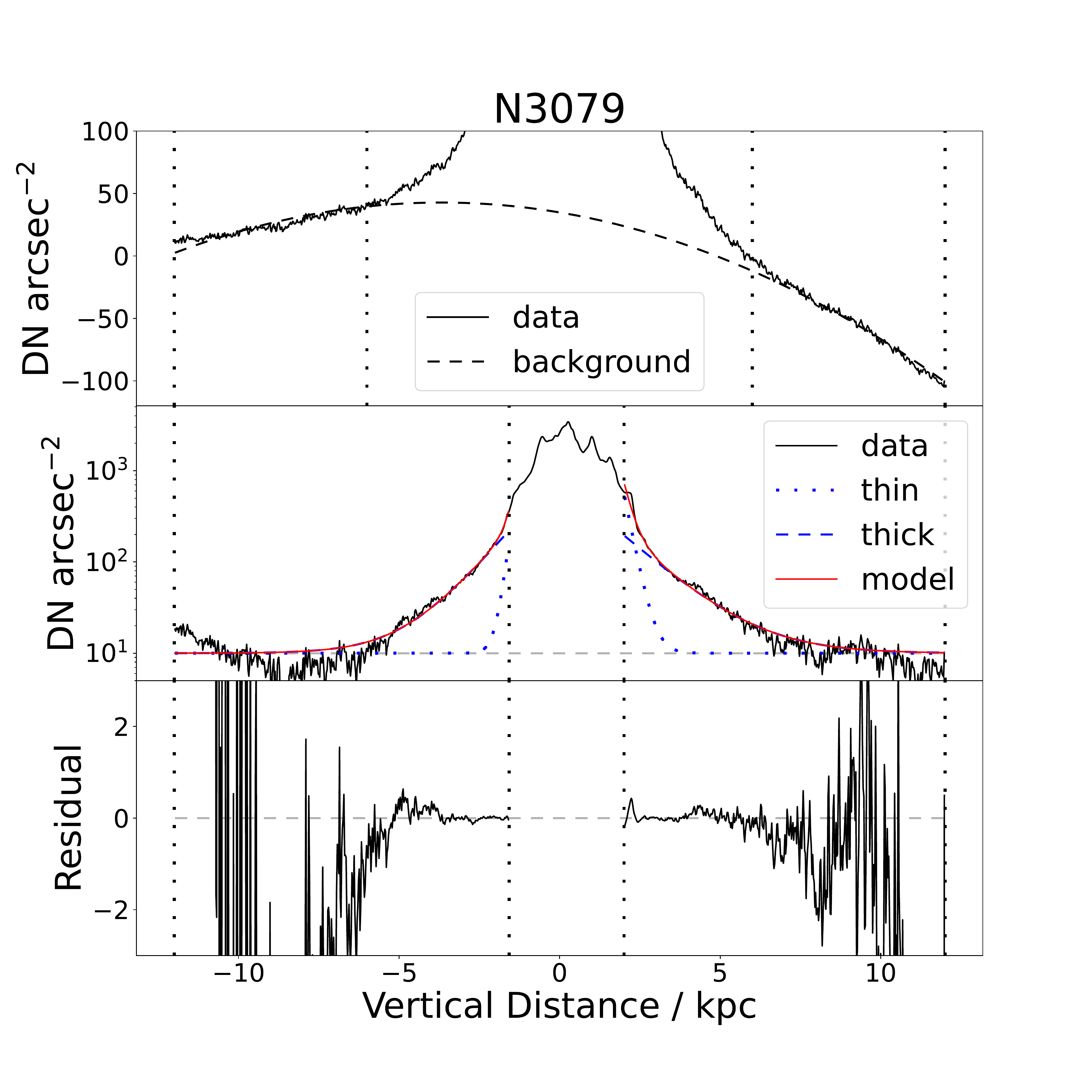}
        }
    \caption{
    }
\end{figure}
\begin{figure}[!h]
    \centering
    \subfigure[]{
        \includegraphics[width=0.401\textwidth]{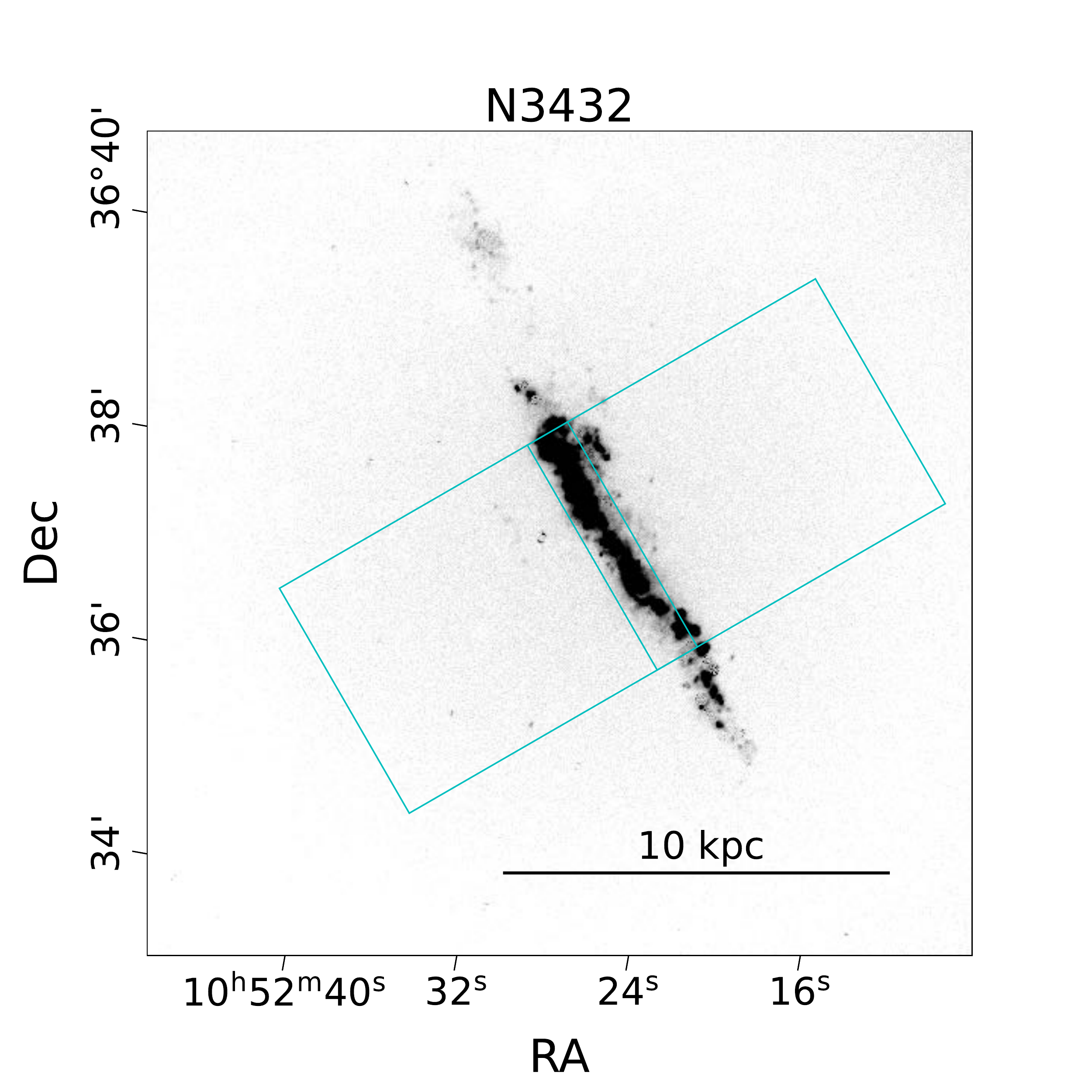}
        }
    \subfigure[]{
        \includegraphics[width=0.561\textwidth]{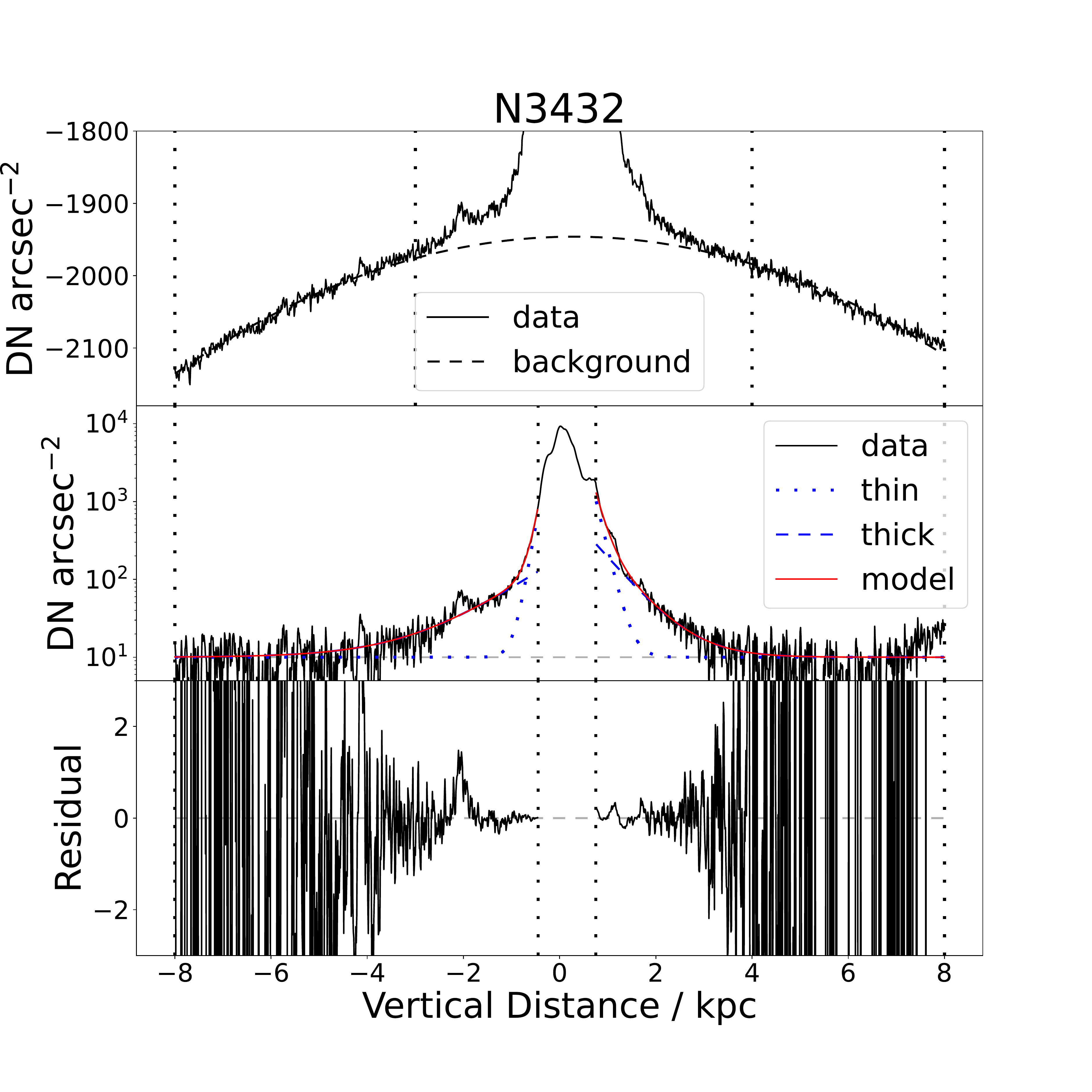}
        }
    \caption{
    }
\end{figure}
\begin{figure}[!h]
    \centering
    \subfigure[]{
        \includegraphics[width=0.401\textwidth]{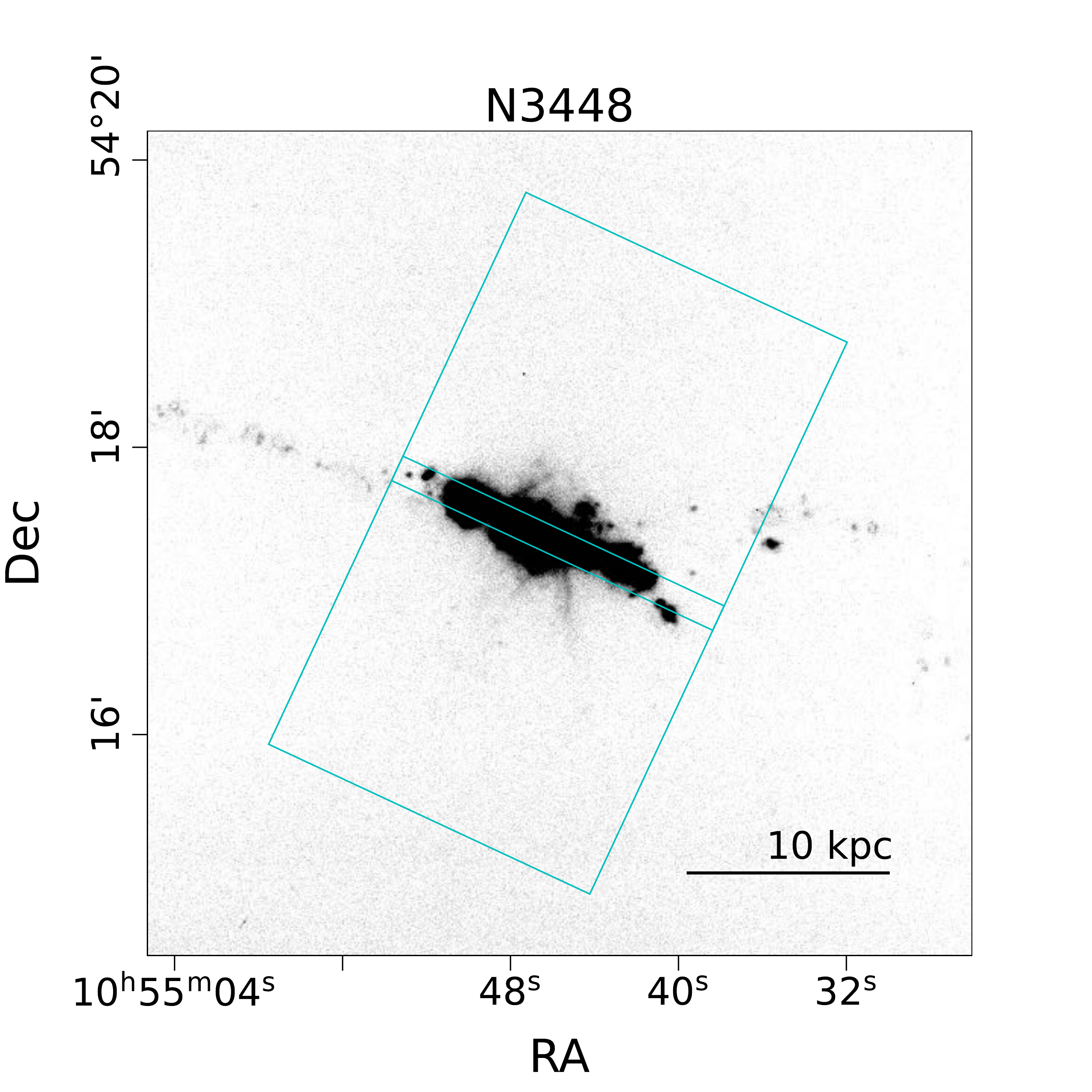}
        }
    \subfigure[]{
        \includegraphics[width=0.561\textwidth]{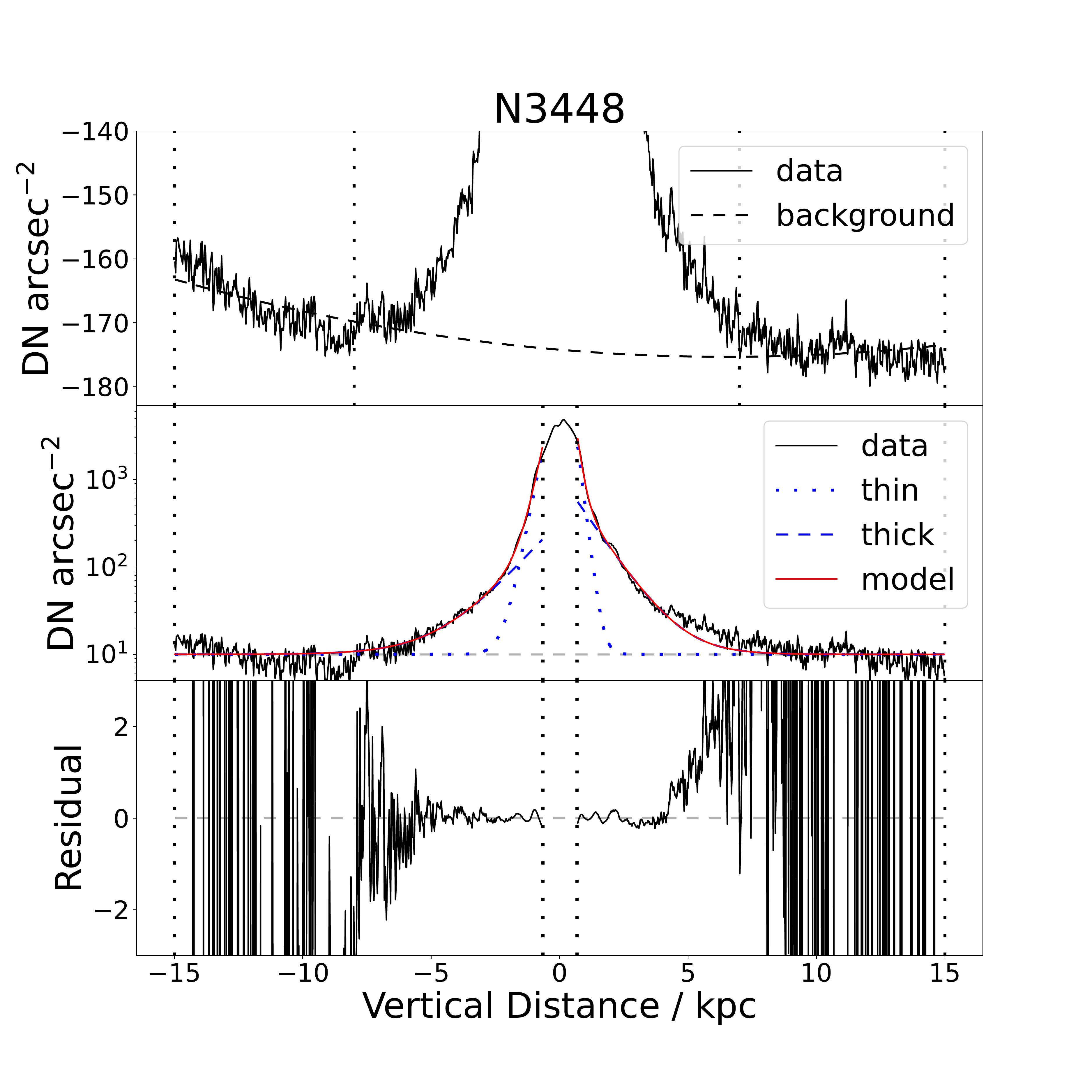}
        }
    \caption{
    }
\end{figure}
\begin{figure}[!h]
    \centering
    \subfigure[]{
        \includegraphics[width=0.401\textwidth]{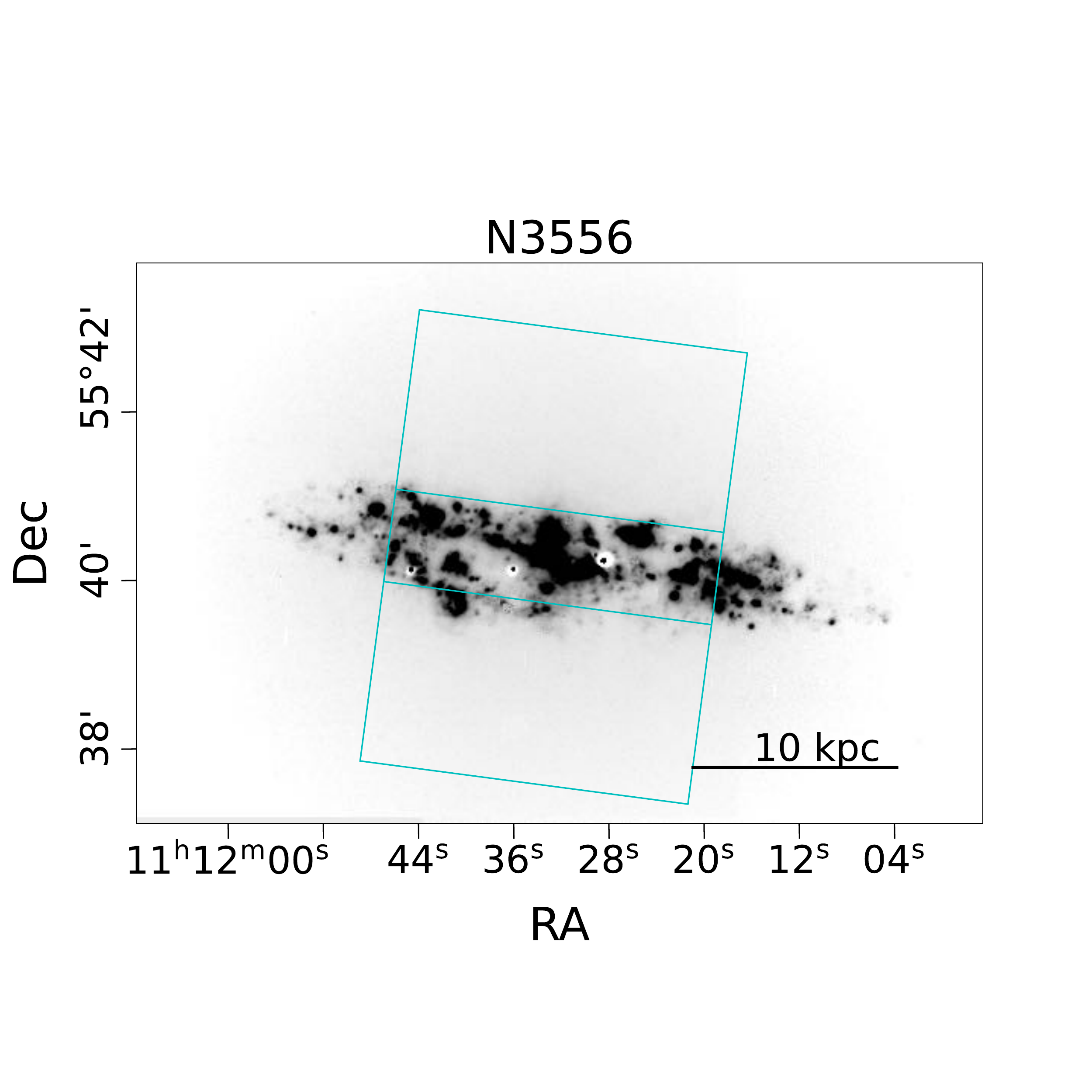}
        }
    \subfigure[]{
        \includegraphics[width=0.561\textwidth]{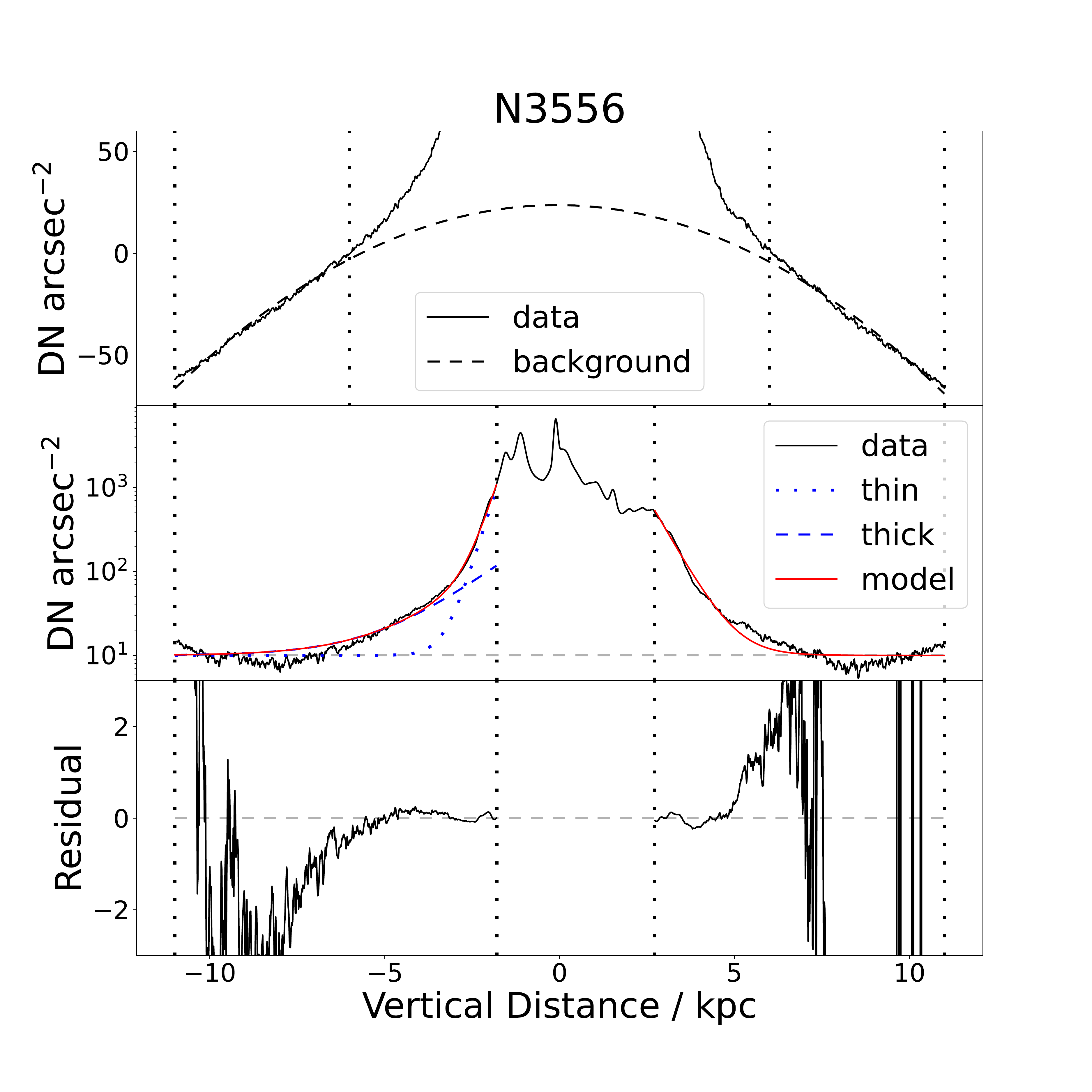}
        }
    \caption{
    }
\end{figure}
\begin{figure}[!h]
    \centering
    \subfigure[]{
        \includegraphics[width=0.401\textwidth]{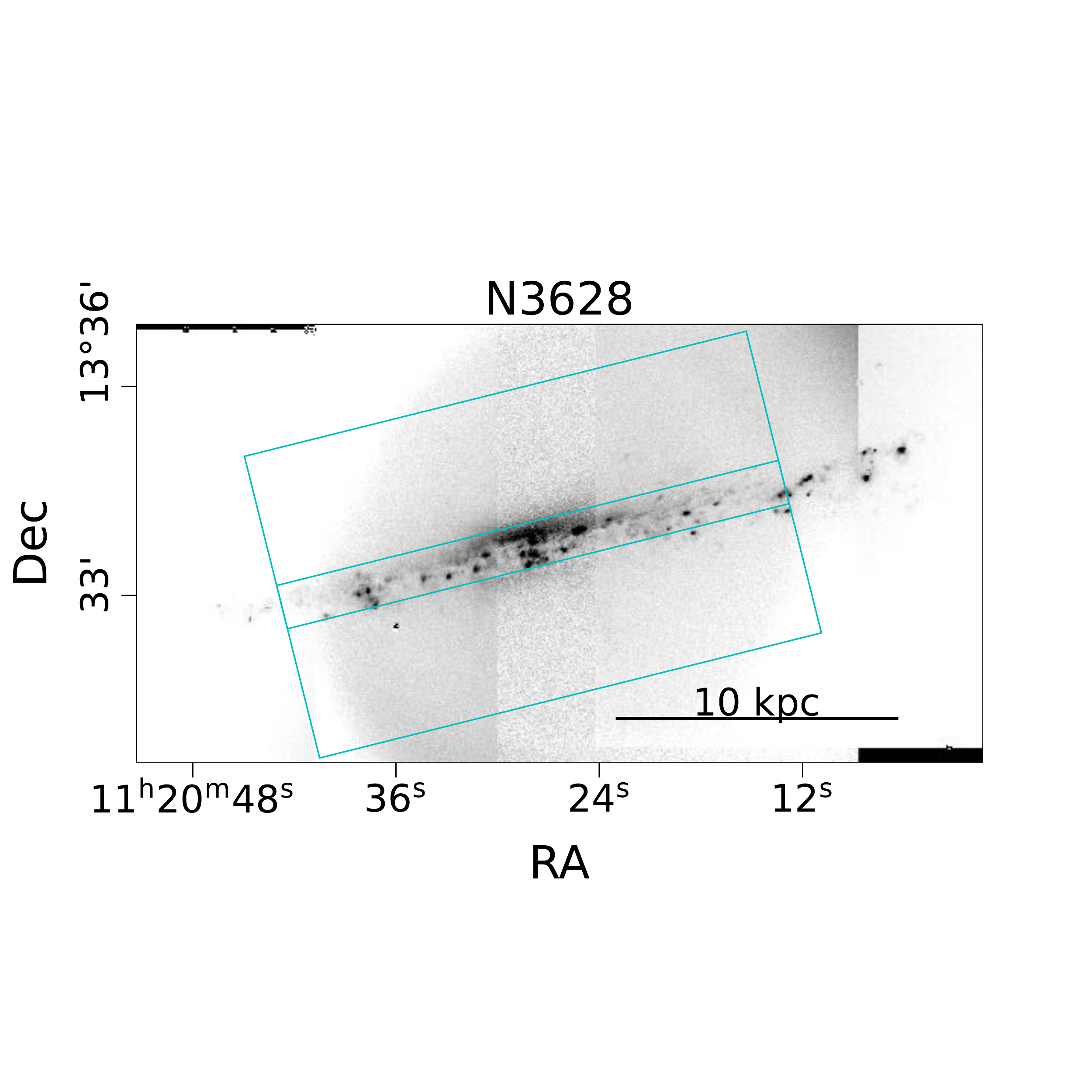}
        }
    \subfigure[]{
        \includegraphics[width=0.561\textwidth]{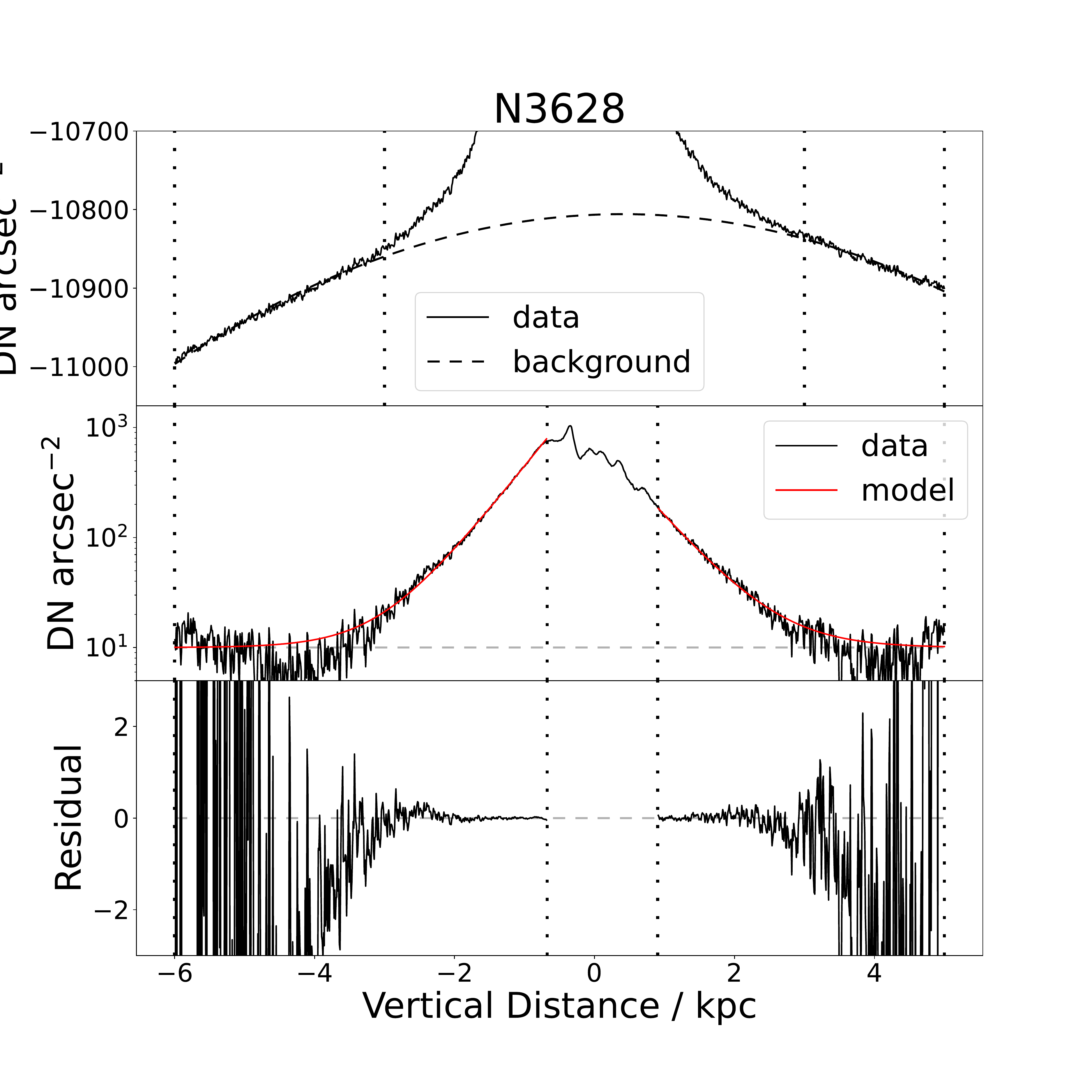}
        }
    \caption{
    }
\end{figure}
\begin{figure}[!h]
    \centering
    \subfigure[]{
        \includegraphics[width=0.401\textwidth]{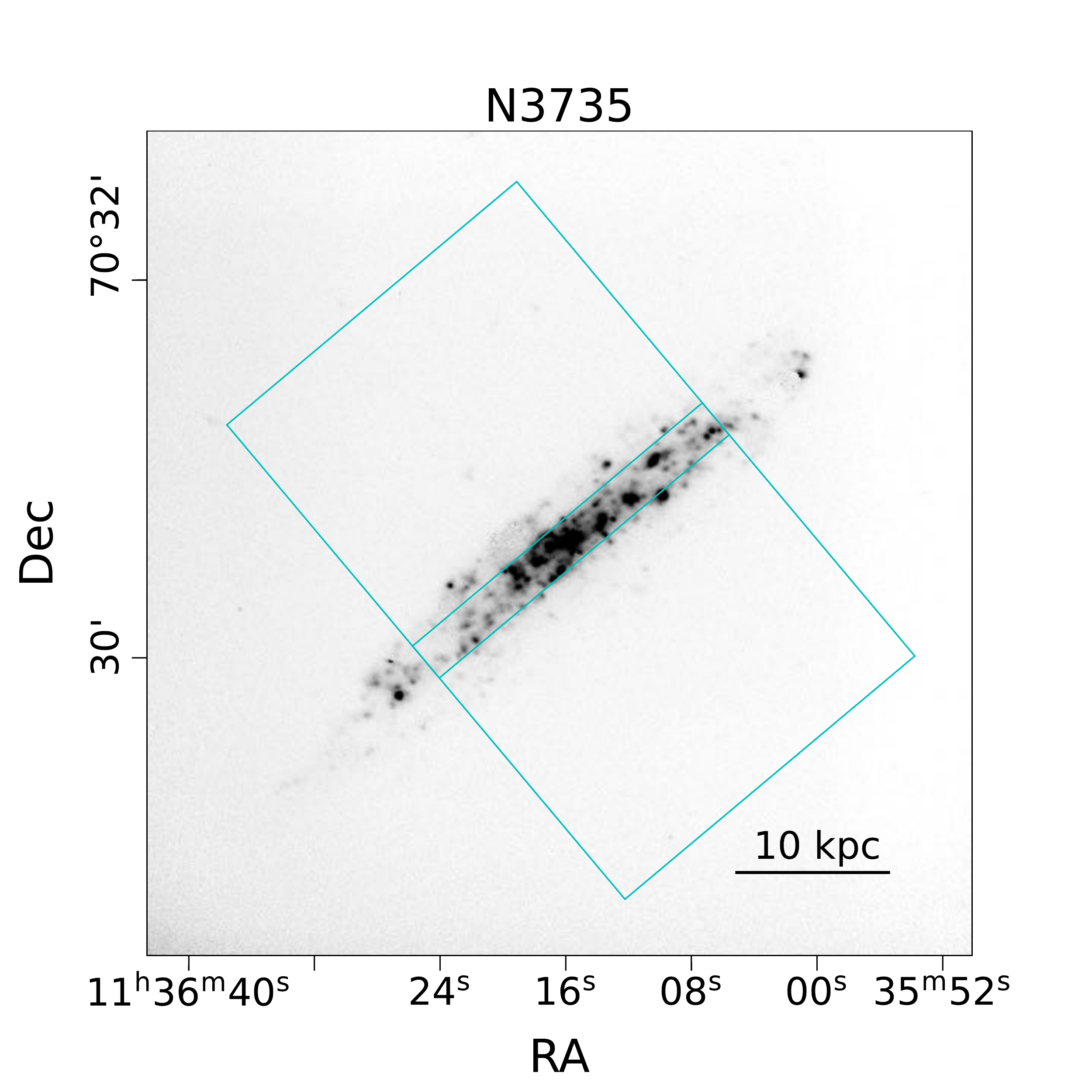}
        }
    \subfigure[]{
        \includegraphics[width=0.561\textwidth]{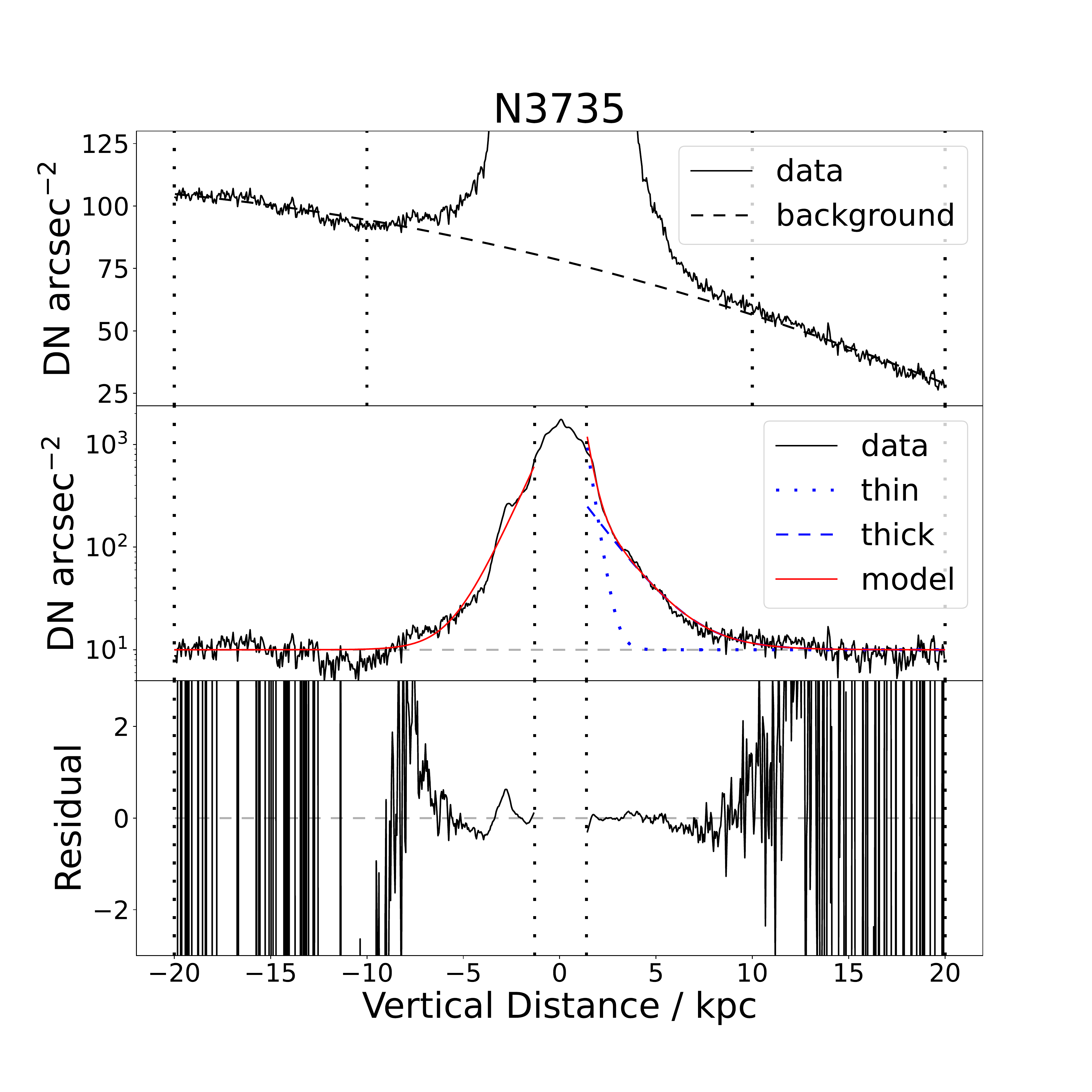}
        }
    \caption{
    }
\end{figure}
\begin{figure}[!h]
    \centering
    \subfigure[]{
        \includegraphics[width=0.401\textwidth]{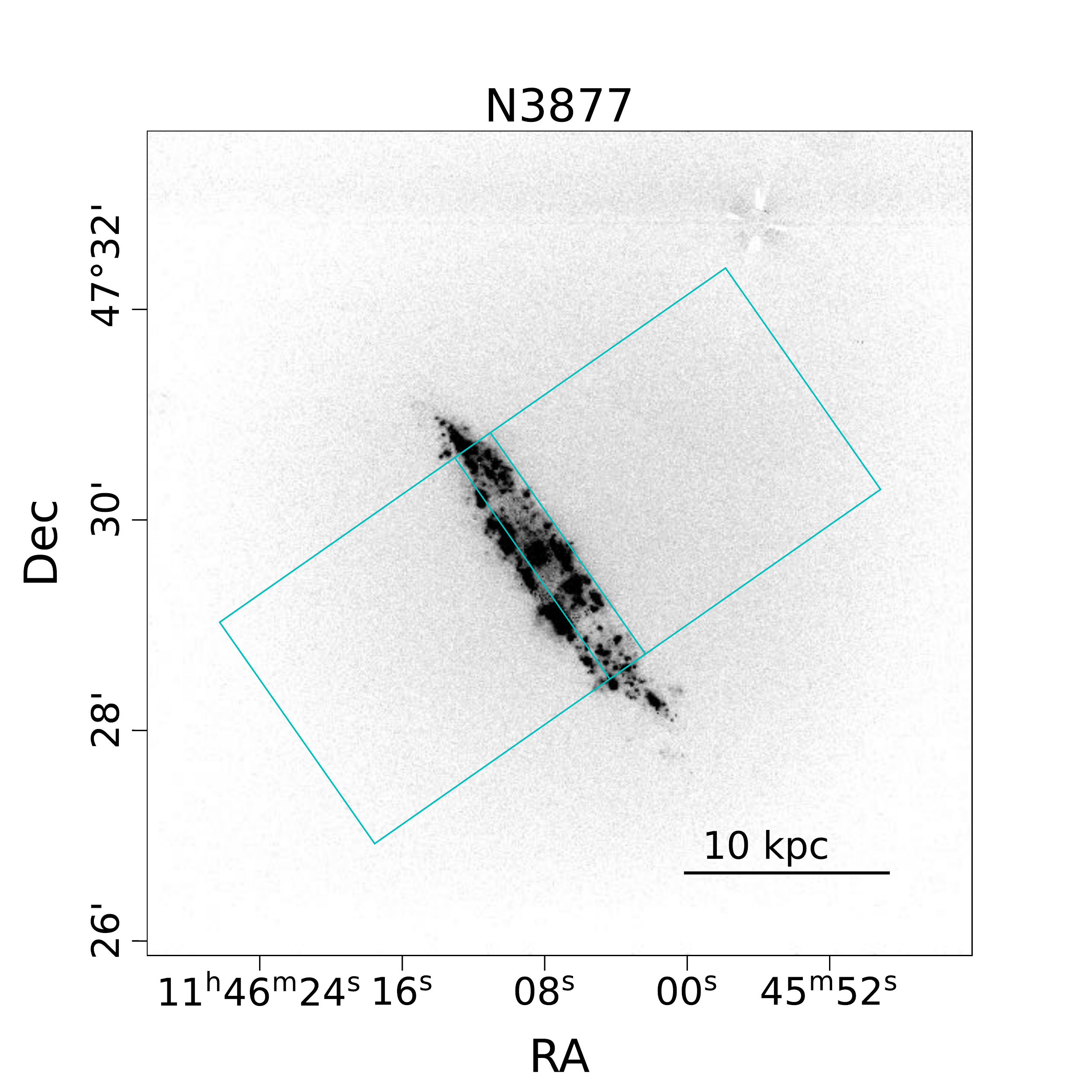}
        }
    \subfigure[]{
        \includegraphics[width=0.561\textwidth]{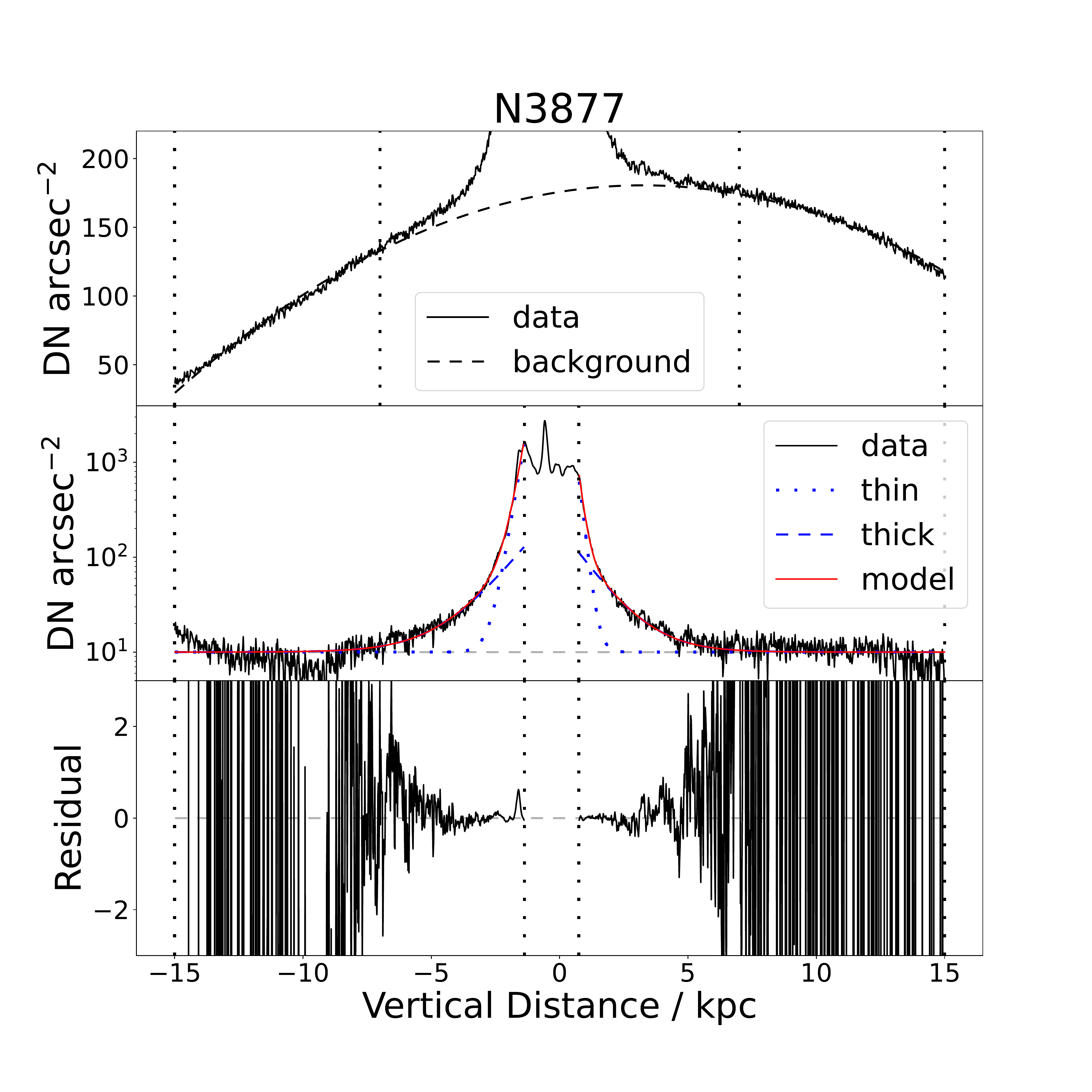}
        }
    \caption{
    }
\end{figure}
\begin{figure}[!h]
    \centering
    \subfigure[]{
        \includegraphics[width=0.401\textwidth]{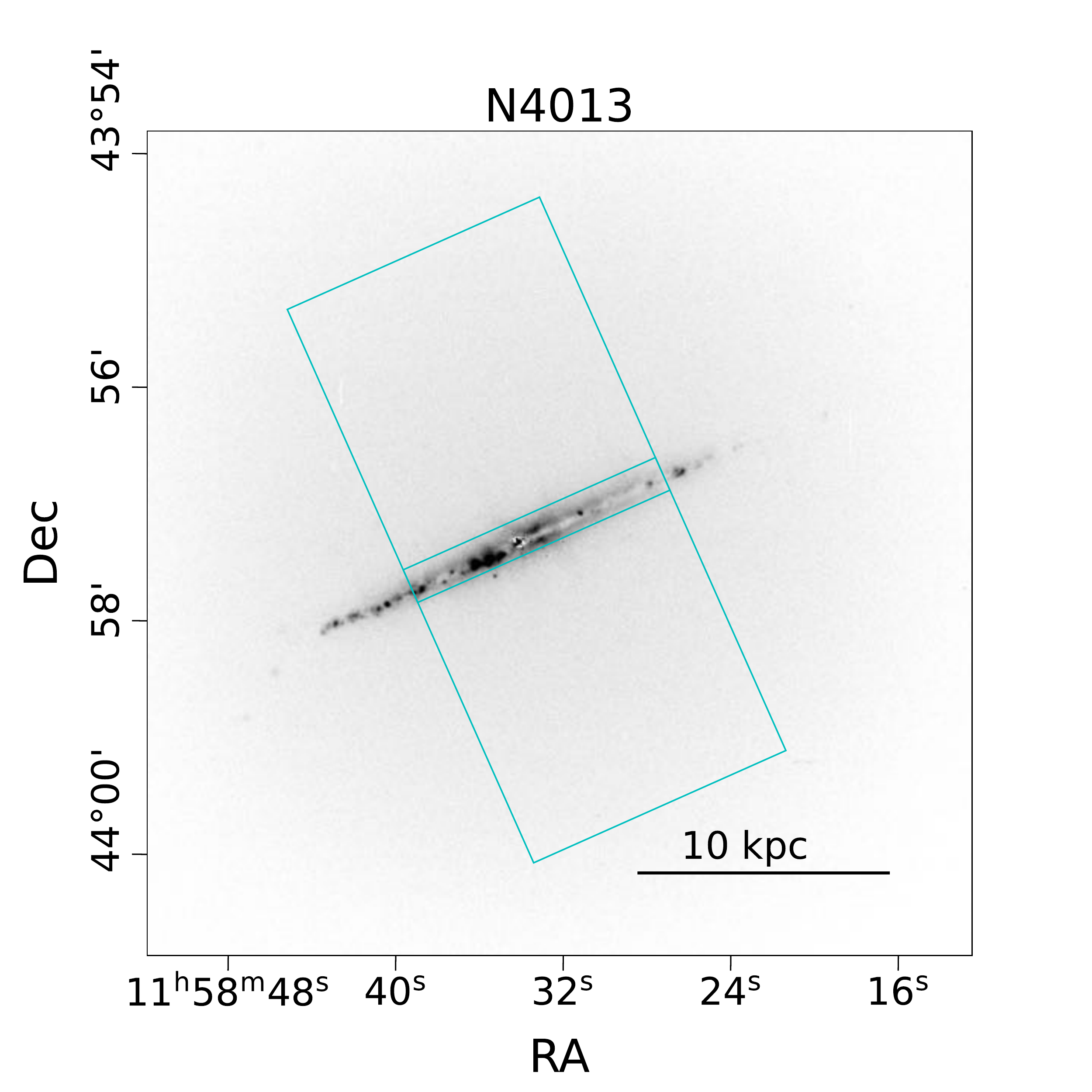}
        }
    \subfigure[]{
        \includegraphics[width=0.561\textwidth]{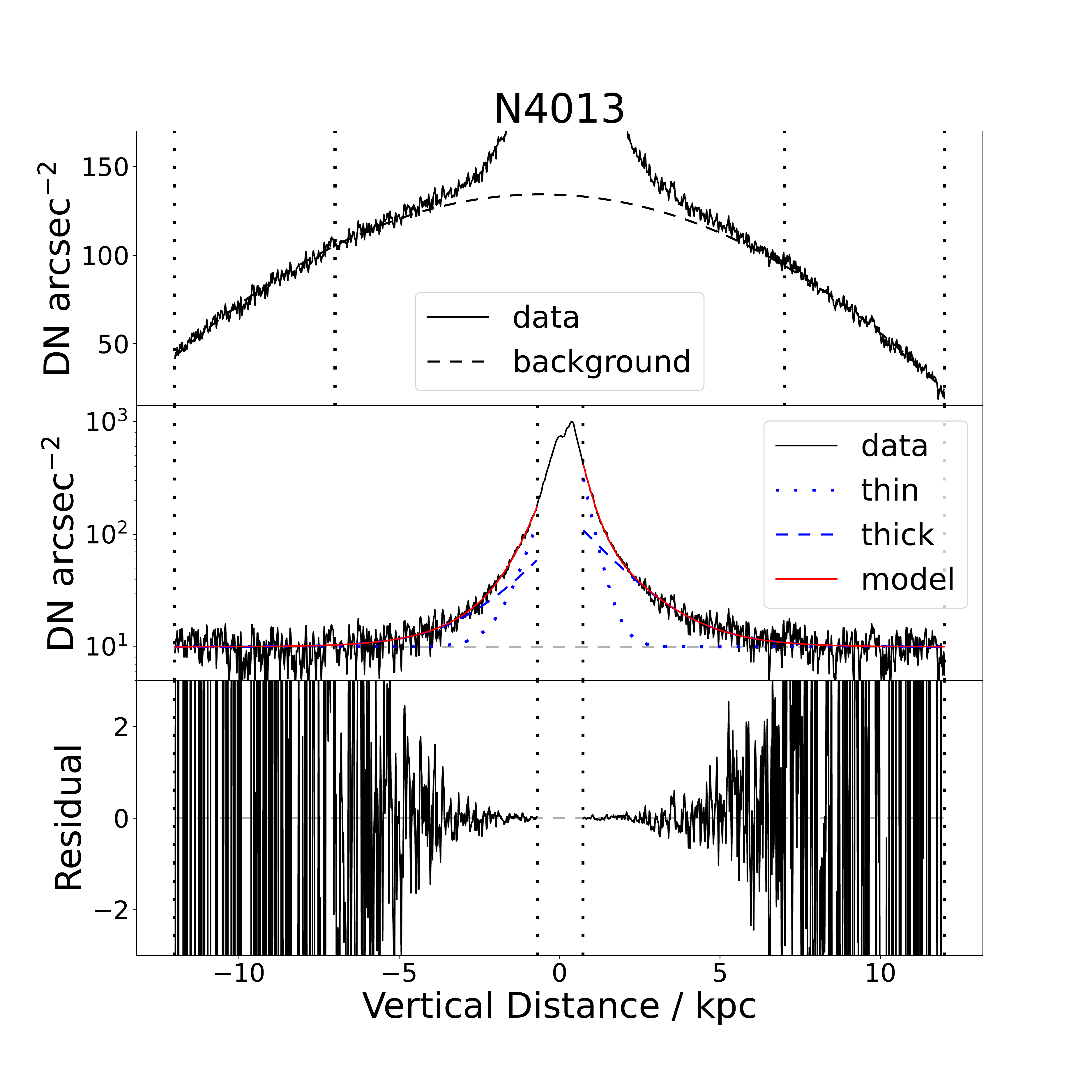}
        }
    \caption{
    }
\end{figure}
\begin{figure}[!h]
    \centering
    \subfigure[]{
        \includegraphics[width=0.401\textwidth]{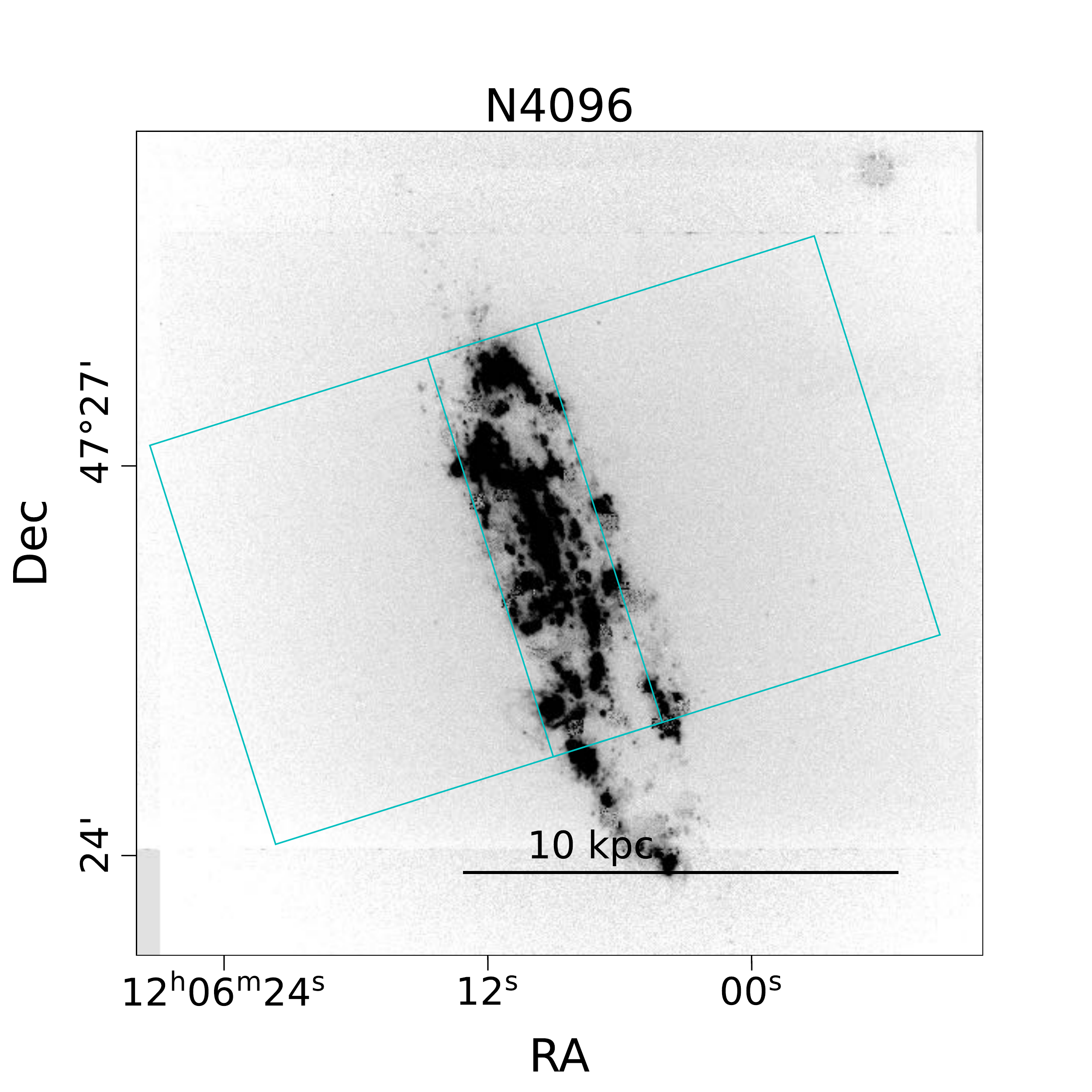}
        }
    \subfigure[]{
        \includegraphics[width=0.561\textwidth]{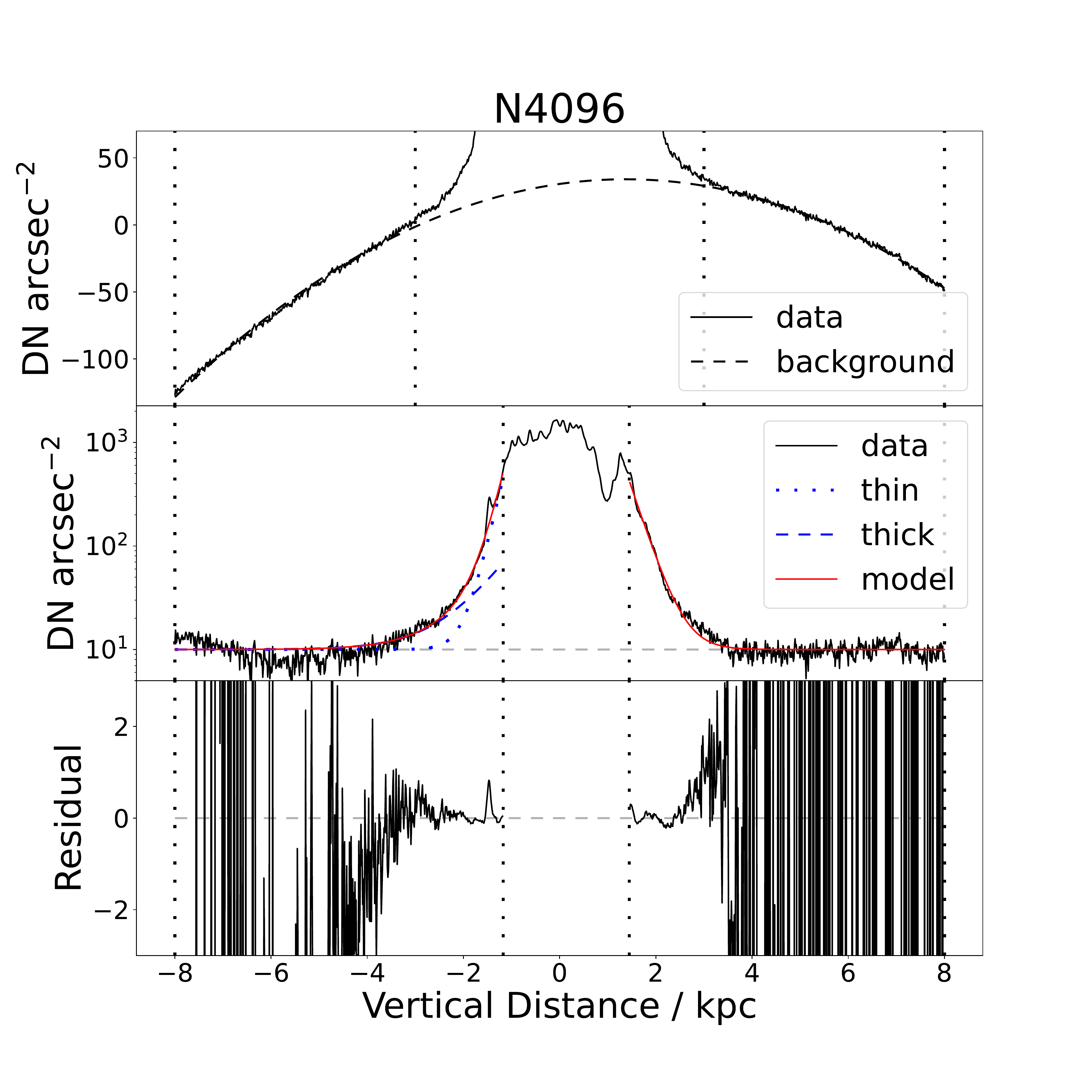}
        }
    \caption{
    }
\end{figure}
\begin{figure}[!h]
    \centering
    \subfigure[]{
        \includegraphics[width=0.401\textwidth]{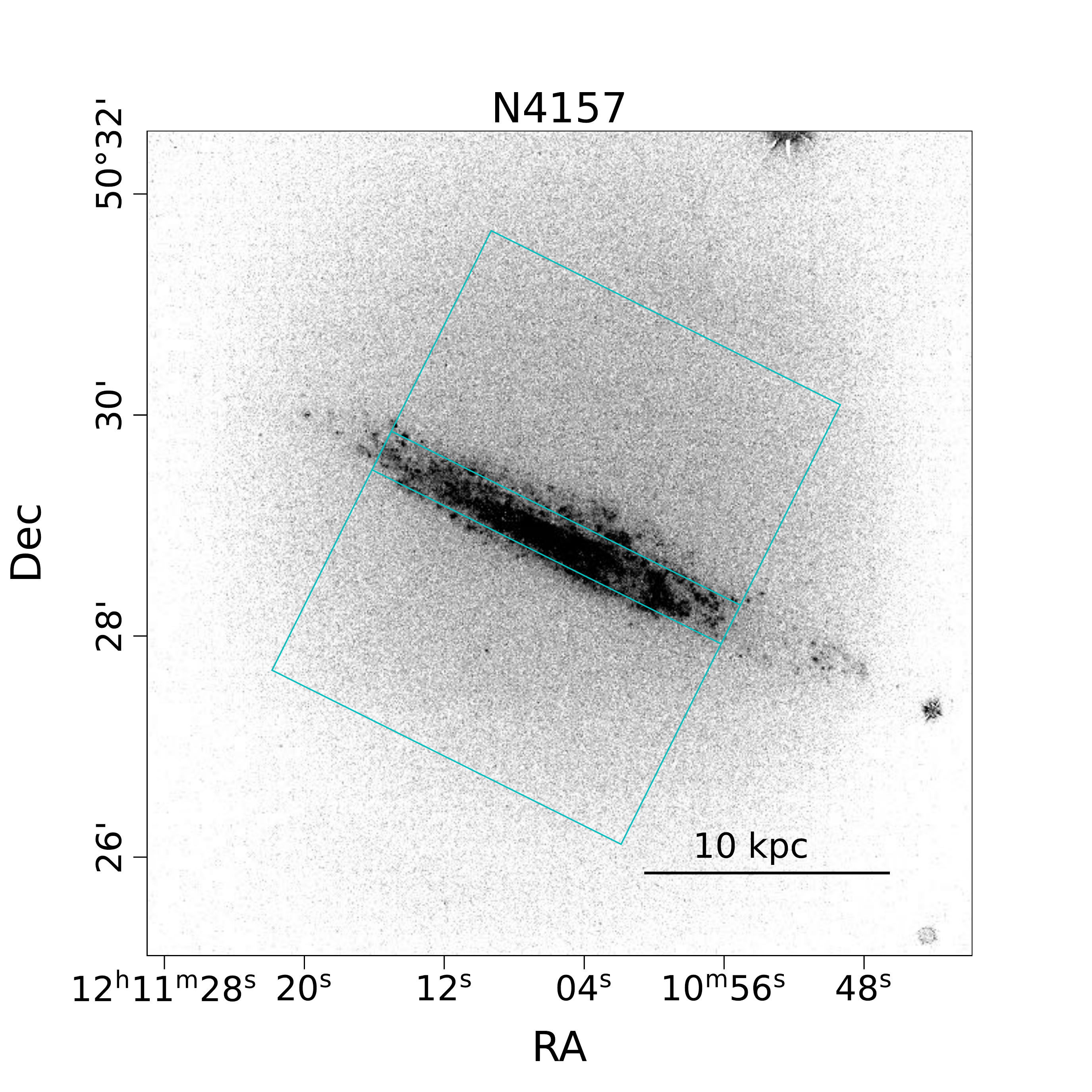}
        }
    \subfigure[]{
        \includegraphics[width=0.561\textwidth]{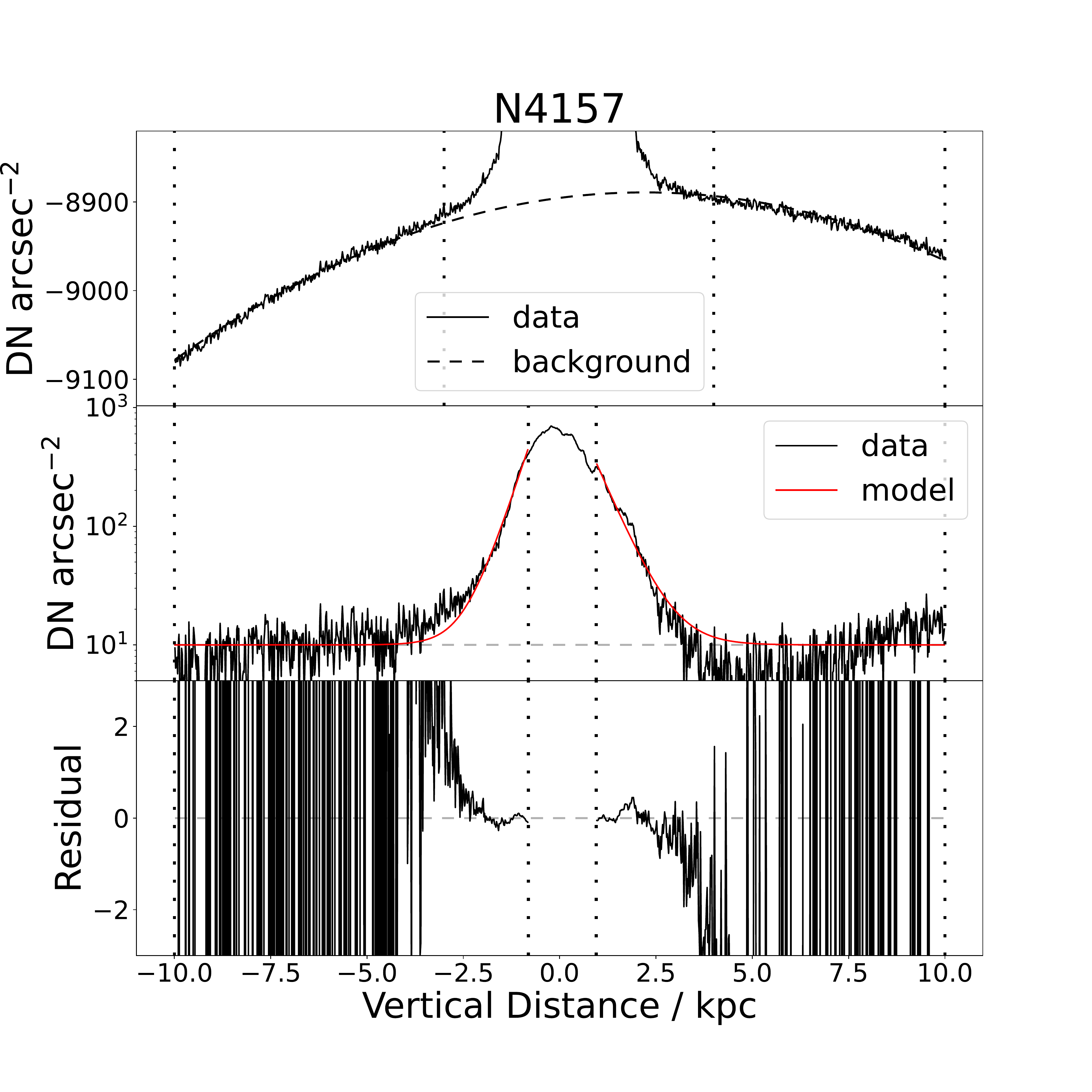}
        }
    \caption{
    }
\end{figure}
\begin{figure}[!h]
    \centering
    \subfigure[]{
        \includegraphics[width=0.401\textwidth]{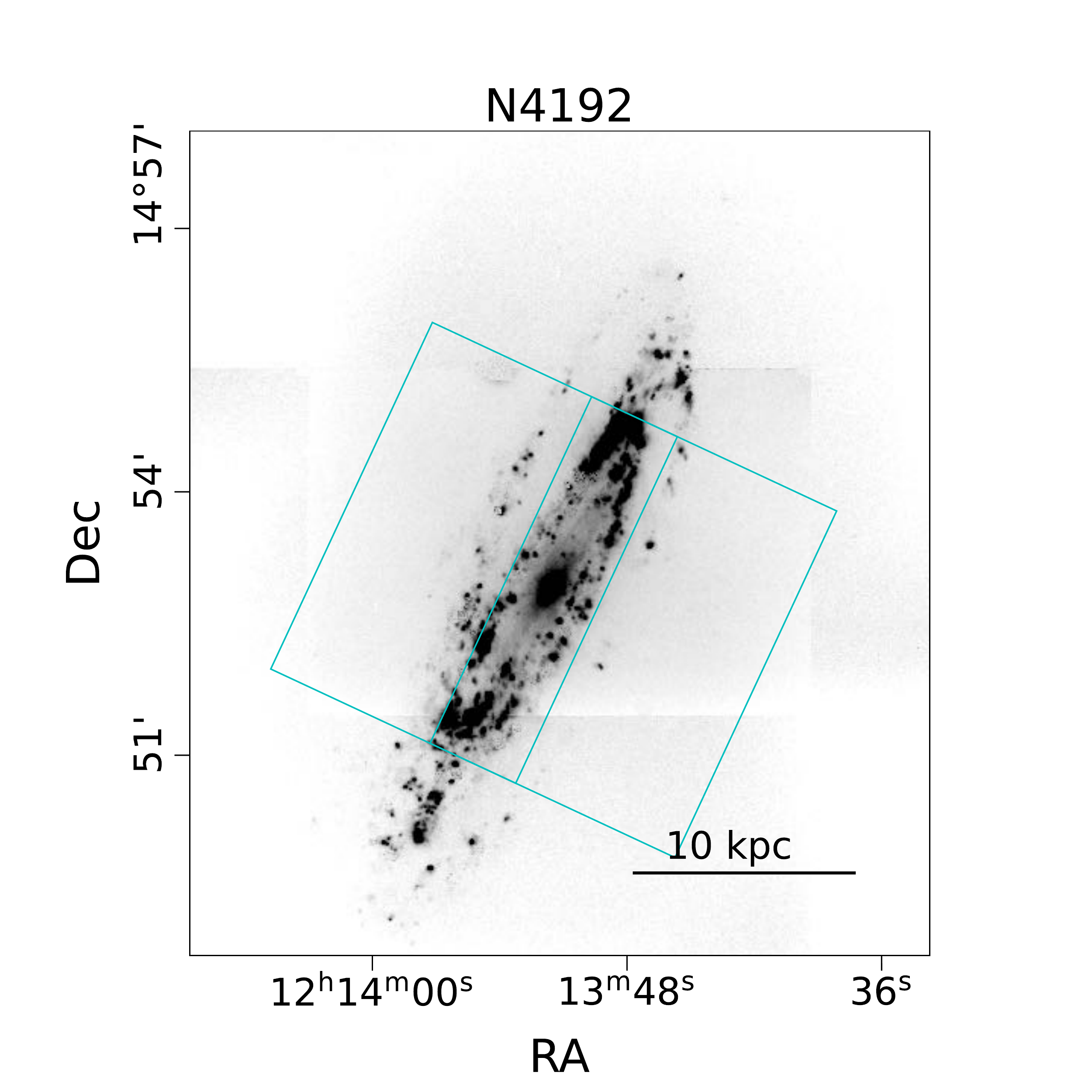}
        }
    \subfigure[]{
        \includegraphics[width=0.561\textwidth]{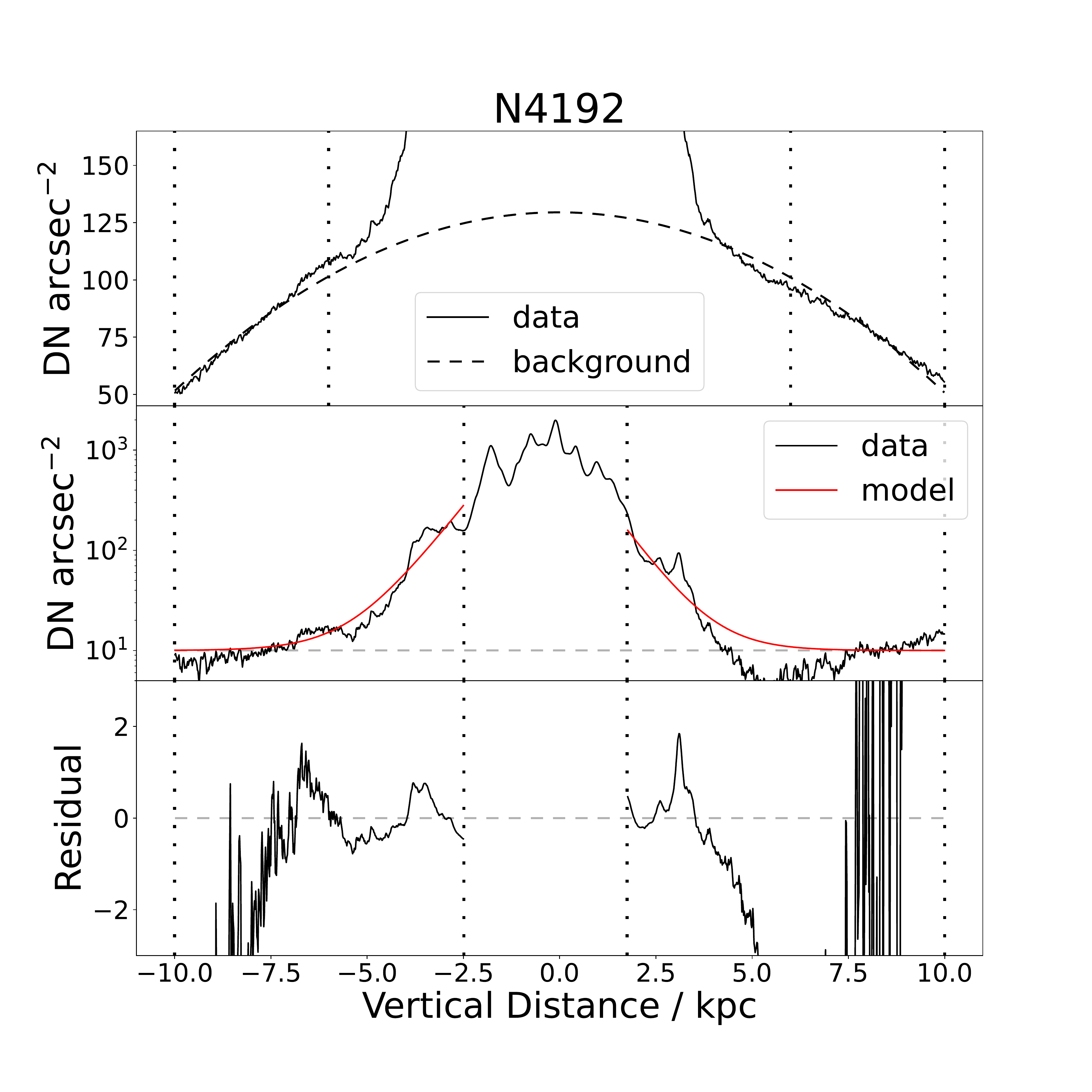}
        }
    \caption{
    }
\end{figure}
\begin{figure}[!h]
    \centering
    \subfigure[]{
        \includegraphics[width=0.401\textwidth]{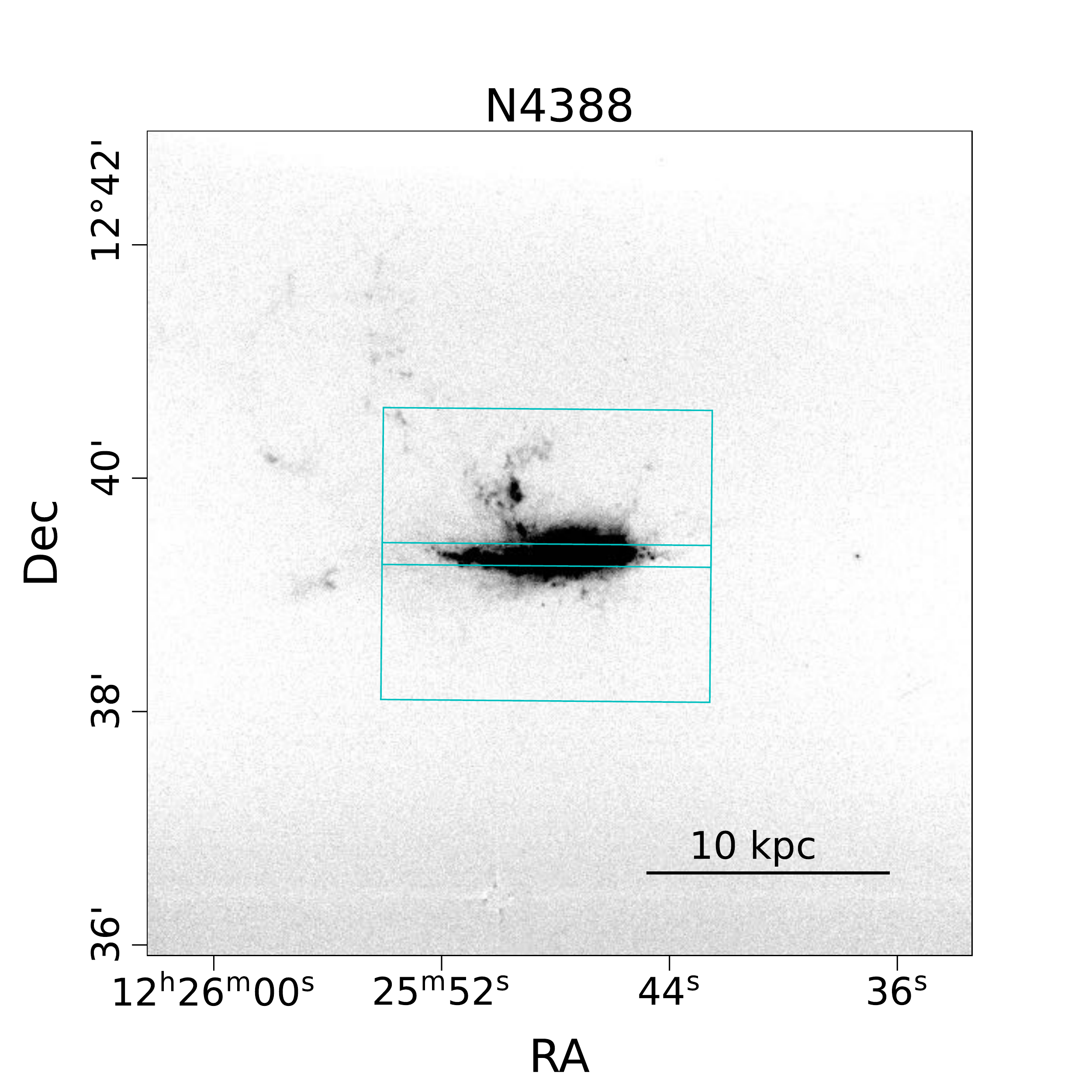}
        }
    \subfigure[]{
        \includegraphics[width=0.561\textwidth]{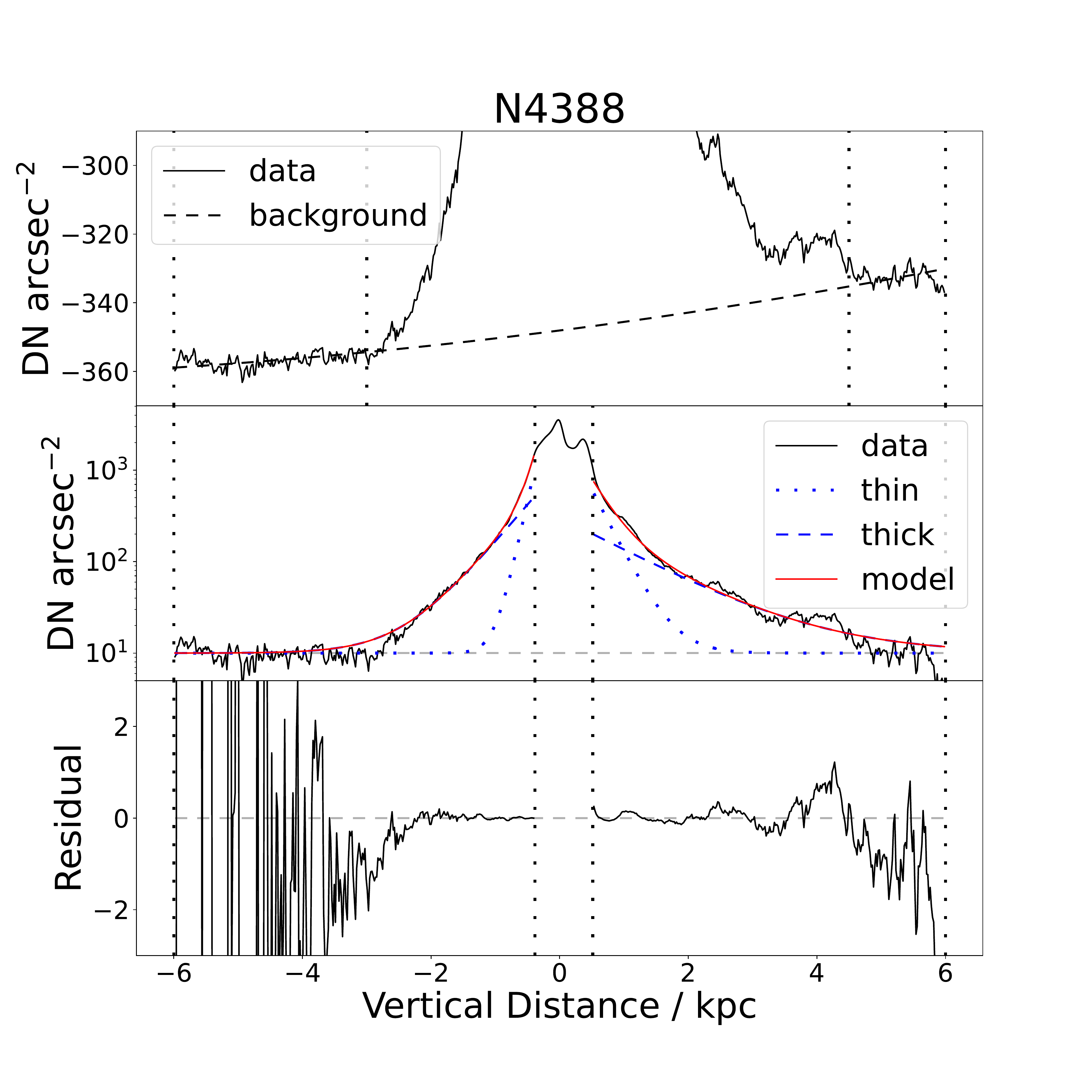}
        }
    \caption{
    }
\end{figure}
\begin{figure}[!h]
    \centering
    \subfigure[]{
        \includegraphics[width=0.401\textwidth]{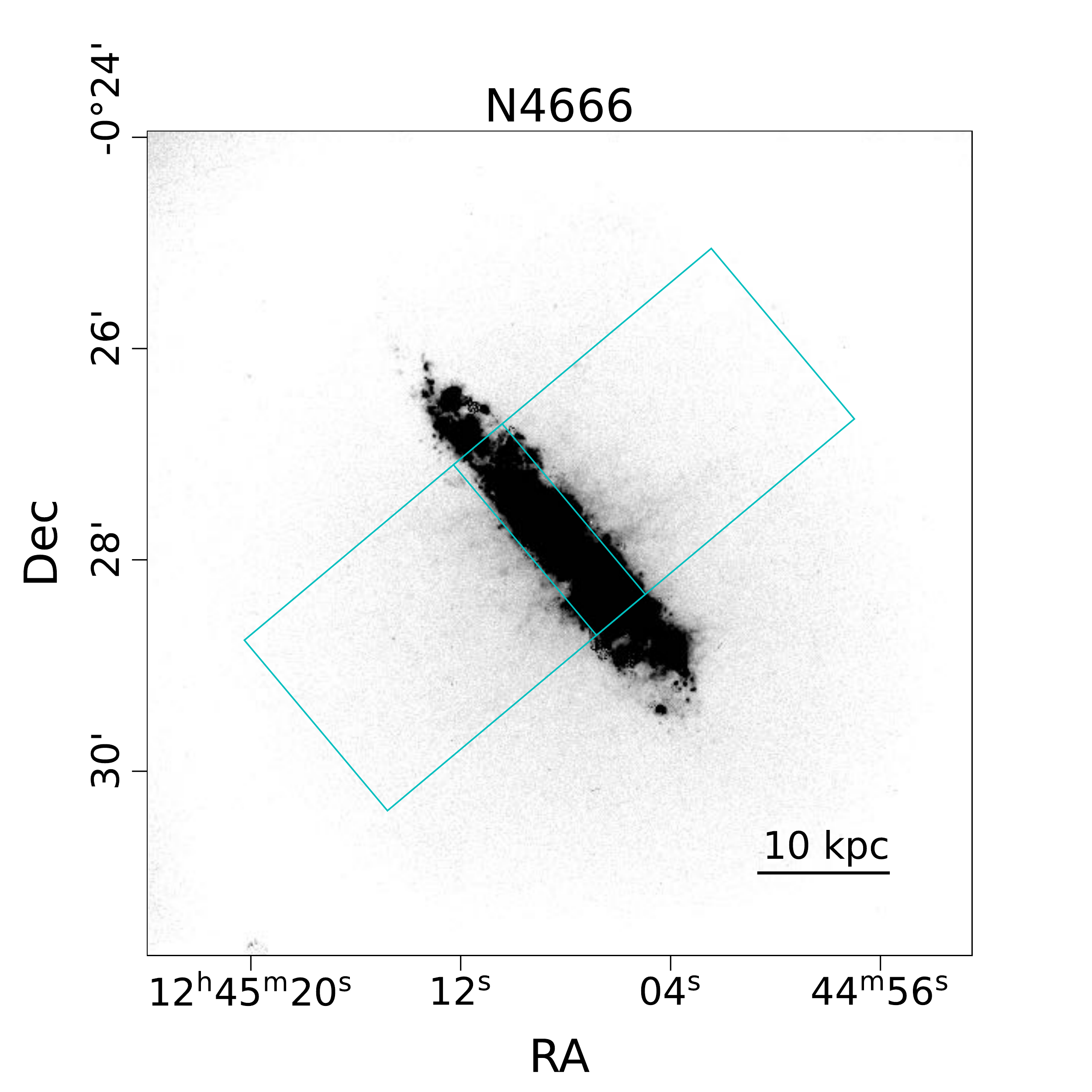}
        }
    \subfigure[]{
        \includegraphics[width=0.561\textwidth]{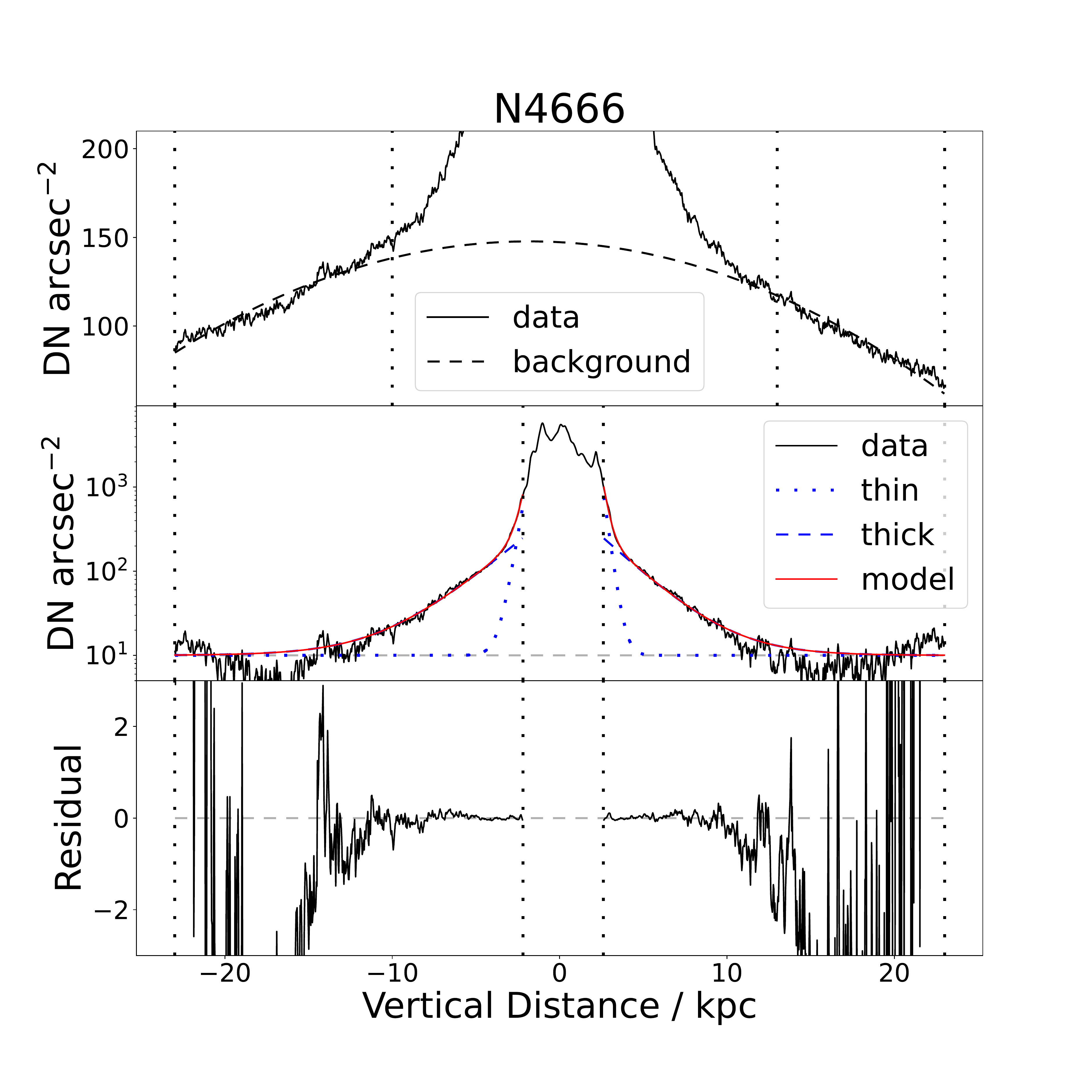}
        }
    \caption{
    }
\end{figure}
\begin{figure}[!h]
    \centering
    \subfigure[]{
        \includegraphics[width=0.401\textwidth]{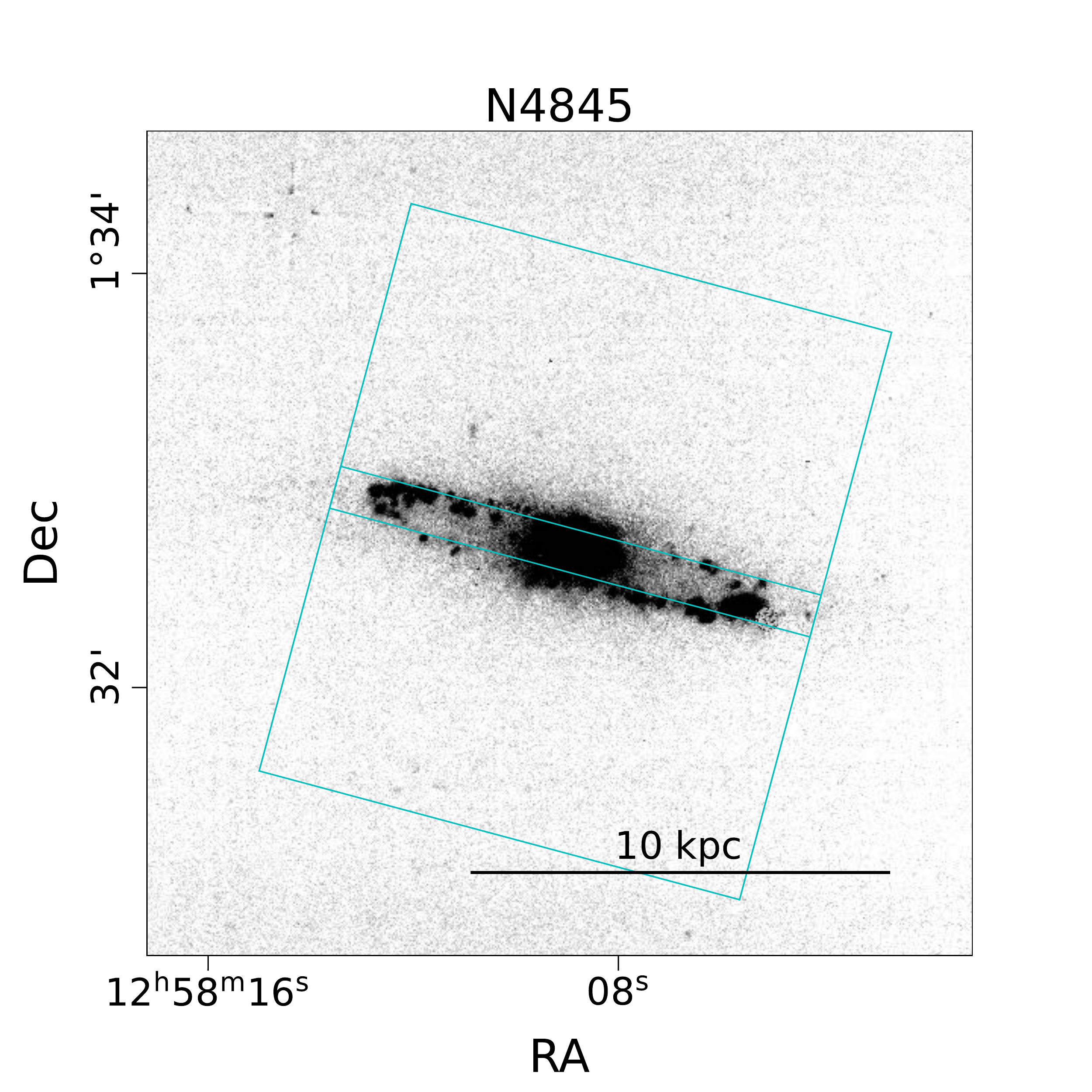}
        }
    \subfigure[]{
        \includegraphics[width=0.561\textwidth]{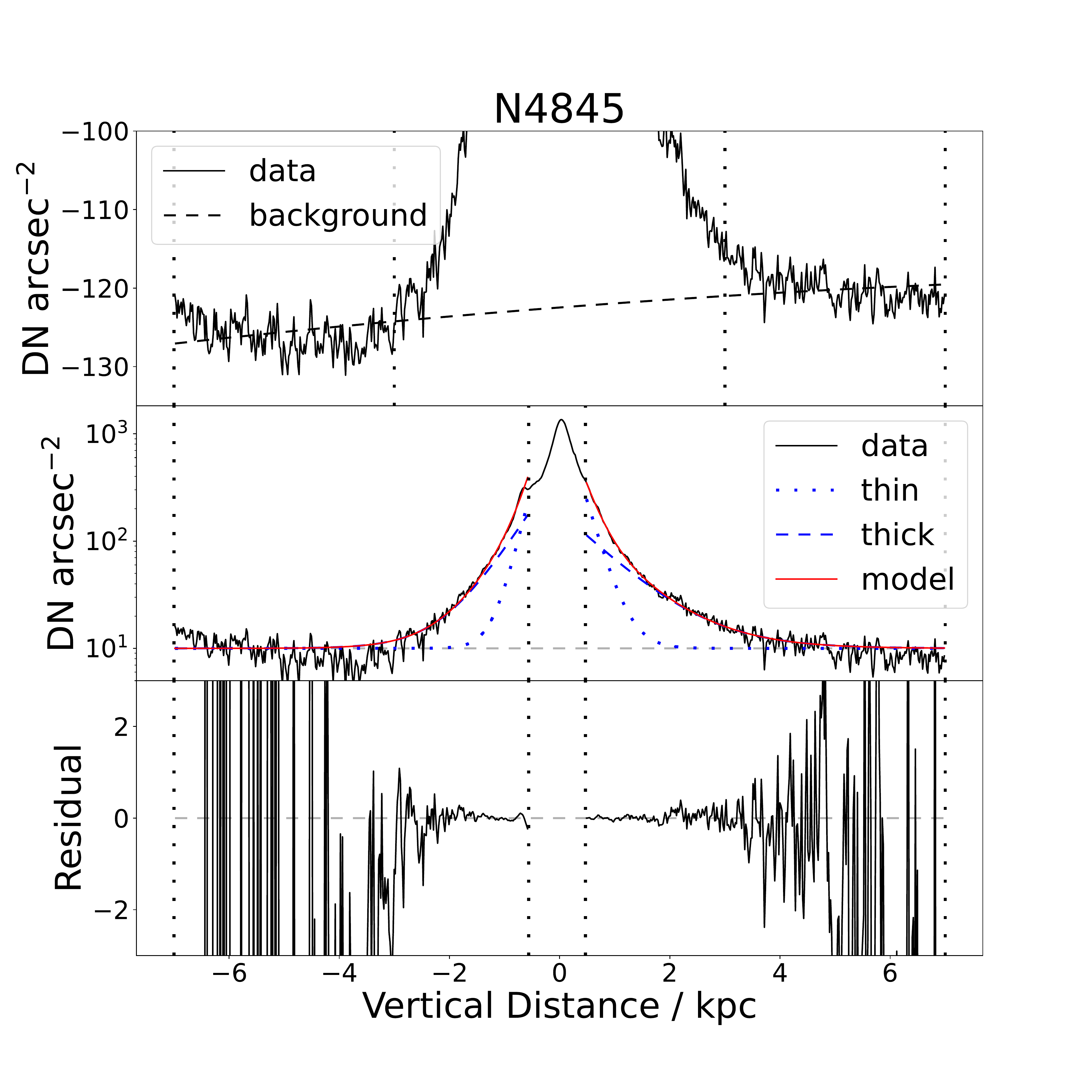}
        }
    \caption{
    }
\end{figure}
\begin{figure}[!h]
    \centering
    \subfigure[]{
        \includegraphics[width=0.401\textwidth]{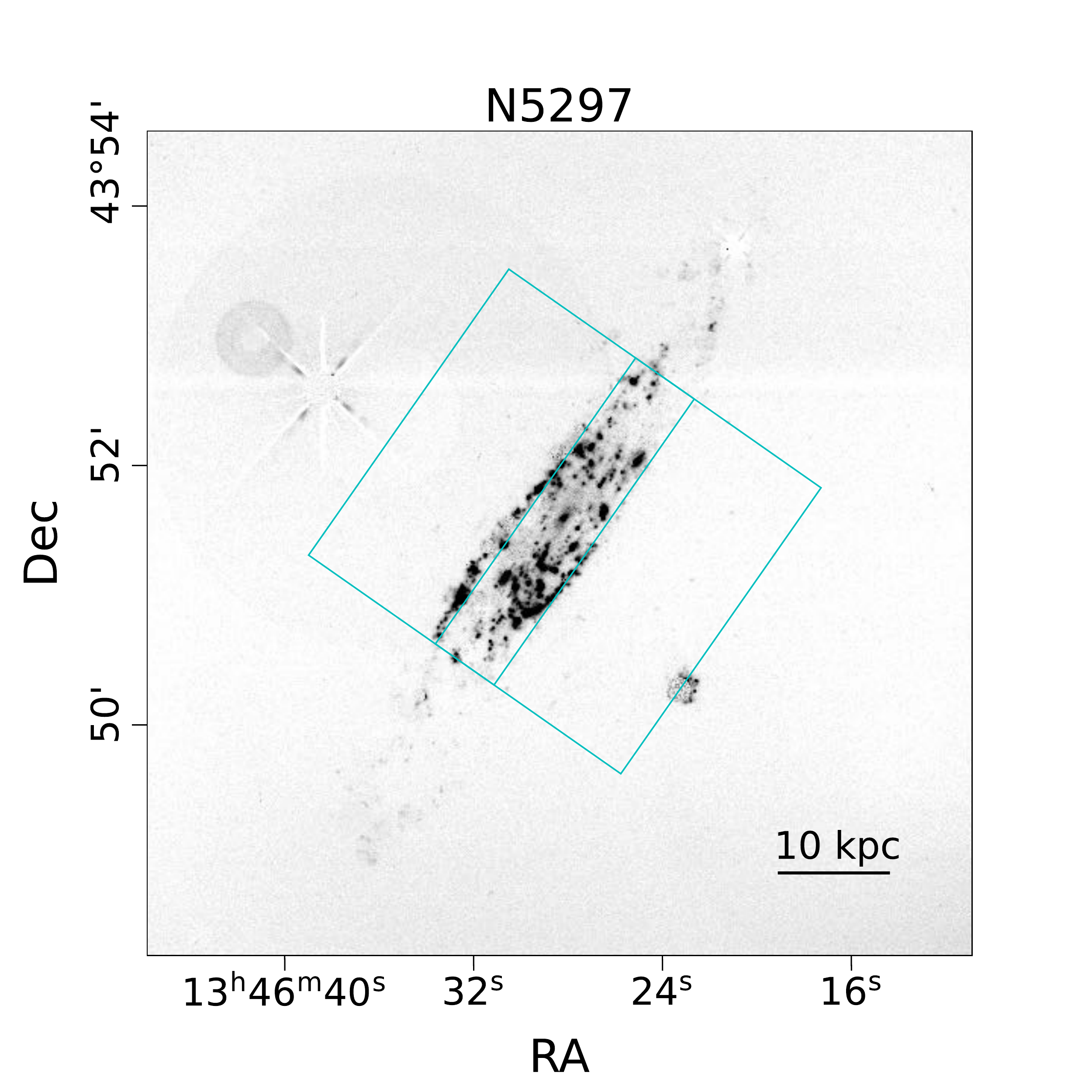}
        }
    \subfigure[]{
        \includegraphics[width=0.561\textwidth]{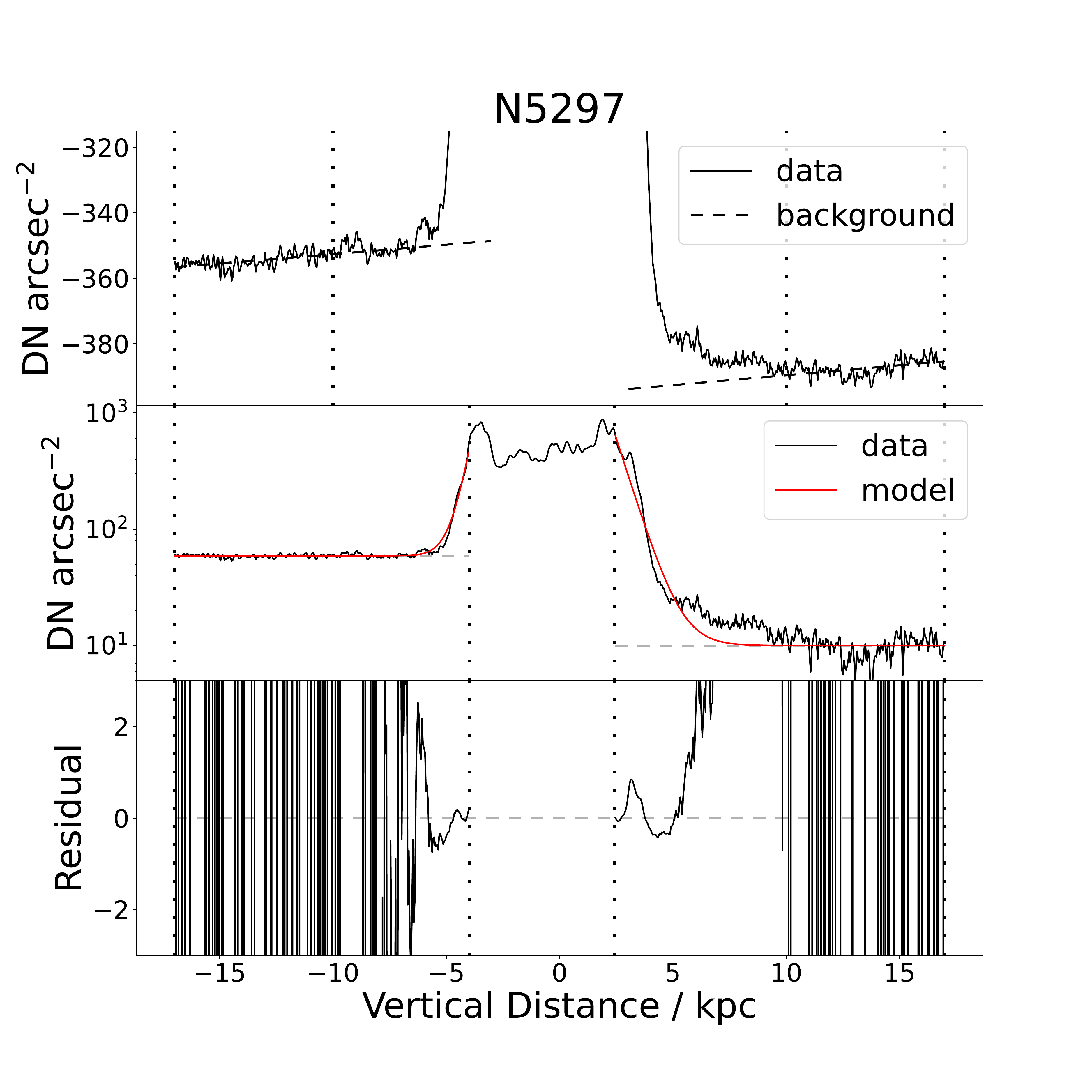}
        }
    \caption{The huge halo on the top-left of the image results in an offset on the 1-D background. Despite of the offset, the 1-D background is fitted by a same second-order polynomial model.
    }
\end{figure}
\begin{figure}[!h]
    \centering
    \subfigure[]{
        \includegraphics[width=0.401\textwidth]{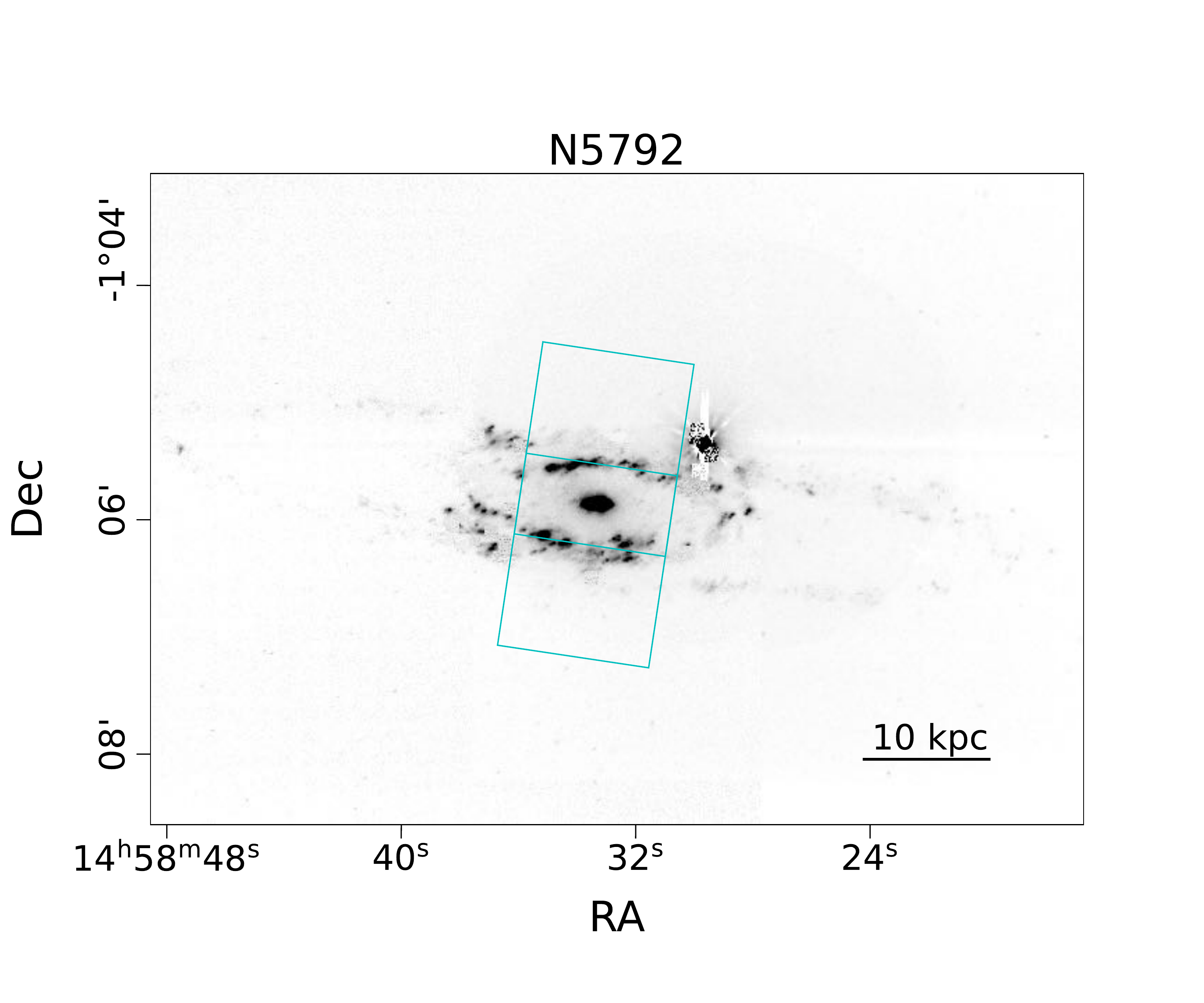}
        }
    \subfigure[]{
        \includegraphics[width=0.561\textwidth]{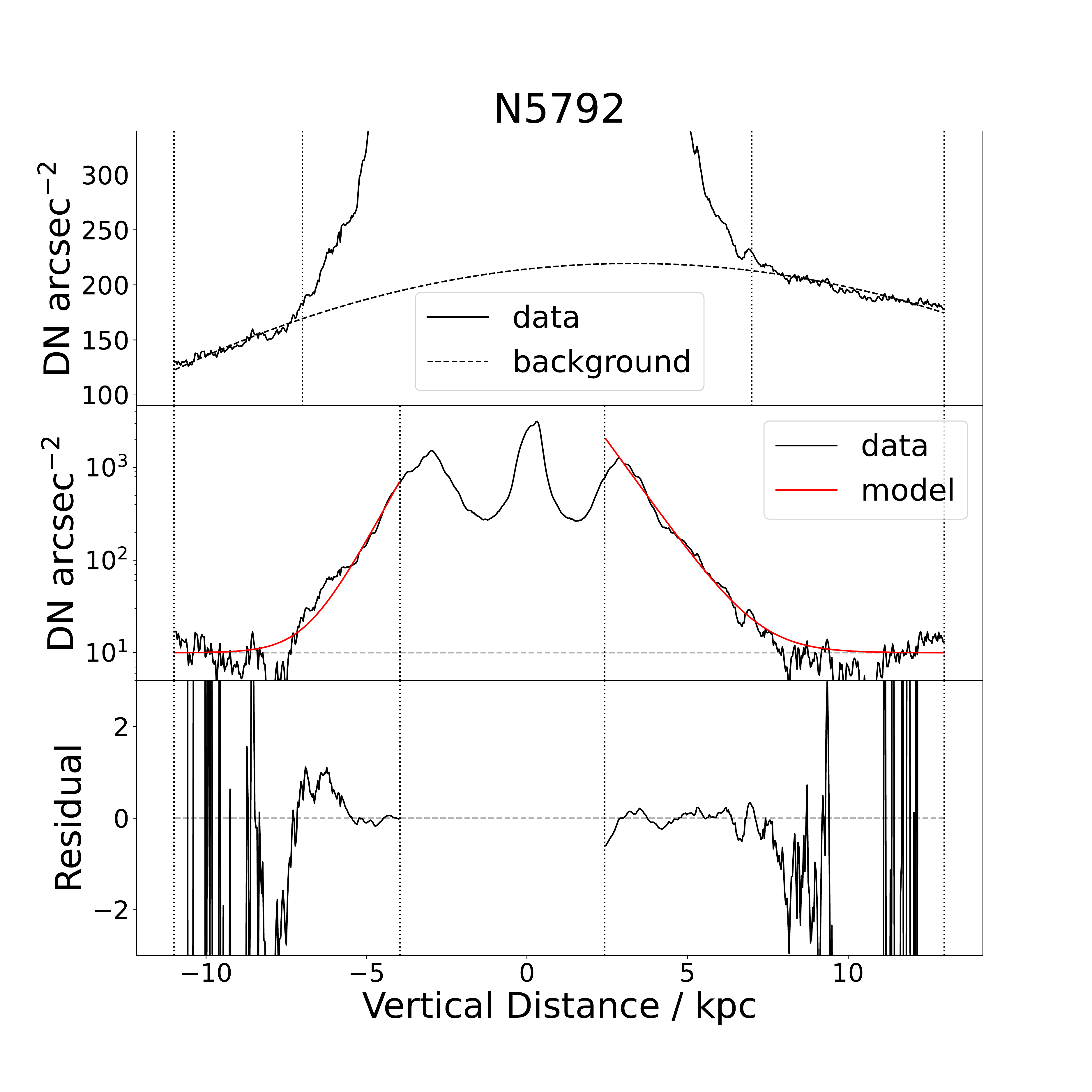}
        }
    \caption{
    }
\end{figure}
\begin{figure}[!h]
    \centering
    \subfigure[]{
        \includegraphics[width=0.401\textwidth]{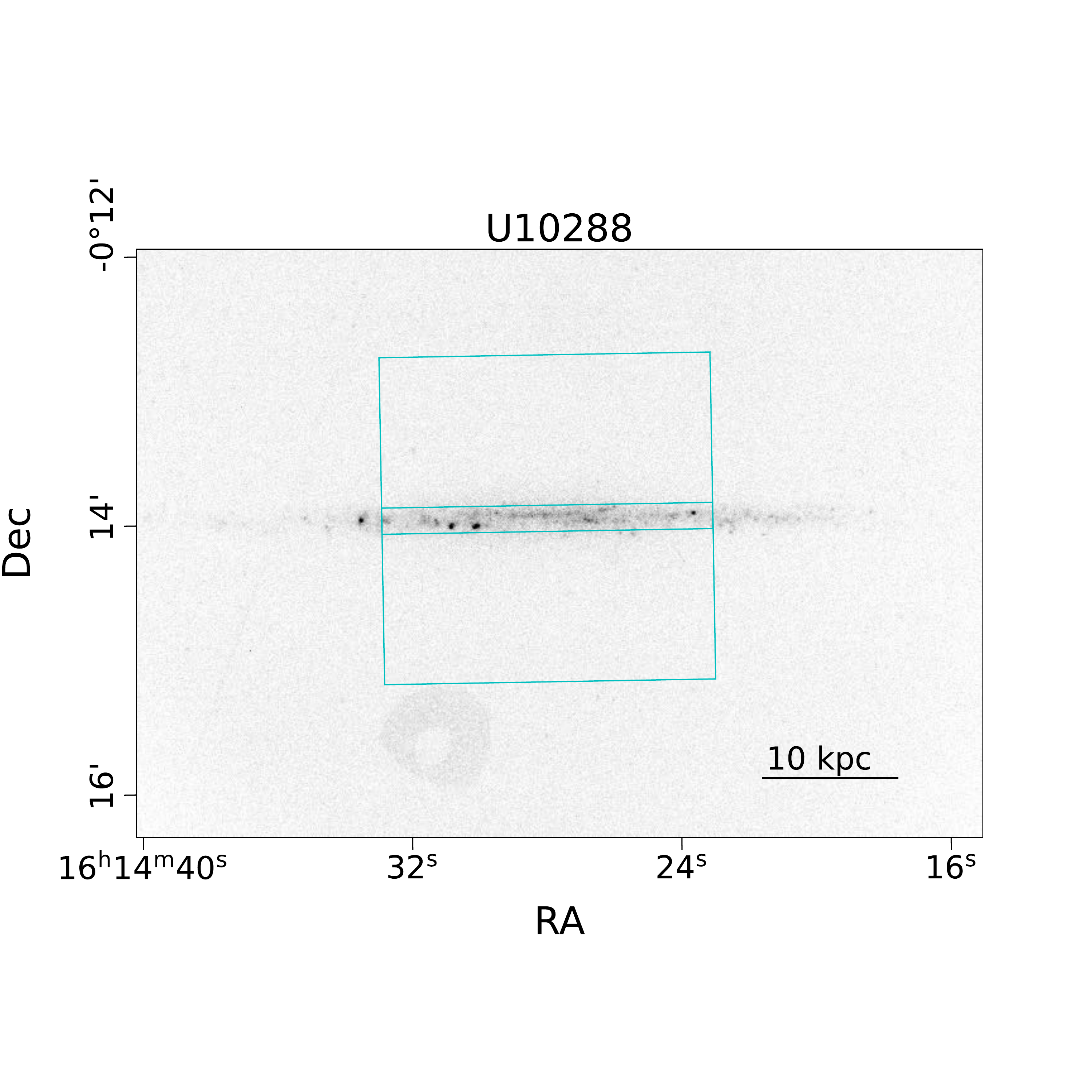}
        }
    \subfigure[]{
        \includegraphics[width=0.561\textwidth]{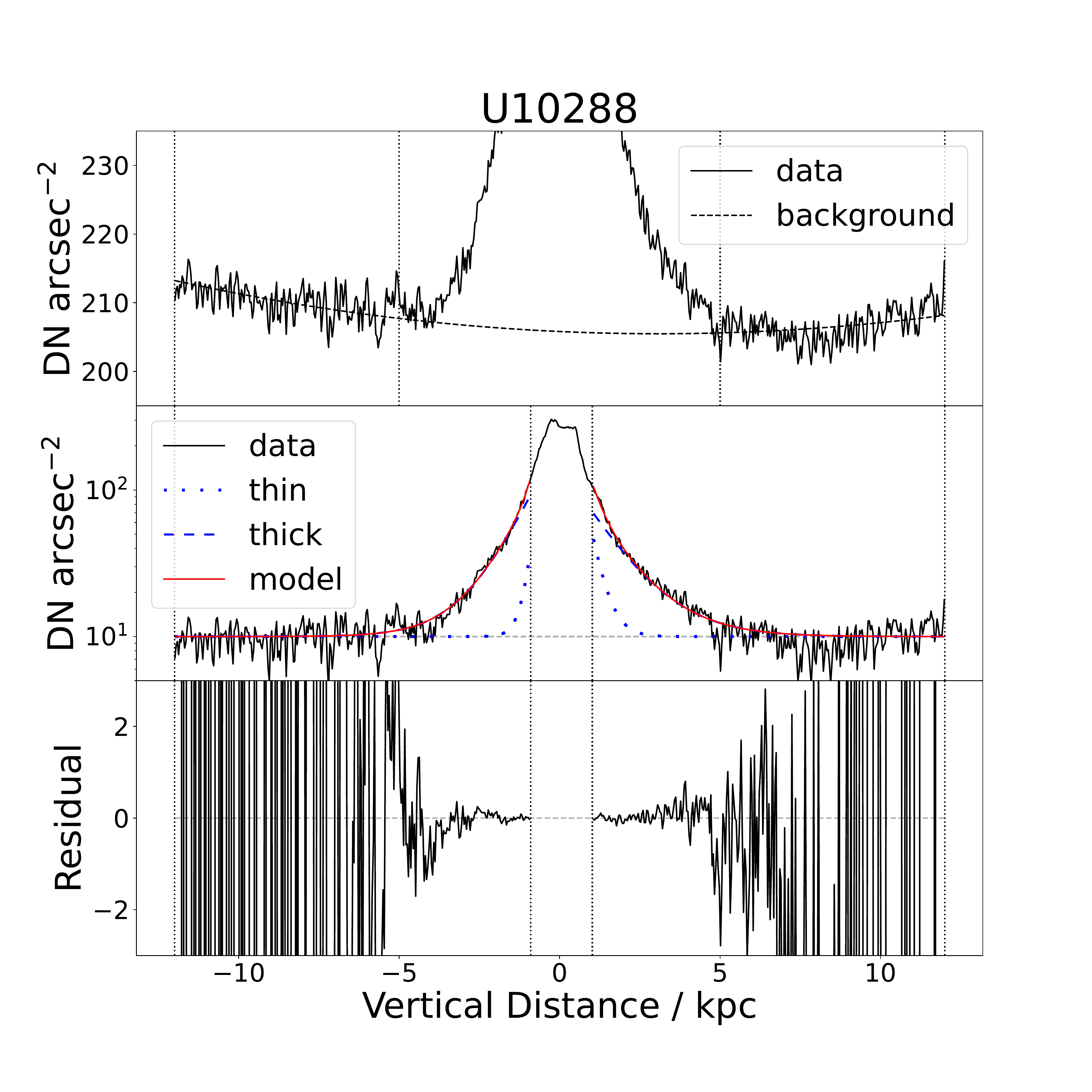}
        }
    \caption{
    }
\end{figure}
\end{appendix}

\end{CJK*}
\end{document}